\documentclass{article}

\PassOptionsToPackage{numbers, compress}{natbib}
\usepackage[nonatbib, final]{neurips_2023}

\usepackage[utf8]{inputenc}
\usepackage[T1]{fontenc}
\usepackage{xurl}
\usepackage[colorlinks, allcolors=blue]{hyperref}
\usepackage{booktabs}
\usepackage{amsfonts}
\usepackage{nicefrac}
\usepackage{microtype}
\usepackage{xcolor}
\usepackage{graphicx}
\usepackage{tabularx}
\usepackage{enumitem}
\usepackage{multirow}
\usepackage{array}
\usepackage{apacite}
\usepackage{orcidlink}

\title{Risk assessment at AGI companies: \\
A review of popular risk assessment techniques \\
from other safety-critical industries}

\author{
  Leonie Koessler\thanks{Corresponding author: \url{leonie.koessler@kcl.ac.uk}. Leonie Koessler worked on the project as part of the 2023 GovAI Winter Research Fellowship.} \hspace{0.2mm} \orcidlink{0009-0006-6863-5477} \\
  Centre for the Governance of AI \\
  \And
  Jonas Schuett \orcidlink{0000-0001-7154-5049} \\
  Centre for the Governance of AI
}

\begin{document}

\maketitle

\begin{abstract}
Companies like OpenAI, Google DeepMind, and Anthropic have the stated goal of building artificial general intelligence (AGI) – AI systems that perform as well as or better than humans on a wide variety of cognitive tasks. However, there are increasing concerns that AGI would pose catastrophic risks. In light of this, AGI companies need to drastically improve their risk management practices. To support such efforts, this paper reviews popular risk assessment techniques from other safety-critical industries and suggests ways in which AGI companies could use them to assess catastrophic risks from AI. The paper discusses three risk identification techniques (scenario analysis, fishbone method, and risk typologies and taxonomies), five risk analysis techniques (causal mapping, Delphi technique, cross-impact analysis, bow tie analysis, and system-theoretic process analysis), and two risk evaluation techniques (checklists and risk matrices). For each of them, the paper explains how they work, suggests ways in which AGI companies could use them, discusses their benefits and limitations, and makes recommendations. Finally, the paper discusses when to conduct risk assessments, when to use which technique, and how to use any of them. The reviewed techniques will be obvious to risk management professionals in other industries. And they will not be sufficient to assess catastrophic risks from AI. However, AGI companies should not skip the straightforward step of reviewing best practices from other industries.
\end{abstract}
\vspace{-1em}

\begin{figure}[ht!]
    \centering
    \includegraphics[width=\textwidth]{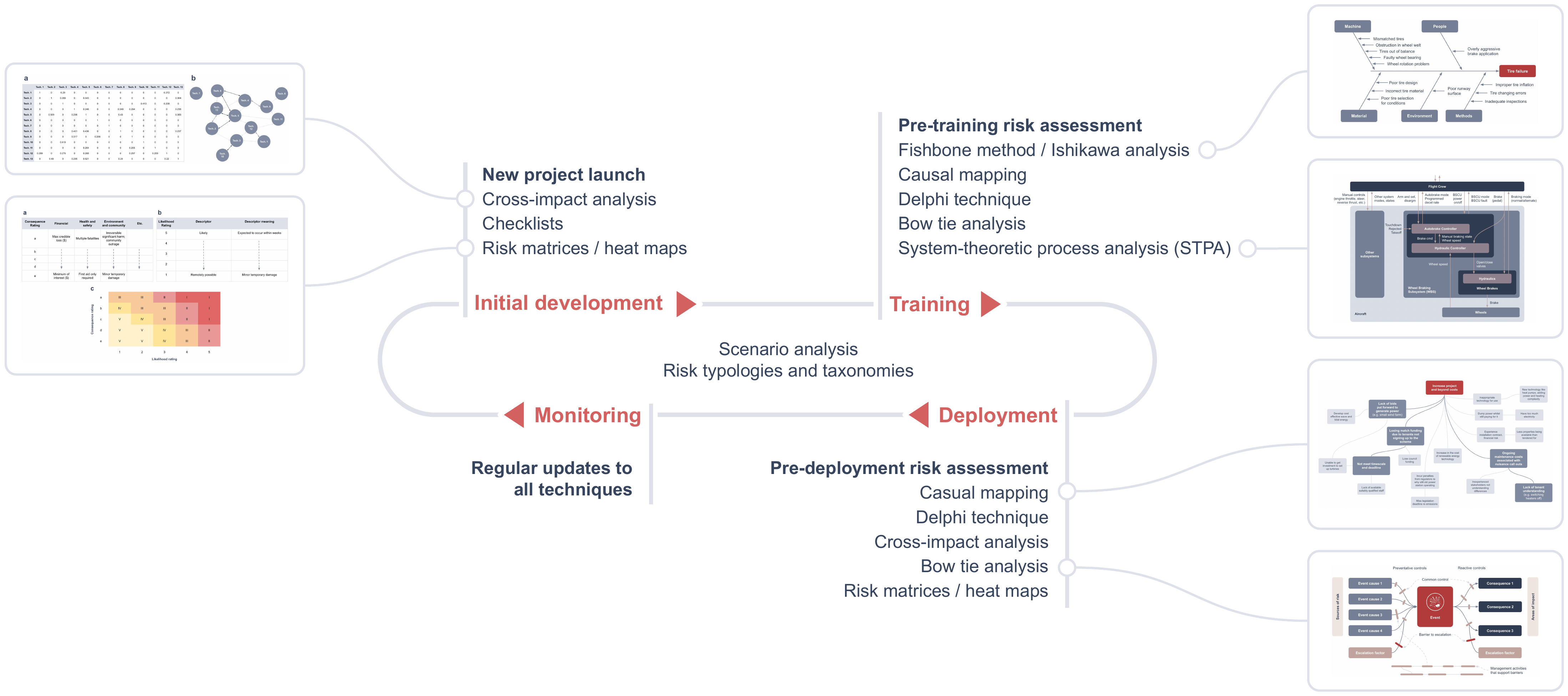}
    \caption{Exemplary use of risk assessment techniques across the AI system lifecycle}
    \label{fig1}
\end{figure}

\section*{Executive summary}

Future AI systems may pose catastrophic risks. Companies that develop and deploy such systems must take adequate measures to manage these and other risks. This paper reviews popular risk assessment techniques from other industries – namely, finance, aviation, nuclear, and biolabs – and discusses how they can be used to assess catastrophic risks from AI. Although some of these techniques may be helpful, they are by no means sufficient. In particular, AI companies should also use techniques like model evaluations for dangerous capabilities and propensities.

\textbf{Selection criteria.} To identify techniques, we used IEC 31010:2019, a leading risk assessment standard, as a starting point and added popular techniques from other industries based on a literature review. We then defined criteria for excluding and prioritizing techniques based on the particularities of catastrophic risks from AI. First, we only included techniques that are applicable to societal risks, not just business risks. Second, we focused on techniques that are able to account for low-probability, high-impact events. Third, we excluded techniques that aim to ensure that humans perform reliably on routine tasks (which, however, might become more relevant in the future). Fourth, we also excluded techniques that were originally developed to assess hardware reliability (which might also become more relevant in the future). Finally, we prioritized techniques that can deal with complex interactions between events, help combine the viewpoints of a variety of people, and provide clarity on future developments. [\hyperref[3]{more}]

\textbf{Selected techniques.} Based on these criteria, we selected three risk identification techniques, five risk analysis techniques, and two risk evaluation techniques.

\begin{itemize}[leftmargin=2em]
    \item \textbf{Risk identification} aims to identify risks and their sources. To that end, \emph{scenario analysis} uses forward reasoning to develop future scenarios which are then examined for the risks they entail. By contrast, the \emph{fishbone method} uses backward reasoning from a risk to its sources. \emph{Risk typologies and taxonomies} structure the risk universe and can identify additional risks. [\hyperref[4]{more}]
    
    \item \textbf{Risk analysis} aims to facilitate a deep understanding of the causes, consequences, and likelihood of risks. \emph{Causal mapping} helps to establish causal relationships between events associated with risks. The \emph{Delphi technique} collates expert forecasts to assess the likelihood of events or scenarios. \emph{Cross-impact analysis} combines the functions of the previous two techniques by analyzing expert forecasts on the correlations between events. Both \emph{bow tie analysis} and \emph{system-theoretic process analysis (STPA)} focus on controls, i.e. mechanisms supposed to impede risks from materializing. [\hyperref[5]{more}]
    
    \item \textbf{Risk evaluation} aims to establish whether a risk is acceptable or whether its treatment is warranted. \emph{Checklists} are useful to decentralize risk evaluation for routine decisions that may add up and increase risks, while \emph{risk matrices} are overall frameworks for deciding on the necessity of treating risks based on consequence and likelihood or vulnerability. [\hyperref[6]{more}]
\end{itemize}

\textbf{When and how to conduct risk assessments.} AGI companies should continuously and iteratively assess risks. Pre-deployment risk assessments are particularly important, but AGI companies should also conduct pre-training risk assessments. There is no specific trigger for each of the techniques discussed in this paper. Instead, they can all be applied in various situations, each technique with a different focus or function. In a given situation, AGI companies should avoid relying on a single technique but use several techniques to gain a more complete understanding of risks. Finally, AGI companies need to set up structures and processes to ensure that the results of risk assessments actually have a bearing on decisions. [\hyperref[7]{more}]
\newpage

\renewcommand{\arraystretch}{1.5}
\begin{table}[t!]
  \centering
  \begin{tabularx}{\textwidth}{>{\raggedright}p{0.3\linewidth}>{\raggedright}p{0.3\linewidth}>{\raggedright\arraybackslash}p{0.3\linewidth}}
    \toprule
        \textbf{Technique} & \textbf{Explanation} & \textbf{Reasons for including} \\
    \midrule
        Scenario analysis (\emph{recommended}) & Involves developing future scenarios and analyzing them for risks & Focuses on future developments; uses forward reasoning \\
        Fishbone method (\emph{recommended}) & Involves drawing a diagram from a risk to its sources by repeatedly asking "why?" or "how might that occur?" & Provides structure and visualization for brainstorming; uses backward reasoning; simple \\
        Risk typologies and taxonomies (\emph{strongly recommended}) & Categorizations of risks; conceptually vs. empirically derived & Organizes the risk universe; can provide helpful input to many other techniques \\
        Causal mapping (\emph{recommended}) & Involves drawing a map of causes and consequences regarding a specific issue, including their interactions & Helps to combine different viewpoints; focuses on future developments; takes into account interactions between events; allows remote and anonymous participation \\
        Delphi technique (\emph{strongly recommended}) & Procedure to collect and collate expert judgments, mostly forecasts & Helps to combine different viewpoints; focuses on future developments; can be quantitative; allows remote \& anonymous participation \\
        Cross-impact analysis (\emph{encouraged}) & Involves breaking an issue down into contributing events, gathering and analyzing expert opinions on their likelihood; can yield possible and likely future scenarios & Helps to combine different viewpoints; focuses on future developments; takes into account interactions between events; can be quantitative; allows remote and anonymous participation \\
        Bow tie analysis (\emph{recommended}) & Involves drawing a diagram of causes and consequences of a risk, as well as preventive and reactive controls & Focuses on controls; simple \\
        System-theoretic process analysis (STPA) (\emph{encouraged}) & Examines a complex system for safe and unsafe states; involves backward reasoning from undesired events to how controls may not have the desired effect & Focuses on controls; comprehensive system-theoretic perspective \\
        Checklists (\emph{encouraged}) & Standardized questionnaires containing a list of open or closed questions & Allow decentralization of risk evaluation \\ Risk matrices (\emph{strongly recommended}) & Matrices that combine consequence and likelihood of risks / consequence of risks and vulnerability of the system at stake & Enable comparisons and prioritization \\
    \bottomrule
  \end{tabularx}
  \vspace{0.5em}
  \caption{Overview of the selected risk assessment techniques} \label{table1}
\end{table}

\section{Introduction}\label{1}

Companies like OpenAI, Google DeepMind, and Anthropic have the stated goal of building artificial general intelligence (AGI) – AI systems that perform as well as or better than humans on a wide variety of cognitive tasks. While it remains unclear when, if at all, AGI will be built and forecasting AI progress is inherently difficult,\footnote[1]{Some attempts include expert surveys \shortcite<e.g.>{muller_future_2016, grace_when_2018, stein-perlman_2022_2022, zhang_forecasting_2022}, comparisons with the development of intelligence in humans \cite{cotra_draft_2020}, and comparisons with developments of other technologies \cite{davidson_report_2021}.} the prospect of AGI is increasingly taken seriously. AGI has reached the mainstream academic discourse \shortcite<e.g.>{pei_towards_2019, fei_towards_2022, mahler_regulating_2022, roli_how_2022, salmon_managing_2023}, is widely covered in the news \shortcite<e.g.>{yudkowsky_pausing_2023, hogarth_we_2023, klein_surprising_2023, metz_godfather_2023}, and on the agenda of governments around the world \shortcite<e.g.>{the_white_house_national_2023, hm_government_national_2021, niti_aayog_national_2018}.

AI systems already cause significant harm. For example, some facial recognition systems discriminate against women and people of color \shortcite{buolamwini_gender_2018, raji_actionable_2019}. Language models can produce racist, sexist, and homophobic outputs \shortcite{bolukbasi_man_2016, bender_dangers_2021, weidinger_taxonomy_2022}, or can be used for disinformation campaigns \shortcite{buchanan_truth_2021} and cyberattacks \shortcite{brundage_malicious_2018, hazell_large_2023}, while image generation systems can be used to create harmful content, such as non-consensual deepfake pornography \shortcite{westerlund_emergence_2019}. These and other risks must be taken seriously and warrant significant attention. In addition to existing risks, there are increasing concerns about future catastrophic risks, including human extinction. In a recent statement, hundreds of leading AI scientists (e.g. Turing Prize winners Geoffrey Hinton and Yoshua Bengio), the CEOs of leading AI companies (e.g. Sam Altman and Demis Hassabis), and other prominent figures (e.g. Bill Gates) claim that “mitigating the risk of extinction from AI should be a global priority alongside other societal-scale risks such as pandemics and nuclear war” \shortcite{center_for_ai_safety_statement_2023}.

AGI companies need to take adequate measures to manage these risks. Relevant risks need to be assessed and adequate mitigations and controls need to be put in place. This paper focuses on risk assessment, which according to ISO 31000:2018 includes the identification, analysis, and evaluation of risks. The results of risk assessments inform decisions about risk treatment \shortcite{iso_31000_2018}. In a recent survey of leading experts from AGI companies, academia, and civil society (\emph{N} = 51), 98\% of respondents somewhat or strongly agreed that AGI companies should conduct pre-deployment risk assessments, while 94\% thought that they should also conduct pre-training risk assessments \shortcite{schuett_towards_2023}. Risk assessments could also become an essential part of regulatory frameworks \shortcite{anderljung_frontier_2023, shevlane_model_2023}. Our paper aims to provide guidance on how AGI companies could conduct risk assessments with regard to catastrophic risks from AI. We review popular risk assessment techniques from other industries and suggest ways in which AGI companies could use them. It is worth noting that many of the reviewed techniques will be obvious to risk management professionals in other industries. This is a “feature, not a bug”. The goal of this paper is to make sure that AGI companies are aware of best practices in other industries. We assume that some AGI companies already use some of the techniques in some situations, but we also expect there to be gaps. We also emphasize that these techniques will not be sufficient to assess catastrophic risks from AI.

The paper has four areas of focus. First, it focuses on catastrophic risks from AI. However, the selected techniques can also be used to assess other risks. Importantly, catastrophic risks may be intertwined with other risks, such that assessing them in isolation may not provide the full picture. Second, the paper focuses on how AGI companies can use these techniques. Needless to say, other actors like academics or governments may also use them to assess catastrophic risks from AI. Third, the paper only reviews existing techniques. We do not develop new techniques and only slightly adapt existing techniques. This would require significant effort and is beyond the scope of this paper. Finally, the paper focuses on general techniques. We do not review techniques that have been developed specifically for AI (e.g. dangerous model capabilities and propensities evaluations, or “evals”).\footnote{For example, before the release of GPT-4 and Claude, OpenAI and Anthropic gave the Alignment Research Center early access to the models to evaluate their ability to autonomously replicate and acquire resources \cite{openai_gpt-4_2023, anthropic_ai_2023, alignment_research_center_update_2023}. For more information on model evaluations, see \citeA{shevlane_model_2023}.} We also excluded established techniques from the field of information security (e.g. red teaming and bug bounties).\footnote{AGI companies increasingly engage red teams which adopt an attacker’s mindset to test the organization’s security \shortcite{applebaum_intelligent_2016, brundage_toward_2020, openai_gpt-4_2023}. Google DeepMind is part of Google’s and Alphabet’s “Vulnerability Reward Program” \cite{google_google_nodate}, and OpenAI recently announced its “Bug Bounty Program” \cite{openai_announcing_2023}. Under the term “red teaming”, AGI companies also try eliciting harmful outputs from their AI systems \shortcite{ganguli_red_2022, mishkin_dalle_2022, perez_red_2022}, efforts that fit better under terms like “boundary or stress testing” \shortcite{khlaaf_toward_2023}.}

The paper reviews techniques that AGI companies could use to assess catastrophic risks from AI. Let us define each of these terms. By “AGI”, we mean AI systems that perform as well as or better than humans on a wide variety of cognitive tasks.\footnote{The term “AGI” lacks a universally accepted definition. According to \citeA{goertzel_who_2011}, \citeA{gubrud_nanotechnology_1997} was the first to use the term, defining it as “AI systems that rival or surpass the human brain in complexity and speed, that can acquire, manipulate and reason with general knowledge, and that are usable in essentially any phase of industrial or military operations where a human intelligence would otherwise be needed”. The term “AGI” is related to concepts of “strong AI” \cite{searle_minds_1980}, “superintelligence” \cite{bostrom_how_1998, bostrom_superintelligence_2014}, and “transformative AI” \shortcite{dafoe_ai_2018, gruetzemacher_forecasting_2019, karnofsky_background_2016}.} By “AGI company”, we mean companies that have the stated goal of building AGI \shortcite{schuett_towards_2023}. “Risk” can be defined as the possibility that an undesired event will occur \shortcite{coso_guidance_2017}.\footnote{Other common definitions of the term “risk” in risk management comprise “effect of uncertainty on objectives” \cite{iso_31000_2018} as well as “combination of the probability of occurrence of harm and the severity of that harm” \cite{iso_guide_2014}.} “Risk sources” are the causes that alone or in combination give rise to risks \shortcite{iso_31000_2018}.\footnote{A risk source may have its own sources and so on. There can also be complex interactions between different risk sources. As a result, it often makes sense to assess risk sources as intermediate risks themselves. In the following, we generally use the term “risks” to mean both ultimate risks and risk sources as intermediate risks. If we refer to either ultimate risks or risk sources only, we make this explicit.} “Risk assessment” involves the identification, analysis, and evaluation of risks \shortcite{iso_31000_2018}. Note that our use of risk-related terminology follows leading risk management standards, especially ISO 31000:2018 \shortcite{iso_31000_2018}, IEC 31010:2019 \shortcite{iec_31010_2019}, and COSO ERM 2017 \shortcite{coso_guidance_2017}. By the term “catastrophic risk” we loosely mean the risk of widespread and significant harm, such as several million fatalities or severe disruption to the social and political global order \shortcite{bostrom_global_2008, posner_catastrophe_2004, rees_our_2004, shevlane_model_2023}. This includes “existential risks”, i.e. the risk of human extinction or permanent civilizational collapse \shortcite{bostrom_existential_2002, bostrom_existential_2013, ord_precipice_2020}.\footnote{There are various ways in which AI might cause or contribute to catastrophes. A basic distinction can be drawn between cases where humans intend to cause harm (misuse risks), cases where humans do not intend to cause harm (accident risks), and risks involving complex interactions of economic, political, societal, and other forces (structural risks) \cite{zwetsloot_thinking_2019}. First, malicious actors like terrorists could use AI systems to gain access to or create toxins, pathogens, or weapons \shortcite<e.g.>{brundage_malicious_2018, urbina_dual_2022, anderljung_protecting_2023}. Second, AI systems do not necessarily do what their developers or operators want them to do. When this issue is related to the goals they pursue, it is referred to as “misalignment”. If a highly capable AI system was power-seeking, i.e. aiming to accumulate resources, it might eventually take over control and sideline or even completely neglect human concerns \shortcite<e.g.>{bostrom_artificial_2008, bostrom_superintelligence_2014, russell_human_2019}. Third, many structural catastrophic risk scenarios from AI have been outlined by, for example, \citeA{dafoe_ai_2020}, \shortciteA{bucknall_current_2022}, and \shortciteA{clarke_survey_2022}. Particularly important risk sources may be competitive dynamics that lead AGI companies to cut corners on safety and other desirable features \shortcite{hendrycks_overview_2023} and various dangerous model capabilities and propensities \cite{shevlane_model_2023}, such as “agenticness” \shortcite{chan_harms_2023}. For existential risks from AI in particular, \shortciteA{kokotajlo_main_2019} have brainstormed an extensive but unstructured list of risk sources.}

The paper proceeds as follows. Section \ref{2} reviews related work. Section \ref{3} describes our method for selecting risk assessment techniques. Section \ref{4} discusses three risk identification techniques: scenario analysis, fishbone method, and risk typologies and taxonomies. Section \ref{5} discusses five risk analysis techniques: causal mapping, Delphi technique, cross-impact analysis, bow tie analysis, and system-theoretic process analysis (STPA). Section \ref{6} discusses two risk evaluation techniques: checklists and risk matrices. Section \ref{7} discusses when to conduct risk assessments, when to use which technique, and how to use any risk assessment technique. Section \ref{8} concludes with a summary of our main contributions, limitations, and suggestions for further research. The \hyperref[Appendix]{Appendix} lists popular risk assessment techniques we did not select.

\section{Related work}\label{2}

In the following, we give an overview of previous work that reviews or applies established risk assessment techniques to catastrophic risks from AI.

\textbf{Risk identification techniques.} Many expert surveys have been conducted about the trajectory of AI progress and the associated risks, including catastrophic ones \shortcite<e.g.>{baum_how_2011, clarke_survey_2021, grace_when_2018, michael_what_2022, muller_future_2016, stein-perlman_2022_2022, zhang_forecasting_2022}. \shortciteA{clarke_classifying_2022} and researchers from Google DeepMind \shortcite{kenton_clarifying_2022} have developed risk typologies for existential risks from AI. Several other typologies involve, but are not limited to, catastrophic risks from AI \shortcite{avin_classifying_2018, cotton-barratt_defence_2020, liu_governing_2018}. \shortciteA{yampolskiy_taxonomy_2015} and \shortciteA{critch_tasra_2023} have developed risk taxonomies for the development of a dangerous AI system, while \shortciteA{critch_ai_2020} have done the same for the deployment of a misaligned power-seeking AI. \shortciteA{rao_tricking_2023} have come up with a risk taxonomy for user prompts that generate model outputs unwanted by its developers. Otherwise, not for catastrophic, but more broadly for any societal risks from AI, many taxonomies have been developed by independent researchers \shortcite<e.g.>{center_for_security_and_emerging_technology_cset_taxonomy_nodate, khlaaf_toward_2023, newman_taxonomy_2023, raji_fallacy_2022, suresh_framework_2021} as well as researchers from different AGI companies \shortcite<e.g.>{microsoft_assessing_2020, shelby_identifying_2023, weidinger_taxonomy_2022}. Failure modes and effects analysis (FMEA) has been adapted and applied to risks from AI by \shortciteA{li_fmea-ai_2022} and \shortciteA{raji_closing_2020} hazard and operability study (HAZOP) by \shortciteA{zendel_cv-hazop_2015}, and preliminary hazard analysis (PHA) by \shortciteA{khlaaf_hazard_2022}.

\textbf{Risk analysis techniques.} Established risk analysis techniques that have been applied to the superintelligence takeover scenario include influence diagrams \shortcite{barrett_model_2017}, fault tree analysis (FTA) \shortcite{barrett_model_2017, callaghan_risk_2017}, and event tree analysis (ETA) \shortcite{callaghan_risk_2017}. Furthermore, causal mapping has been used by \shortciteA{cremer_artificial_2021} to determine warning signs of transformative AI, as well as by \shortciteA{losi_system_2023} to identify feedback loops when regulating AI. \shortciteA{kilian_examining_2023} have applied cross-impact analysis to establish correlations between relevant factors in and generate future scenarios of the global socio-technical AI landscape. \shortciteA{chin_embedding_2022} has used bow tie analysis to map controls with regard to the deployment of a harmful AI system.

\textbf{Risk evaluation techniques.} \shortciteA[Annex C]{hendrycks_x-risk_2022} have suggested a checklist for assessing the impact of a potential research project on existential risks from AI. Moreover, \shortciteA[Section 3.2.2.1.1]{barrett_actionable_2023} have provided a starting point for a pre-development or pre-deployment checklist for catastrophic risks from AI. Apart from these, only impact assessment checklists for any societal risks from AI have been developed, for example, on trustworthy AI by the \shortciteA{eu_high-level_expert_group_on_ai_assessment_2020} on AI fairness by \shortciteA{madaio_ai_2020}, and on responsible AI by \shortciteA{microsoft_responsible_2022}. Again, not for catastrophic, but more broadly for any societal risks from AI, several risk matrices have been developed \shortcite<e.g.>{khlaaf_toward_2023, microsoft_harms_2022}.

\textbf{Review of several risk assessment techniques.} The work most similar to this paper is a blog post which provides a great but much more shallow review of several risk assessment techniques \shortcite{chin_what_2022}. In the academic literature, only \shortciteA{callaghan_risk_2017} have reviewed several risk assessment techniques. They suggest FTA and ETA for assessing catastrophic risks from AI. We believe that FTA and ETA on their own may be too simplistic to provide deep insights into catastrophic risks from AI, although variations of them may be very useful. Catastrophic risks from AI involve not only failures of individual components of systems, but complex technical, economic, political, and societal factors, events, and their interactions \shortcite{khlaaf_toward_2023, leveson_stpa_2018}, see also Section \ref{3} and Section \ref{5.5}). In contrast, FTA and ETA involve drawing logical diagrams of causes and consequences of risks, which simplifies risk sources and their interactions into binary events and linear causal chains. Nevertheless, we mention event trees as a tool for developing scenarios (Section \ref{4.1}), and include the fishbone method (Section \ref{4.2}) which can be said to be a special type of FTA, as well as bow tie analysis (Section \ref{5.4}) which can be understood as involving both fault and event trees. However, these techniques go beyond mere FTA and ETA (scenario analysis can employ other methods to develop scenarios; fishbone method provides structure by establishing categories of causes; bow tie analysis focuses on controls and includes ongoing practices), and we also highlight their simplicity as a major limitation. Moreover, \shortciteA{callaghan_risk_2017} do not provide the criteria which they used to select techniques, and which other techniques they considered. Their paper is also limited to the (at the time predominant) superintelligence takeover scenario, where a single extremely capable AI gets out of hand and destroys humanity. Finally, it is not specifically targeted at AGI companies, but decision-makers in general.

\textbf{Gap in the literature.} Overall, we observe extensive interest in the topic of risk assessment techniques for catastrophic risks from AI. International organizations have started to issue frameworks for risk assessment at AI companies, but they are not tailored to catastrophic risks and lack concreteness \shortcite{xia_towards_2023}. More actionable efforts so far have mostly focused on developing novel techniques specifically for catastrophic risks from AI, such as evals, or applying a single established risk assessment technique to the context of catastrophic risks from AI. Only \shortciteA{callaghan_risk_2017} have attempted a review of several established risk assessment techniques, but their paper has several limitations (see previous paragraph). In conclusion, there is no comprehensive and up-to-date review of which established techniques could be useful for AGI companies to assess catastrophic risks from AI.

\section{Methodology}\label{3}

This section describes our method for identifying and selecting risk assessment techniques. We used a leading risk assessment standard as a starting point and added popular techniques from other industries. We then defined criteria for excluding and prioritizing techniques which we used to narrow down the list. The techniques we selected can be found in Table 1. The techniques we excluded can be found in the Appendix \ref{Appendix}.

\textbf{Popular risk assessment techniques.} The standard IEC 31010:2019 contains a list of some of the most popular risk assessment techniques among different industries \shortcite{iec_31010_2019}. To find additional popular techniques, we reviewed risk assessment techniques in finance, aviation, nuclear, and biolabs (e.g. by investigating guidelines from international agencies and textbooks, and consulting with risk management experts). We chose these four industries because they have a long tradition of risk assessment, the latter three being safety-critical industries. While this approach found more than 100 techniques, most of the popular ones had already been contained in IEC 31010:2019.

\textbf{Criteria for excluding techniques.} Next, we used four criteria to narrow down the list. First, catastrophic risks affect society as a whole, not only the organization itself. Techniques therefore need to be applicable to societal risks, not just business risks. Second, catastrophic risks from AI are generally considered low-probability, high-impact events \shortcite<see>{muller_future_2016, grace_when_2018, stein-perlman_2022_2022, zhang_forecasting_2022}. We therefore excluded techniques that neglect tail risks. Third, some techniques from industries like aviation, nuclear, or biolabs focus on human performance reliability when it comes to routine tasks. Currently, at AGI companies, there seem to be no routine tasks of employees which, if performed incorrectly, could lead to catastrophe. Therefore, we excluded techniques that aim to ensure that humans perform reliably on routine tasks. However, if this changes in the future (e.g. if employees are tasked with overseeing a dangerous AI system or its usage), these techniques should be reconsidered. Fourth, many techniques from industries like aviation or nuclear have been developed to assess hardware reliability. We assume hardware failures to be much less critical for catastrophic risks from AI. Most of these techniques can be applied to issues other than hardware reliability. Yet, because AI systems and the risks they pose are highly complex, the techniques may not be well-suited for this context \shortcite{khlaaf_toward_2023, leveson_stpa_2018}. We therefore excluded many of these traditional techniques. However, if hardware becomes more critical in the future (e.g. if hardware-enabled mechanisms are implemented, if one AI system is used to keep in check another dangerous AI system, or if AI systems become part of critical infrastructure), these techniques should be reconsidered. 

\textbf{Criteria for prioritizing techniques.} Then, we used the following criteria to identify techniques which are particularly promising. Catastrophic risks from AI are highly complex. They involve various technical, economic, political, and societal factors, events, and their interactions \shortcite{dobbe_system_2022, khlaaf_toward_2023}. Therefore, we prioritized techniques that can be used to examine complex interactions between events. For this reason, we included system-theoretic process analysis (STPA) (Section \ref{5.5}), even though it is not listed by \shortciteA{iec_31010_2019}. Because of the complexity of catastrophic risks from AI, we also prioritized techniques that help combine the viewpoints of a variety of people with different knowledge and perspectives. This may include experts with different backgrounds (e.g. technical, economic, or political), or employees with different positions in the organization (e.g. researchers, engineers, or managers). Furthermore, catastrophic risks from AI are marked by high uncertainty. They have never happened before and most likely will not happen twice (although some risk sources might). We therefore prioritized techniques that provide clarity on future developments. Quantifying the likelihood of catastrophic risks from AI may be especially challenging \shortcite{baum_quantifying_2020, beard_analysis_2020, beard_existential_2020}. We therefore prioritized qualitative over quantitative techniques. However, quantification may be useful to concretize concerns and enable better comparisons and communication about risks, such that it should at least be attempted \shortcite{baum_quantifying_2020, beard_analysis_2020, beard_existential_2020}, see also Section \ref{2} and Section \ref{5.5}). We therefore included some quantitative techniques, too. Finally, we aimed for a variety of techniques, and selected a representative technique when several very similar techniques existed.\footnote{We followed IEC 31010:2019 in assigning the selected techniques to one of the three risk assessment steps \cite{iec_31010_2019}. However, some of them comprise elements of several risk assessment steps, and some of them can be used for other purposes entirely, too (like general forecasting).}

\section{Risk identification}\label{4}

Risk identification is the first step in the risk assessment process \shortcite{iec_31010_2019, iso_31000_2018}. AGI companies should try to identify the risks of specific models – before deploying them, but also before training them \shortcite{schuett_towards_2023}. In addition to that, they should try to identify all relevant risks in the abstract (so-called “risk universe”). Based on our selection criteria, the following three risk identification techniques seem particularly promising: scenario analysis (Section \ref{4.1}), fishbone method (Section \ref{4.2}), and risk typologies and taxonomies (Section \ref{4.3}).

\subsection{Scenario analysis}\label{4.1}

In a scenario analysis, organizations develop and analyze future scenarios of the environment they operate in and plan accordingly (e.g. how competitive the “AGI market” will be in two years). They develop scenarios, for example, by combining driving forces (e.g. the number of AGI companies, their business models, moats, etc.) and then analyze these scenarios by thinking through their implications on risks. Scenario analysis is often used by organizations for long-term and emergency planning \shortcite{iec_31010_2019}.

\begin{figure}[t!]
    \centering
    \includegraphics[width=0.8\textwidth]{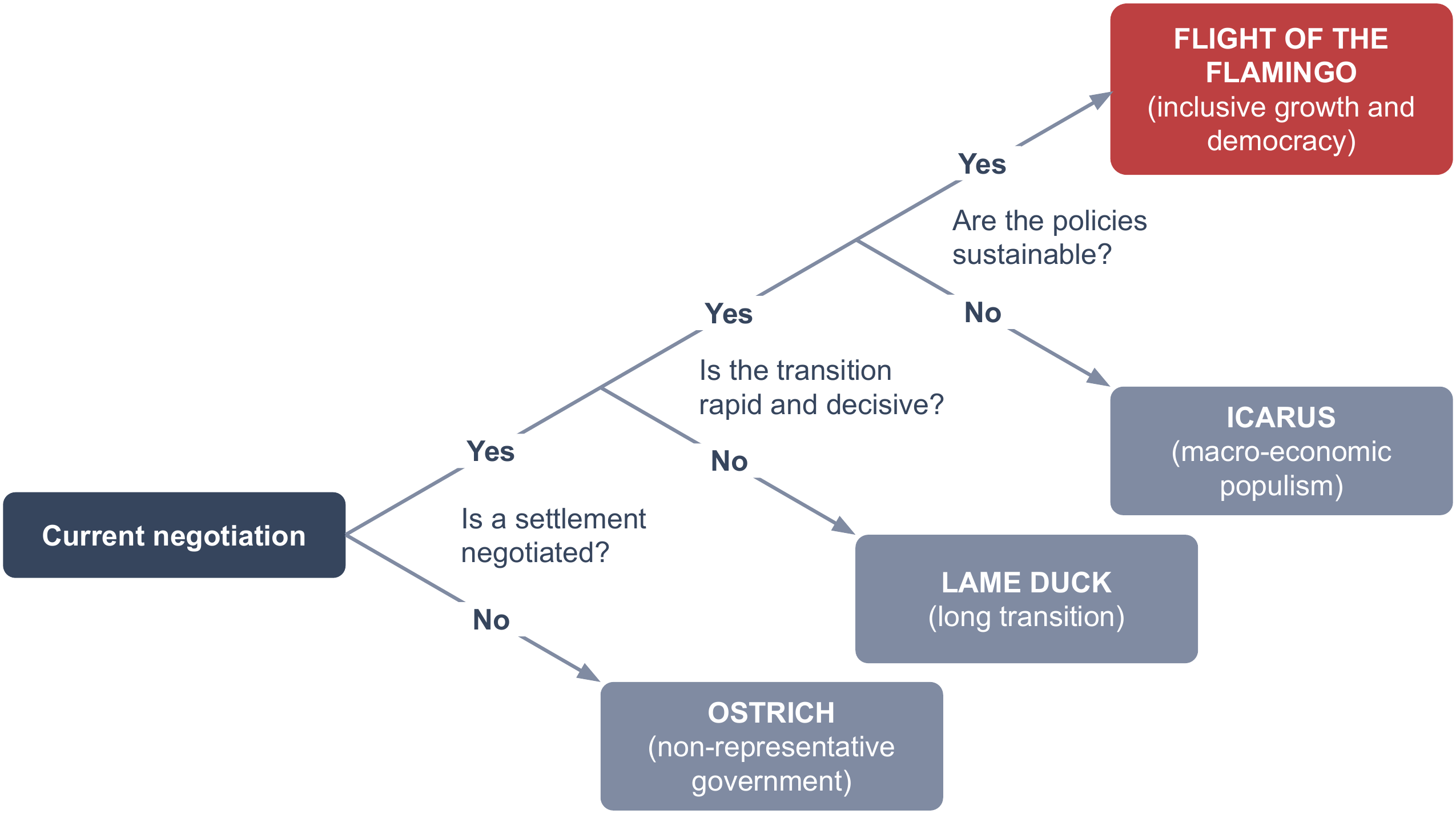}
    \caption{Event tree for the political situation of a country \protect\cite{kosow_methods_2008}}
    \label{fig2}
\end{figure}

\begin{table}[ht!]
  \centering
    \begin{tabular}{cccc}
    \toprule
    \multicolumn{2}{c}{} & \multicolumn{2}{c}{Government approach to energy industry} \\
    \cline{3-4}
    \multicolumn{2}{c}{} & Dirigiste & Laissez-faire \\
    \hline
    \multirow{2}{*}{Natural gas development} & Favorable & Scenario 1 & Scenario 2 \\
    & Unfavorable & Scenario 3 & Scenario 4 \\
    \bottomrule
    \end{tabular}
    \vspace{1em}
    \caption{Scenario matrix for the competitive structure of an industry \protect\cite{kosow_methods_2008}}
    \label{table2}
\end{table}

\textbf{How it works.} Scenario analysis has many variations, but the basic procedure is usually the same \shortcite{iec_31010_2019, chermack_scenario_2011, kosow_methods_2008, van_der_heijden_scenarios_2005}. First, organizations develop one or several plausible future scenarios with regard to a specific issue. This step can involve gathering expert forecasts or using statistical methods, but it can also consist of mere brainstorming and group discussions. A particularly common approach is to identify a small number of driving forces, i.e. key events or trends, and combine them in all possible consistent ways. In the end, organizations should have a list of scenarios and a story for each of them about how it could unfold. For example, with regard to the competitive structures of an industry, the driving forces could be developments in the supply of relevant materials and government approaches \shortcite{kosow_methods_2008}. To combine these driving forces, organizations may draw an event tree (a diagram that depicts a chronological sequence of binary events) or construct a scenario matrix (a matrix that combines the two most important driving forces into all four possible scenarios) \shortcite{kosow_methods_2008}. Second, in the same or in a separate workshop, organizations discuss each scenario for the risks it creates or exacerbates. Again, this can be done through more or less structured brainstorming and group discussions. Third, moving beyond risk assessment into risk treatment, organizations discuss strategies and emergency response plans.

\textbf{How AGI companies could use it.} AGI companies could develop and analyze scenarios, for example, of the trajectory of AI progress, the “AGI market”, and the geopolitical AI landscape over the next few months or several years. With regard to AI progress, driving forces could be major technological breakthroughs, like eliminating model hallucinations. For the trajectory of the “AGI market”, important trends beyond the ones mentioned above may include the open-sourcing of models, the predominance of foundation models, and regulatory approaches. The geopolitical AI landscape could be largely influenced by events in China, the US-China relationship, and trends in the militarization of AI.

The AI Index (\url{https://aiindex.stanford.edu}) as well as websites such as Epoch (\url{https://epochai.org}) and Our World in Data (\url{https://ourworldindata.org/artificial-intelligence}) contain data on some of these driving forces, for example, trends in the number of research publications, the amount of investments into AGI companies, and proposed regulation. AGI companies could pick the most important events and trends, combine them to generate scenarios, and discuss the implications of those for risks. In many cases, AGI companies can also directly analyze previously developed scenarios for the risks they entail \shortcite<e.g.>{davidson_what_2023, leung_who_2019}.

AGI companies could also use scenario analysis to develop and analyze emergency situations. This may involve several group discussions in which participants would first need to come up with a list of worst-case scenarios (e.g. a model being leaked, evals revealing high situational awareness of a model, or a model “escaping” into the internet). Next, participants could analyze these scenarios for the downstream risks they entail (e.g. the model being used for various undesired purposes, the model being more likely to be deceptive, or the model causing various types of harm). Finally, participants could develop emergency response plans (which is part of risk treatment and will thus not be elaborated on further in this paper).

\textbf{Benefits.} The use of driving forces ensures building on existing information, even if it is scarce. This may provide for more realistic scenarios than mere brainstorming. Another benefit of this technique is that it can be used to systematically investigate different futures instead of focusing on a single scenario. Given the high stakes of catastrophic risks, AGI companies should aim to be “better safe than sorry” and plan ahead for a variety of possibilities. The scenarios developed can also be used to monitor whether things are moving in a dangerous direction. If events keep occurring as implied by a scenario, this can be understood as a warning sign \shortcite{iec_31010_2019, etzioni_how_2020, cremer_artificial_2021}.

\textbf{Limitations.} On the downside, there is little evidence that scenarios developed through this technique actually occur \shortcite{iec_31010_2019}. Therefore, they should not be relied on as predictions of how the future will unfold. Another limitation of scenario analysis is that it does not provide guidance on how to choose and combine driving forces. Instead, it largely hinges on the knowledge and expertise of the participants. It is thus crucial to select people with relevant backgrounds and skills.

\textbf{Recommendations.} AGI companies probably already develop and analyze future scenarios to inform business decisions. We recommend them to do so with a focus on catastrophic risks from AI. To that end, AGI companies should take a comprehensive approach, developing and analyzing various different scenarios. In particular, we recommend developing scenarios for different timeframes, from the next few months to several years, as well as for emergency situations. As a starting point, AGI companies could develop scenarios through brainstorming and group discussions. For issues that turn out very complex or important, AGI companies could consider more elaborate versions that involve expert forecasts or statistical methods. These could also be outsourced to external consultancy or research organizations.

\subsection{Fishbone method}\label{4.2}

The fishbone method, also known as Ishikawa analysis or cause-and-effect diagram, helps organizations to identify sources of risks. In contrast to scenario analysis (Section \ref{4.1}), which typically uses forward reasoning, the fishbone method uses backward reasoning from an undesired event to its causes and sub-causes. The causal relationships are visualized in a diagram that resembles a fishbone. The fishbone method is a very popular and comparatively simple technique \shortcite{iec_31010_2019, ishikawa_guide_1976}.

\begin{figure}[t!]
    \centering
    \includegraphics[width=0.8\textwidth]{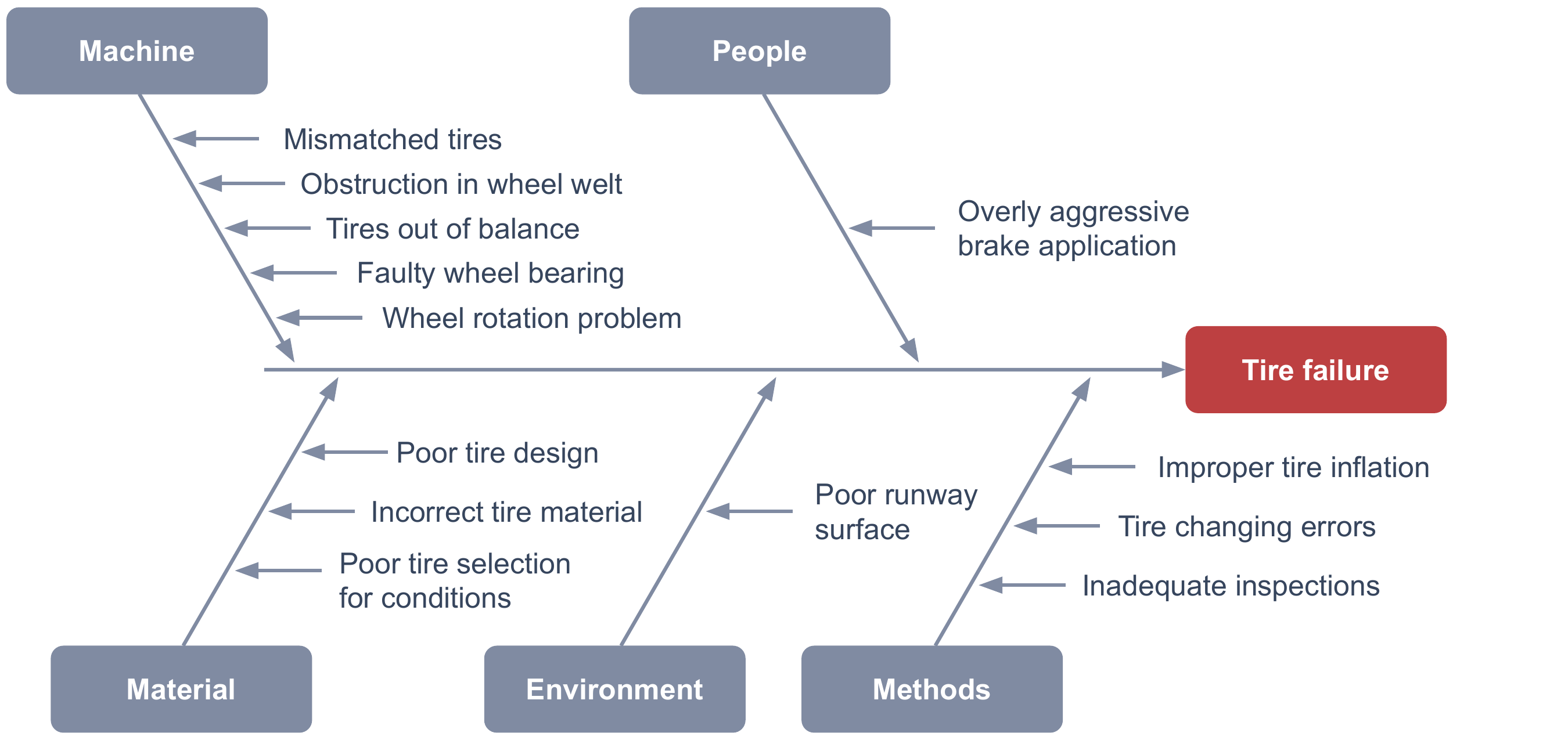}
    \caption{Simple fishbone diagram for a plane tire failure \protect\cite{stolzer_safety_2023}}
    \label{fig3}
\end{figure}

\textbf{How it works.} The technique consists of the following steps \shortcite{iec_31010_2019, ishikawa_guide_1976, rausand_risk_2020}. To begin with, organizations choose a risk to be analyzed and place it as the “head” of the fish-like structure. For example, in aviation, this could be the failure of a plane tire \shortcite{stolzer_safety_2023}. Next, they select the primary categories of causes for the risk and place them as the “bones” branching out from the “spine” of the fish. Some commonly used categories include different combinations of “6Ms”, such as machinery (equipment), manpower (people), milieu (environment), methods and processes, management, and money \shortcite{stolzer_safety_2023}. Then, organizations identify the causes and sub-causes within each category by repeatedly asking questions like “why?” and “how might that occur?”. Organizations repeat these questions until they no longer yield useful information. For example, within the category of machinery, a cause may be mismatched tires, a sub-cause of which may be the replacement of an old tire by a newer version \shortcite{stolzer_safety_2023}. Once the diagram is complete, organizations review it to ensure consistency and comprehensiveness. Categories and causes should cover technical, human, and organizational aspects. Finally, organizations discuss the identified causes to determine which of them are most important.

\textbf{How AGI companies could use it.} AGI companies could use the fishbone method at the level of catastrophic risks or their sources. At the level of the ultimate risks, AGI companies could (literally) flesh out the many risk scenarios that have already been developed at a high level of abstraction (e.g. the “head” could be “takeover by a misaligned power-seeking AI”). At the level of risk sources, for example, they could examine the emergence of dangerous model capabilities and propensities one by one (e.g. the “head” could be “AI system has situational awareness” or “AI system seeks power”). The driving forces determined in scenario analysis (Section \ref{4.1}) as well as the classifications developed for risk typologies and taxonomies (Section \ref{4.3}) may provide or help to identify categories.

When planning training runs or evals, AGI companies could use the fishbone method to investigate why various undesired events might happen in the course of those. For example, they could examine how an “agentic” model could take advantage of the situation to avoid its shut-down. Some categories may be the different dangerous model capabilities and propensities. How could each of them lead to the undesired event? Other categories may be the researchers conducting the evals, and the model evaluation process. How could the model take advantage of the researchers’ cognitive biases and other human characteristics? For instance, could a researcher be convinced by the model to not shut it down? How could the model evaluation process contribute to the undesired event? For instance, are there any points in time when no one is checking what the model is doing, or other critical moments such as shifts between researchers?

The fishbone method was originally developed to ensure product quality. In this case, the categories represent the main steps of the product development process, proceeding chronologically from left to right. Each step is examined for how it contributes to the risk under consideration \shortcite{ishikawa_guide_1976}. AGI companies could use this chronological version of the technique to examine the impacts on risks of the different steps of their model or product development pipeline (e.g. the different steps in training or in the whole AI system lifecycle from planning to monitoring).

\textbf{Benefits.} The use of categories and the iterative questioning down to the roots of risks makes the fishbone method less likely to miss risk sources than mere brainstorming. Its visualization further helps to understand and communicate risks and their sources to relevant stakeholders, such as researchers or leadership \shortcite{iec_31010_2019}. In contrast to most other risk assessment techniques, the fishbone method uses backward reasoning. This makes it a valuable addition when it comes to assessing highly uncertain and complex catastrophic risks from AI (Section \ref{7}). Finally, the fishbone method is simple and less time-consuming than other techniques \shortcite{rausand_risk_2020}.

\textbf{Limitations.} The fishbone method does not account for interactions between risk sources that do not follow a linear causal chain, such as feedback loops or other synergies. It is thus not suitable for analyzing risks that involve complex interactions between events (e.g. competitive dynamics). Moreover, the technique hinges on the selection of categories. If important categories are missed, so are the respective causes \shortcite{iec_31010_2019}. In some cases, the fishbone method may aggregate and visualize information rather than generating information. Simply asking “why?” does not always spark new insights for complex issues people have already spent considerable time reflecting upon.

\textbf{Recommendations.} We recommend AGI companies to use the fishbone method to take a systematic and thorough approach to identifying risk sources. Since the technique is simple and takes little time, AGI companies could simply try it out and see whether it is helpful. For example, they could investigate the sources of dangerous model capabilities and propensities in the abstract, or before pre-training, fine-tuning, or evaluating a new model.

\subsection{Risk typologies and taxonomies}\label{4.3}

Risk typologies and taxonomies are categorizations of risks. They are conceptually or empirically derived. Risk typologies and taxonomies structure the entirety of previously identified risks and can help to identify additional risks by uncovering gaps \shortcite{iec_31010_2019}. Categorizations of risks are considered a must-have by risk management practitioners \shortcite{pritchard_risk_2015, stolzer_safety_2023}.

\begin{figure}[t!]
    \centering
    \includegraphics[width=\textwidth]{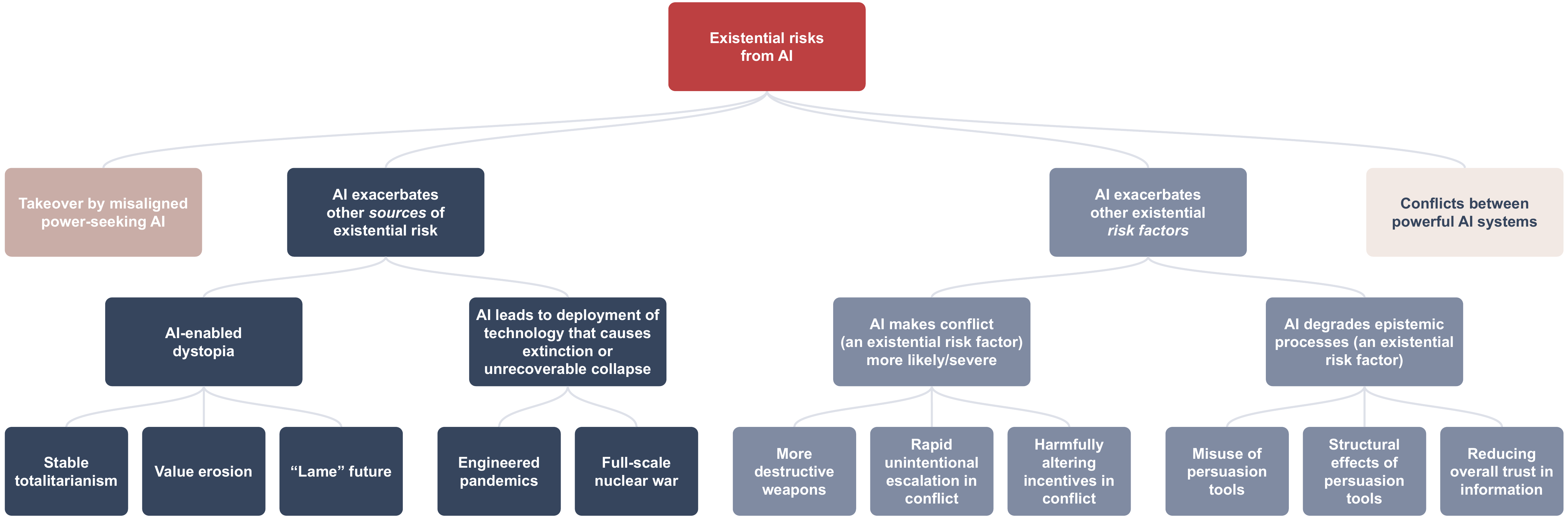}
    \caption{Typology of existential risks from AI \protect\cite{clarke_classifying_2022}}
    \label{fig4}
\end{figure}

\begin{table}[b]
  \small
  \centering
    \begin{tabularx}{\textwidth}{>{\raggedright\arraybackslash}p{0.16\linewidth}>{\raggedright\arraybackslash}p{0.16\linewidth}>{\raggedright\arraybackslash}p{0.16\linewidth}>{\raggedright\arraybackslash}p{0.16\linewidth}>{\raggedright\arraybackslash}p{0.16\linewidth}}
    \toprule
    \multicolumn{2}{c}{} & \multicolumn{3}{c}{Source of misalignment} \\
    \cline{3-5}
    \multicolumn{2}{c}{} & Specification gaming (SG) & SG + GMG & Goal misgeneralization (GMG) \\
    \hline
    \multirow{2}{=}{Path to existential catastrophe} & Misaligned power-seeking (MAPS) & \citeA{cohen_advanced_2022} & \citeA{carlsmith_is_2022}, \citeA[part 2]{christiano_what_2019}, \citeA{cotra_without_2022}, \citeA{ngo_alignment_2023}, \citeA{shah_ai_2022} & \citeA{soares_central_2022}, \citeA{hubinger_how_2022} \\
    & Interaction of multiple systems & \citeA{critch_what_2021}, \citeA[part 1]{christiano_what_2019} & ? & ? \\
    \bottomrule
    \end{tabularx}
    \vspace{0.5em} \caption{Typology of accident risks \protect\cite{kenton_clarifying_2022}} \label{table3}
\end{table}

\textbf{How they work.} While typologies are based on abstract concepts, taxonomies are based on empirical data \shortcite{iec_31010_2019}. As an exemplary risk typology, risk managers in aviation use the Human Factors Analysis and Classification System (HFACS) which splits the causes of human mistakes into latent conditions within the system and unsafe actions by individuals \shortcite{hfacs_hfacs_nodate, reason_human_2000, stolzer_safety_2023}. As an exemplary risk taxonomy, risk management in biolabs makes use of a taxonomy by the International Committee on Taxonomy of Viruses which categorizes viruses based on their genome, structure, and strategy of replication \shortcite{wooley_viral_2017}. In practice, hybrid forms often blend risk typologies and taxonomies and may simply be referred to as “risk taxonomies” \shortcite{iec_31010_2019}. 

To develop a risk typology or taxonomy, organizations usually collect all risks and risk sources that have already been identified. Next, they attempt to structure this list in a useful way and fill potential gaps by ideating the missing items. How to best do this is very context-specific. Categorization schemes for risk typologies and taxonomies are typically intended to be mutually exclusive (i.e. risks can only be in one category) and collectively exhaustive (i.e. they cover all relevant risks). Risk typologies and taxonomies can have subcategories or zoom in on a particular category from another risk typology or taxonomy \shortcite{iec_31010_2019}. For example, unsafe actions by individuals as a cause of human mistakes can be split into errors and violations, which in turn are composed of decision errors, skill-based errors, and perceptual errors, as well as routine and exceptional violations \shortcite{hfacs_hfacs_nodate, reason_human_2000, stolzer_safety_2023}. Risk typologies and taxonomies are often presented as tree diagrams or tables.

\textbf{How AGI companies could use them.} AGI companies already use typologies and taxonomies for some types of risks from AI. For example, researchers from Google DeepMind have created a taxonomy for risks from language models \shortcite{weidinger_taxonomy_2022}. We strongly suspect that OpenAI and Anthropic also have similar risk typologies and taxonomies, even if they do not make them public. (Note that the list of risks in the GPT-4 system card is explicitly not intended as a taxonomy; \citeA{openai_gpt-4_2023}.) However, we could not find any public information about whether AGI companies also have typologies and taxonomies for catastrophic risks from AI. Since there is no empirical data on catastrophic risks from AI, in this context typologies are more suitable for the ultimate risks, while taxonomies are more suitable for risk sources, for which some empirical data exists. 

A number of researchers have already proposed typologies for catastrophic risks from AI. The most high-level typology splits catastrophic risks from AI into accident, misuse, and structural risks. For existential risks, \shortciteA{clarke_classifying_2022} distinguishes between takeover by misaligned power-seeking AI, AI exacerbating other existential risks, AI exacerbating other existential risk factors, and conflict between powerful AI systems (Figure \ref{fig4}). With regard to accident risks, researchers from Google DeepMind have developed a typology that distinguishes between technical sources of misalignment and paths to existential catastrophe. They find that some combinations are missing \shortcite{kenton_threat_2022, kenton_clarifying_2022, krakovna_linkpost_2023}. AGI companies could develop additional typologies for misuse risks as well as risks at the intersection of accident and misuse. 

There are also typologies of catastrophic risks in general. For example, \shortciteA{cotton-barratt_defence_2020} distinguish between origin (e.g. accident), scaling mechanism (e.g. cascading), and endgame (e.g. ubiquity). Similarly, \shortciteA{avin_classifying_2018} distinguish between critical system affected (e.g. food chains), global spread mechanism (e.g. digital), as well as prevention and mitigation failure (e.g. cognitive biases). AGI companies could narrow these two typologies down to catastrophic risks from AI. While they already contain some catastrophic risks from AI, they may also be used to identify additional ones. For instance, \shortciteA{avin_classifying_2018} only mention weaponization as a catastrophic risk from AI. However, AGI companies can use their typology to identify many other plausible combinations. One example would be food chains breaking down once they completely rely on AI systems due to automation bias, i.e. the belief that machines are more accurate or reliable than humans.

\begin{table}
  \small
  \centering
  \begin{tabularx}{0.7\textwidth}{p{3.5cm} p{5.5cm}}
    \toprule
        \textbf{AI system failures} & \textbf{Examples} \\
    \midrule
        Impossible tasks &
            \vspace{-0.7em} \begin{itemize}[leftmargin=*, nosep]
                \item Conceptually impossible
                \item Practically impossible \vspace{-0.2em}
            \end{itemize} \\
        \vspace{-1.4em} Engineering failures &
            \begin{itemize}[leftmargin=*, nosep] \vspace{-1em}
                \item Design failures
                \item Implementation failures
                \item Missing safety features \vspace{-0.2em}
            \end{itemize} \\
        \vspace{-1.4em} Post-deployment failures &
            \begin{itemize}[leftmargin=*, nosep] \vspace{-1em}
                \item Robustness issues
                \item Failure under adversarial attacks
                \item Unanticipated interactions \vspace{-0.2em}
            \end{itemize} \\
        \vspace{-1.4em} Communication failures &
            \begin{itemize}[leftmargin=*, nosep] \vspace{-1em}
                \item Falsified or overstated capabilities
                \item Misrepresented capabilities \vspace{-0.8em}
            \end{itemize} \\
    \bottomrule
  \end{tabularx}
  \vspace{1em} \caption{Taxonomy of AI system failures \protect\cite{raji_fallacy_2022}} \label{table4}
\end{table}

There are also a number of taxonomies for sources of catastrophic risks from AI. \shortciteA{yampolskiy_taxonomy_2015} and \shortciteA{critch_tasra_2023} have developed risk taxonomies for the development of a dangerous AI system. \shortciteA{yampolskiy_taxonomy_2015} distinguishes between timing pre and post deployment, as well as different motivations by the actors involved. \shortciteA{critch_tasra_2023} build on this and include diffuse responsibility among several actors. \shortciteA{critch_ai_2020} have developed a risk taxonomy for the deployment of misaligned power-seeking AI. They distinguish between uncoordinated deployment (relevant for risks caused by the interactions of several different systems by several different companies), unrecognized prepotence (understood as extremely high capability), unrecognized misalignment, involuntary deployment, and voluntary (e.g. malicious or indifferent) deployment. \shortciteA{rao_tricking_2023} have come up with a risk taxonomy for ways in which users can prompt models to generate output that was not intended by the developers. They distinguish between types of attacks, namely instruction-based or non-instruction based, the intent of the attack, which can be goal hijacking, prompt leaking, or denial of service, and the manner of the attack, that is whether the attack is conducted by a user or by a separate “man-in-the-middle” who alters the input of a user.

AGI companies could develop further taxonomies of sources of catastrophic risks from AI. In order to cover all risk sources, AGI companies should also consider the whole AI system lifecycle \shortcite{oecd_advancing_2023, newman_taxonomy_2023, suresh_framework_2021}. Useful categories may be: risks stemming from data, compute, model architecture, pre-training, fine-tuning, evaluating, testing, deployment, and monitoring. Some more fine-grained categories may be: risks related to different AI training techniques (e.g. supervised learning, self-supervised learning, reinforcement learning), AI model modalities (e.g. language, vision, or multimodality), and AI system applications (e.g. chat, search, or code, research, or image generation). AGI companies should also include risks associated with all the main actors involved, such as engineers, researchers, managers, downstream developers, users, other AGI companies, governments, or the general public. Alternatively, AGI companies may start thinking backwards from different types of harm AI systems might cause (e.g. financial cost, damage to critical infrastructure, or fatalities) as has been done for any risks from AI by researchers from Google DeepMind \shortcite{weidinger_taxonomy_2022}, Google \shortcite{shelby_identifying_2023}, and \shortcite{microsoft_assessing_2020}. Generally, the various taxonomies developed with regard to any societal risks from AI may serve as blueprints or starting points for taxonomies focused on catastrophic risks from AI \shortcite<e.g.>{center_for_security_and_emerging_technology_cset_taxonomy_nodate, khlaaf_toward_2023, raji_fallacy_2022}.

\textbf{Benefits.} Risk typologies and taxonomies have three main benefits. First, they can help to avoid blind spots \shortcite{iec_31010_2019}. Without a structured approach to risk identification, organizations will likely miss risks. Second, risk typologies and taxonomies can help to build a common understanding of the risk landscape among different people within the AGI company \shortcite{iec_31010_2019}. This is important because, for instance, managers and researchers may need to be involved to effectively address risks. Therefore, they need to be on the same page about what constitutes a risk. Third, risk typologies and taxonomies can support other risk assessment techniques. For example, they are necessary to generate checklists (Section \ref{6.1}), and they can provide or help to identify the driving forces of an issue as part of scenario analyses (Section \ref{4.1}) and the causes of a risk as part of the fishbone method (Section \ref{4.2}).

\textbf{Limitations.} The main limitation of risk typologies and taxonomies is that creating them is very time-consuming. As an illustration, Google DeepMind’s taxonomy for risks from language models has 23 authors \shortcite{weidinger_taxonomy_2022}. When it comes to catastrophic risks from AI, complexity and a lack of empirical data may further complicate the development of risk typologies and taxonomies. Another limitation is that, in practice, it is rarely possible to create risk typologies and taxonomies that are fully comprehensive. They will most likely miss some risks or risk sources. AGI companies should therefore avoid relying on a single risk typology or taxonomy.

\textbf{Recommendations.} We strongly recommend AGI companies to use risk typologies and taxonomies if this is not already the case. They should create and maintain at least two complementary typologies of catastrophic risks because a single risk typology will likely miss important risks. For the same reason, they should also maintain several taxonomies for each individual risk source. AGI companies may want to zoom in on a number of risk sources that seem particularly important (e.g. dangerous model capabilities and propensities). All risk typologies and taxonomies should be reviewed and updated on a regular basis (e.g. every three or six months).

\section{Risk analysis}\label{5}

Risk analysis is the second step in the risk assessment process. It aims to facilitate a deep understanding of the causes, consequences, and likelihood of risks \shortcite{iec_31010_2019}. In the following, we discuss five risk analysis techniques: causal mapping (Section \ref{5.1}), Delphi technique (Section \ref{5.2}), cross-impact analysis (Section \ref{5.3}), bow tie analysis (Section \ref{5.4}), and system-theoretic process analysis (STPA) (Section \ref{5.5}).

\subsection{Causal mapping}\label{5.1}

Causal mapping is an exploratory technique that helps organizations to better understand complex interactions between different causes and consequences of risks. It involves a group of people collectively drawing a map of events and their relationships \shortcite{iec_31010_2019}.

\begin{figure}[t!]
    \centering
    \includegraphics[width=\textwidth]{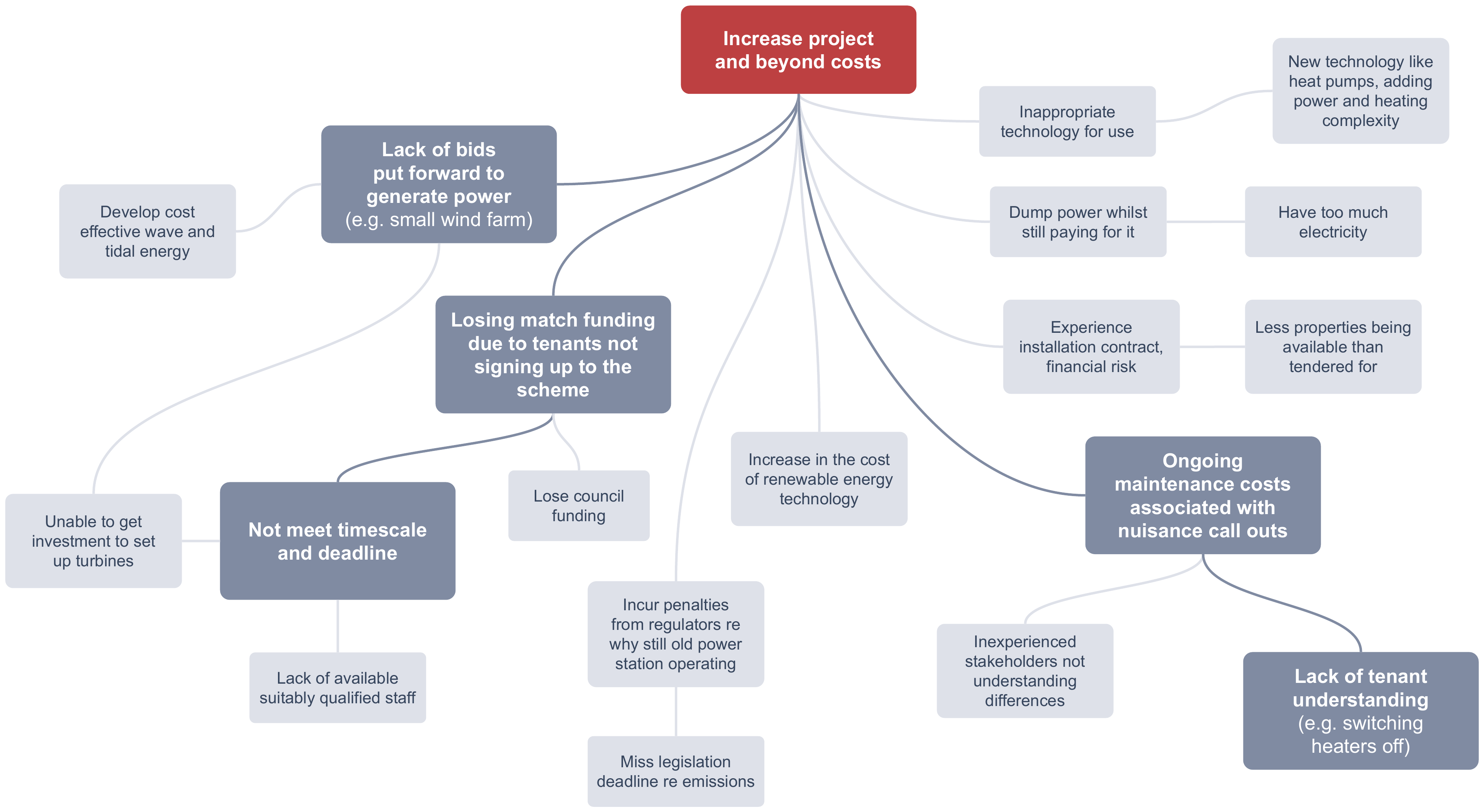}
    \caption{Segment of a causal map for risks when connecting renewables to a local electricity grid \protect\cite{ackermann_systemic_2014}}
    \label{fig5}
\end{figure}

\textbf{How it works.} To develop a causal map, organizations follow a particular sequence of steps \shortcite{iec_31010_2019, ackermann_systemic_2014, bryson_visible_2004}. First, they brainstorm events related to an important issue. For instance, relevant events with regard to risks when connecting renewables to a local electricity grid could be missed deadlines, an increase in energy prices, and tenants not signing up to the scheme \shortcite{ackermann_systemic_2014}. Second, organizations cluster the identified events based on similarities and relationships. For example, events may be related to suppliers, tenants, or project management \shortcite{ackermann_systemic_2014}. Third, organizations draw arrows between the events to show how they influence each other. For instance, the aforementioned risks may all contribute to higher project costs \shortcite{ackermann_systemic_2014}. Finally, organizations analyze the entire causal map for central events, clusters, feedback loops, and other patterns. For example, when connecting renewables to a local electricity grid, one major concern may be project cost overruns \shortcite{ackermann_systemic_2014}. To draw causal maps, organizations can use whiteboards and sticky notes, or software that has been developed specifically for that purpose \shortcite{wikipedia_list_nodate}.

\begin{figure}[b]
    \centering
    \includegraphics[width=\textwidth]{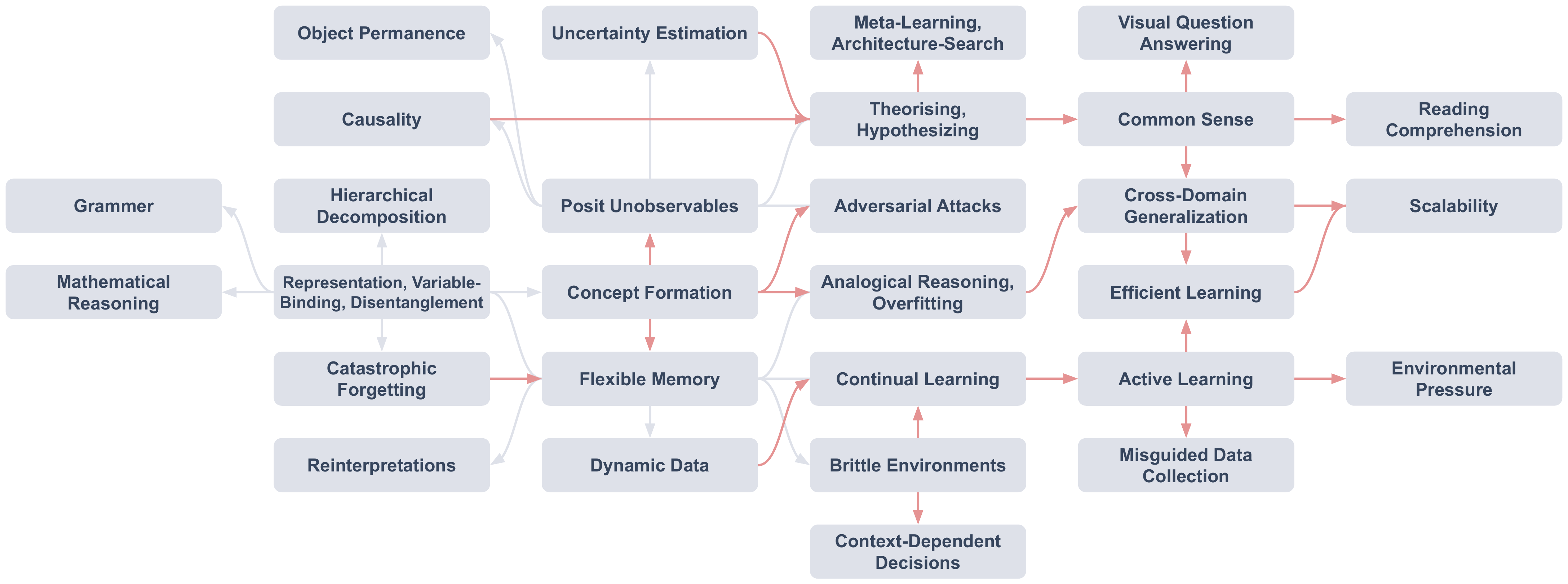}
    \caption{Causal map of warning signs of transformative AI \protect\cite{cremer_artificial_2021}}
    \label{fig6}
\end{figure}

\textbf{How AGI companies could use it.} AGI companies could use causal mapping to build on other techniques which identify risks but neglect their interactions, such as scenario analysis (Section \ref{4.1}) or the fishbone method (Section \ref{4.2}). For example, AGI companies could use causal mapping to explore interdependencies of previously identified dangerous model capabilities and propensities. This may reveal how some of them enable, contribute to, or hinder the development of others and could guide the focus of evals and safety research efforts in general. Similarly, \shortciteA{cremer_artificial_2021} have used causal mapping to investigate milestones for transformative AI. They found that internal representations, memory, and the ability to account for unobservable phenomena may be particularly crucial enablers, and thus warning signs (Figure \ref{fig6}).

\begin{figure}[t!]
    \centering
    \includegraphics[width=\textwidth]{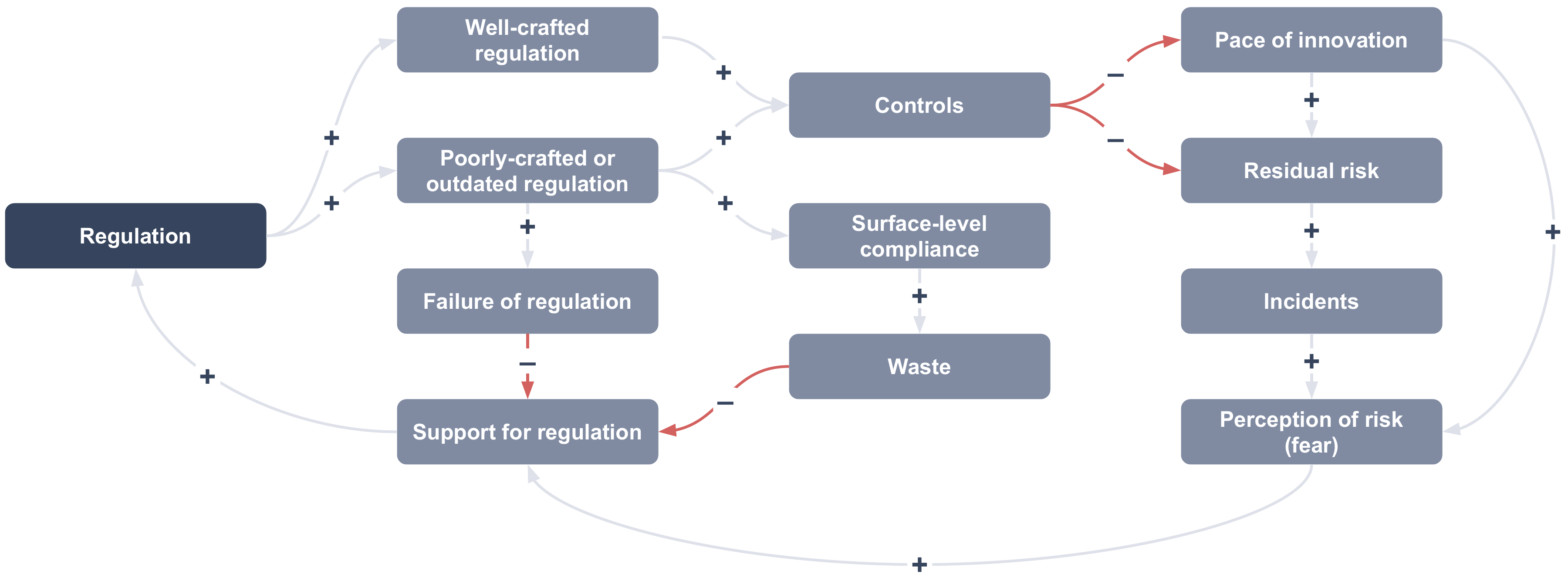}
    \caption{Causal map of feedback loops with regard to regulating risks from AI \protect\cite{losi_system_2023}}
    \label{fig7}
\end{figure}

Causal mapping may be particularly helpful for exploring competitive dynamics, which involve many complex interactions. For example, AGI companies could draw a map of the effects of an external audit framework. Among other things, this may provide information on the importance of standards or regulations, whether AGI companies should announce external audits (before or after they take place), and whether auditors should publish their findings. On a similar topic but higher level of abstraction, \shortciteA{losi_system_2023} has used causal mapping to identify feedback loops with regard to regulating risks from AI. For instance, the diagram shows how ineffective regulation may decrease the support for new regulation (Figure \ref{fig7}).

\textbf{Benefits.} Causal mapping reveals interactions between risks. Without a deliberate attempt to identify these, AGI companies will likely miss some (e.g. how situational awareness affects power-seeking behavior). This would be bad because interactions may be important for efficient and effective risk mitigation (e.g. which dangerous model capabilities or propensities to focus on with evals). The technique is relatively simple and not very time-consuming. Using software for causal mapping has a number of additional benefits. First, it enables anonymous and remote participation \shortcite{iec_31010_2019}. This lowers the barriers for leading experts to work together, and might even reduce concerns around business secrets in cases where these experts are employed by different AGI companies. Second, specialized software streamlines the whole process \shortcite{ackermann_systemic_2014}. Third, causal maps developed with the help of software can be more easily updated as new knowledge is gained \shortcite{cremer_artificial_2021}.

\textbf{Limitations.} Directing a causal mapping exercise with a group can be somewhat challenging for the facilitator. The use of software can help with this to some extent \shortcite{iec_31010_2019}. It should, however, be noted that most software costs money (usually a couple of hundred or thousand dollars). But there is often a free software version AGI companies could try out. Finally, causal mapping is a qualitative technique, meaning it does not provide the likelihood of events. There are specific variations of it that can be quantified, such as fuzzy causal mapping \shortcite{cremer_artificial_2021}. But this is not how the technique is typically used.

\textbf{Recommendations.} We assume that AGI companies already use whiteboards to do some types of causal mapping exercises. We recommend them to try out causal mapping as described in this paper. To that end, we specifically recommend trying out software tools. In order to reap as many benefits as possible from a causal mapping exercise, AGI companies may want to update maps they have developed as new knowledge is gained.

\subsection{Delphi technique}\label{5.2}

The Delphi technique is a particular process of collecting and collating expert judgment. The technique, which was originally developed for military purposes \shortcite{rand_delphi_nodate}, is typically employed in forecasting \shortcite{iec_31010_2019, chapelle_operational_2019, pritchard_risk_2015}, including superforecasting \shortcite{good_judgment_delphineo_nodate}. In the risk analysis context, the Delphi technique can be used to assess the likelihood of risks.

\textbf{How it works.} The detailed procedure is as follows \shortcite{iec_31010_2019, chapelle_operational_2019, pritchard_risk_2015}. First, organizations develop a set of questions that warrant expert input and foresight. For instance, questions could be about the probability or extent of specific events occurring at a given time horizon. As a concrete example, the Delphi technique has been used to estimate the percentage of electric cars within ten years \shortcite{johnson_ten-year_1976}. Second, organizations send the questions to a number of experts (from a handful to a hundred or even more). This can be done through questionnaires or software developed specifically for that purpose \shortcite<e.g.>{good_judgment_delphineo_nodate}. Third, the experts provide their answers to the questions independently, along with their reasoning. Fourth, organizations collect and share the responses among the experts, without revealing which response belongs to whom. This may take the form of sharing all responses or synthesizing them into a summary, but should always include the reasoning behind responses. Fifth, the experts can reassess their initial answers based on the information provided. Organizations continue this cycle of sharing and revising responses until a consensus is reached, or until no further changes in opinions are observed. Typically, about two to four rounds are conducted.

\textbf{How AGI companies could use it.} AGI companies could use the Delphi technique to inform particularly important decisions. For example, before deploying a model, AGI companies could use the Delphi technique to obtain estimates on the likelihood of this provoking specific actions by competitors. \shortciteA{openai_gpt-4_2023} has already relied on professional forecasters in this situation. They predicted, among other things, that delaying the launch of GPT-4 would reduce competitive dynamics. In this situation, the Delphi technique could be used to obtain probabilities for specific risks, like competitors releasing similar models, such as Google’s Bard \shortcite{hsiao_try_2023}, receiving more investments \shortcite<e.g.>{wiggers_anthropic_2023}, or even new AGI companies being founded, such as the European start-up “Mistral AI” \shortcite{bradshaw_four-week-old_2023}. Moving beyond risk assessment into risk treatment, a natural addition may be to ask the experts to what extent measures like delaying the launch by different amounts of time would reduce the probabilities of these risks.

Before pre-training or fine-tuning a model, AGI companies could use the Delphi technique to forecast the likelihood of the emergence of various dangerous model capabilities and propensities. To that end, the experts could be provided with the model architecture, the intended process (e.g. dataset, compute, loss function, optimizer, hyperparameters, etc.), dangerous capabilities and propensities of similar models, and other information deemed relevant. They could then be asked to predict which dangerous model capabilities and propensities may appear or be reinforced by pre-training or fine-tuning as it currently is envisioned. The results could inform changes to the process or safeguards to be put in place in order to keep the model in check.

\textbf{Benefits.} The Delphi technique is a potentially fairly accurate way of estimating the likelihood of risks. For known quantitative values and near-term forecasting, it is usually more accurate than both the average of individual judgments and the results of unstructured group discussions \shortcite{rowe_delphi_1999}. Yet, its accuracy may depend to a large extent on the level of detail of the information exchange between rounds, and the subject matter knowledge as well as forecasting skills of the experts \shortcite{rowe_delphi_1999}. There seems to be only one study on the long-term accuracy of the Delphi technique. This involved a forecasting exercise on the mental health profession in the US spanning 30 years. Out of 18 scenarios that were suggested by the facilitators, participants correctly predicted whether they would occur in 14 instances, and for those accurately predicted the time course within about 1 to 5 years \shortcite{parente_case_2011}. Other benefits are the possibility of anonymous and remote participation \shortcite{iec_31010_2019}.

\textbf{Limitations.} A major limitation of the Delphi technique is that expert forecasting in general has a number of inherent limitations, and is especially hard with regard to unprecedented events \shortcite{armstrong_errors_2014, de_neufville_forecasting_2023, morgan_use_2014}. Therefore, the results of the Delphi technique should not be taken at face value. Another limitation is that if external experts are engaged, they often need proprietary information. Otherwise, the technique is similar to prediction markets where forecasters exchange arguments. While external experts could sign non-disclosure agreements (NDAs), some of them may not want to do so, and AGI companies may still not entrust them with the most sensitive information. Hiring external experts can also be costly. Furthermore, the Delphi technique can be very time-consuming. Depending on the number of experts and cycles, as well as the difficulty of the questions, it can take from a couple of days to weeks or even months. Overall, different versions of the technique are possible, but there is a certain trade-off between effort and thoroughness.

\textbf{Recommendations.} We strongly recommend AGI companies to use the Delphi technique to estimate the likelihood of key risks. The technique seems most warranted in especially important situations, such as before deploying, and potentially before pre-training, fine-tuning, or evaluating a new model \shortcite{schuett_towards_2023}. We recommend AGI companies to involve external forecasters that have a strong track-record, such as Samotsvety (\url{https://samotsvety.org}).

\subsection{Cross-impact analysis}\label{5.3}

Cross-impact analysis is another technique that helps organizations to better understand interactions between different events that contribute to a risk. In contrast to causal mapping (Section \ref{5.1}), which determines causal influences, cross-impact analysis establishes correlations. Cross-impact analysis gathers expert forecasts on the likelihood of events – similar to the Delphi technique (Section \ref{5.2}), but also takes into account the effects of other events that might occur. It can also be considered an elaborate way to develop scenarios for scenario analysis (Section \ref{4.1}) \shortcite{iec_31010_2019, european_foresight_platform_cross-impact_nodate, gordon_cross-impact_1994}.

\begin{figure}[t!]
    \centering
    \includegraphics[width=\textwidth]{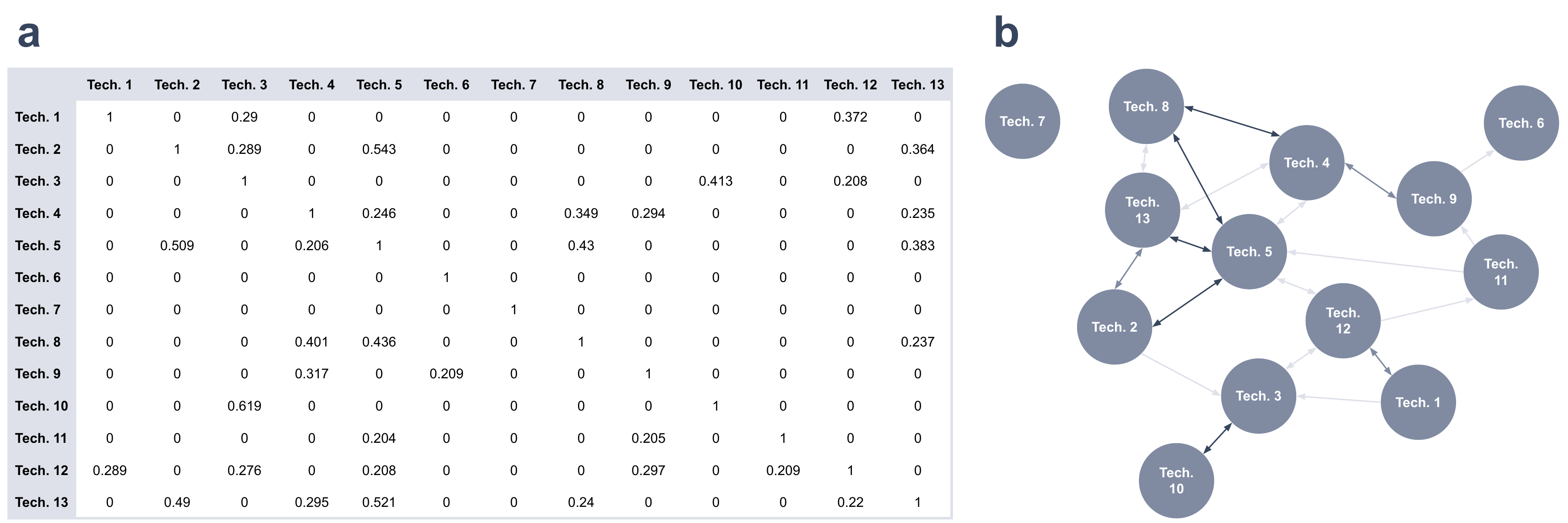}
    \caption{Cross-impact matrix (a) and interdependency map (b) of the sub-technologies of humanoids \protect\cite{kim_hybrid_2016}}
    \label{fig8}
\end{figure}

\textbf{How it works.} Cross-impact analysis is a very complex technique. Basically, it starts with breaking down an issue (e.g. advances in AI) into potentially contributing events (e.g. advances in hardware or algorithms). It then involves gathering expert forecasts on the likelihood of the occurrence of these events, independent and conditional on each of the other events occurring. Through running a computer analysis on these estimates, a map that depicts the strength of the relationship between the different events, and future scenarios of how the whole issue may develop can be generated. Different versions of the technique that require different levels of effort are possible \shortcite{iec_31010_2019}. The simplest option is to rely on software developed specifically for that purpose. While some software is available for free \shortcite<e.g.>{centre_for_interdisciplinary_risk_and_innovation_research_cross-impact_nodate}, consulting firms may also provide software for purchase.

The detailed steps of cross-impact analysis are as follows. To begin with, organizations create a matrix of events they assume to be contributing to an important issue. The matrix should contain all events on both axes. In an example from the literature, relevant events for the creation of humanoids, i.e. robots that resemble a human body, were deemed advancements in the necessary sub-technologies \shortcite{kim_hybrid_2016}. Then, to fill the matrix, organizations ask experts about the likelihood of each event occurring independently and conditional on each of the other events occurring at a given time horizon. These estimates can be quantitative (i.e. probabilities) or qualitative (i.e. indications on a scale, e.g. from -3 to 3) \shortcite{iec_31010_2019, european_foresight_platform_cross-impact_nodate, gordon_cross-impact_1994}.

Afterwards, organizations use software or statistical methods, usually Monte Carlo simulations, to confirm the consistency of the estimates provided by the experts. The resulting “cross-impact matrix” displays stronger and weaker correlations between events and can be visualized in an “interdependency map”. For the creation of humanoids, some sub-technologies were found to be much more correlated than others \shortcite{kim_hybrid_2016}. Finally, again with the help of software or statistical methods, organizations can generate likely future scenarios. They can also perform so-called “sensitivity analyses”. To that end, they need to change specific inputs, i.e. aggregated expert opinions on independent or conditional likelihood, and observe how this affects the cross-impact matrix, the interdependency map, and the scenarios. This can help identify the most influential events – those whose changes have the most significant overall effects \shortcite{iec_31010_2019, european_foresight_platform_cross-impact_nodate, gordon_cross-impact_1994}.

\textbf{How AGI companies could use it.} AGI companies could use cross-impact analysis to inform particularly important decisions. For example, when planning their business strategy for the next one or several years, AGI companies could use cross-impact analysis to gain a better understanding of potential competitive dynamics over this period of time. First, AGI companies would need to assemble potentially contributing events. Such events may include, among other things, new model deployments by other AGI companies, the opening of funding rounds by existing AGI companies, and the founding of new AGI companies. Second, AGI companies could ask experts about the likelihood of each of these events occurring, and about the effect of one event happening on the likelihood of each of the other events occurring. Based on the estimates, they could create a cross-impact matrix and draw an interdependency map. Finally, generating future scenarios and performing sensitivity analysis may be valuable. For instance, AGI companies could test how the decision to release a new model would influence all other events and the likely scenarios. 

Similarly, \shortciteA{kilian_examining_2023} have used cross-impact analysis to generate future scenarios of the global socio-technical AI landscape. They inferred four clusters of possible futures – slow progress and decentralized diffusion of AI technology in an environment of national protectionism and isolation, moderate progress and multipolar diffusion leading to a transition of power from nation states to leading AI companies, moderate progress and decentralized diffusion of various low-capability AI systems in different parts of the economy, and fast, centralized progress originating in a non-Western country. The authors find that each cluster of scenarios implies different risks.

\textbf{Benefits.} Cross-impact analysis is a sophisticated technique that can provide advanced insights on mutual influences of events, possible and likely scenarios, and the impacts of decisions and changes of circumstances. In other words, it provides organizations with three helpful things: expert forecasts on the likelihood of events occurring conditional on other events occurring, scenarios of how the future may unfold, and a method to observe the impact of specific decisions or external changes on all other events and the overall scenarios \shortcite{iec_31010_2019}. Cross-impact analysis might improve the accuracy of forecasts by ensuring their internal consistency \shortcite<see>{schweizer_reflections_2020} and requiring forecasters to consider the impact of different events on each other \shortcite{european_foresight_platform_cross-impact_nodate}. Cross-impact analysis also allows for anonymous and remote participation \shortcite{iec_31010_2019}.

\textbf{Limitations.} While there are theoretical arguments in favor of the accuracy of cross-impact analysis, we could not find empirical studies. As with any expert forecasting technique, its results should be taken with a grain of salt \shortcite{armstrong_errors_2014, de_neufville_forecasting_2023, morgan_use_2014}. The sophistication of cross-impact analysis also comes with the downside of it being very complicated and time-consuming. Gathering expert opinions and analyzing them can take up to several months – depending on the number of experts and events, and the choice of software or statistical method. Generally, the use of any software simplifies the use of the technique to some extent. Another limitation is that Monte Carlo simulations, which are also embedded in most software mentioned above, focus on the most likely scenarios, potentially neglecting extreme outcomes like catastrophic risks \shortcite{iec_31010_2019}.

\textbf{Recommendations.} We encourage AGI companies to use cross-impact analysis before particularly important decisions, such as the deployment of a new model \shortcite{schuett_towards_2023}. AGI companies could start with a group of internal experts and rely on software \shortcite{centre_for_interdisciplinary_risk_and_innovation_research_cross-impact_nodate}. If this proves useful, they may consider engaging external experts, or commissioning research or consultancy organizations with conducting a more thorough analysis. We recommend AGI companies to attempt quantification (i.e. asking for probabilities instead of indications on a scale). 

\subsection{Bow tie analysis}\label{5.4}

Bow tie analysis helps organizations to examine the effectiveness of their controls with respect to different risks. In risk management, controls are mechanisms that are supposed to reduce the likelihood or impact of undesired events. Bow tie analysis involves mapping causes, consequences, and controls of an undesired event in a diagram that resembles a bow tie. It is a very popular and comparatively simple technique \shortcite{iec_31010_2019}.

\begin{figure}[t!]
    \centering
    \includegraphics[width=\textwidth]{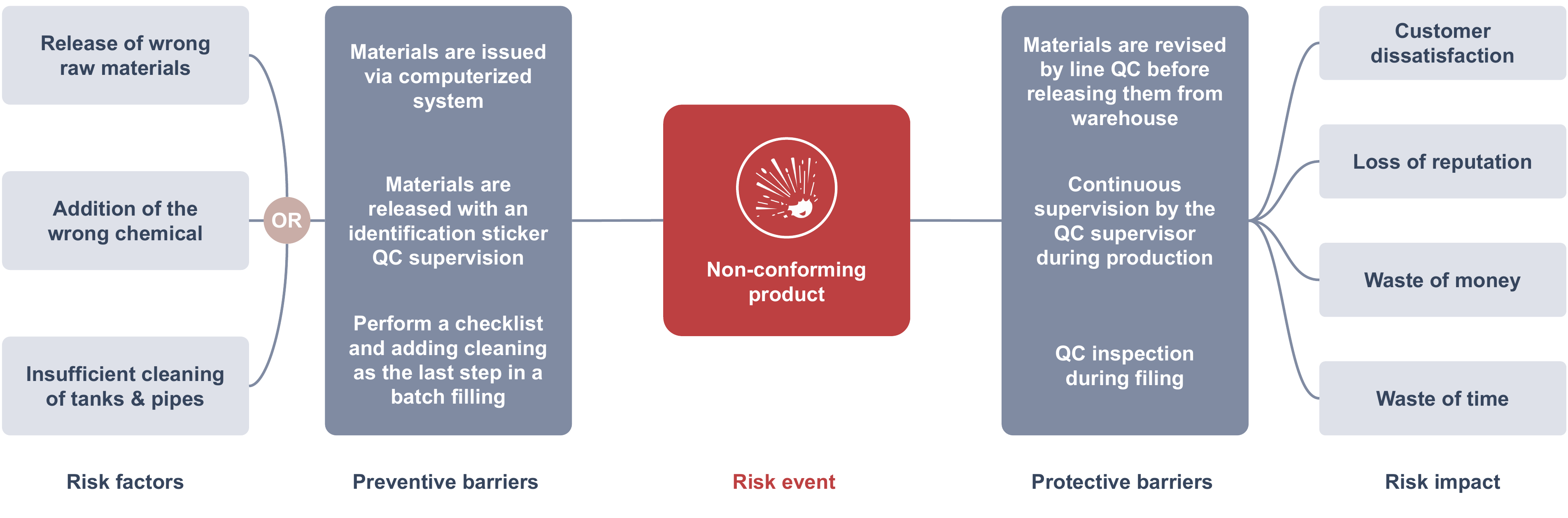}
    \caption{Bow tie diagram for a non-conforming product \protect\cite{aqlan_integrating_2014}}
    \label{fig9}
\end{figure}

\textbf{How it works.} Bow tie analysis consists of the following steps \shortcite{iec_31010_2019, book_lessons_2012, mcconnell_safety_2006}. First, organizations choose an undesired event and place it at the center of the diagram. For example, this could be a product that does not fulfill its requirements \shortcite{aqlan_integrating_2014}. Next, they collect the causes that could lead to this event and position them on the diagram's left side. A cause for a non-conforming product may be the accidental use of wrong materials \shortcite{aqlan_integrating_2014}. Organizations then identify preventive controls, which are measures that aim to avert the undesired event, and place them between the causes and the central knot. For instance, this could include labeling materials with stickers \shortcite{aqlan_integrating_2014}. Subsequently, they determine the possible consequences of the undesired event and situate them on the diagram's right side. A consequence of a non-conforming product may be dissatisfaction of customers \shortcite{aqlan_integrating_2014}. Organizations then identify reactive controls, which are measures supposed to minimize the event's impact after it has occurred, and place them between the consequences and the central knot. For instance, this could be reviewing products before they leave the warehouse \shortcite{aqlan_integrating_2014}.

In more sophisticated versions, organizations also determine escalation factors or conditions that could cause the controls to fail or become less effective, as well as controls that address these escalation factors. Finally, they incorporate ongoing activities that contribute to maintaining and enhancing the effectiveness of all controls. For example, these may include design improvements, maintenance, verifications, procedures, checklists, guidelines, trainings, oversight, audits, and inspections \shortcite{iec_31010_2019}.

\textbf{How AGI companies could use it.} AGI companies could draw bow tie diagrams to examine their controls both with regard to undesired actions by an “agentic” model as well as malicious actors. To that end, they could build on the causes identified with the fishbone method (Section \ref{4.2}), as well as the causes and consequences developed through causal mapping (Section \ref{5.1}).

For example, AGI companies could use bow tie analysis to depict the event of the model copying itself during or right after training. Relevant causes may include the model having access to the training software or additional hardware, or being able to convince humans to provide such access. To impede these causes, preventive controls may comprise access barriers for both the model and humans interacting with the model (e.g. air gapping or multi-party authentication). Potential consequences of the model copying itself may include the model or copies of it evading off-switch mechanisms, uncertainty about whether all copies of the model have been found, and the whole event going unnoticed. To mitigate these, reactive controls may encompass, again, physical access controls for both the model and humans interacting with it, automated mechanisms to detect model copies, and incident response plans.

\begin{figure}[t!]
    \centering
    \includegraphics[width=\textwidth]{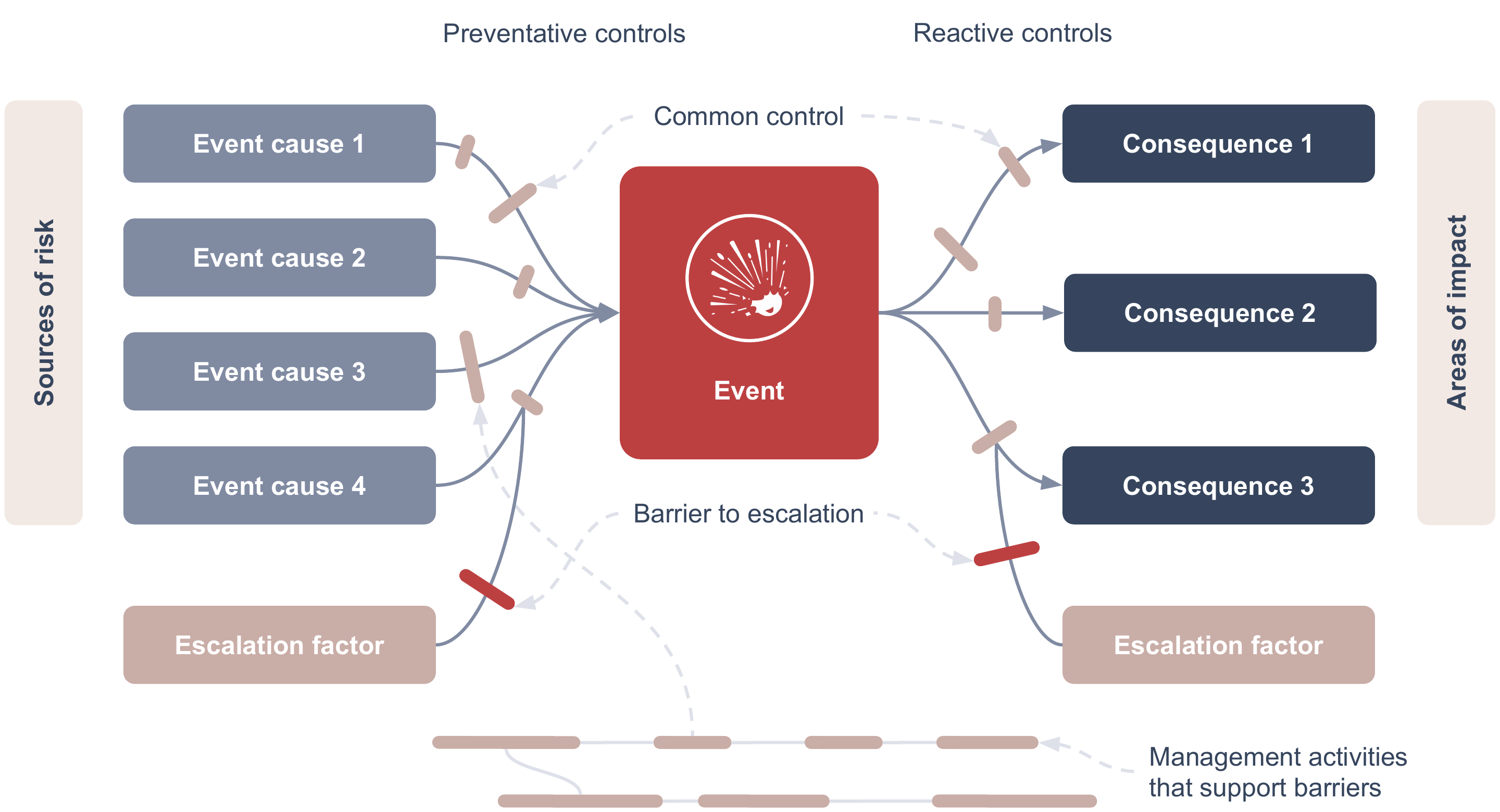}
    \caption{Bow tie diagram template \protect\cite{iec_31010_2019}}
    \label{fig10}
\end{figure}

While monitoring a deployed model, AGI companies could use bow tie analysis to picture the undesired event of users bypassing API safeguards, like rate limits. Potential causes for this event may include users creating multiple accounts and thus increasing the number of their API keys, hacking into other users’ accounts and stealing API keys, or using the same API keys on several devices at the same time. Preventive controls that might hinder these causes to result in the bypassing of rate limits may include various authentication mechanisms, IP-based rate limiting, and limits on the number of devices that can use a single API key at the same time. Should users still manage to bypass rate limits, an exemplary consequence may be users training other models on a large number of outputs. Already, researchers have done this in a presumably legal way \shortcite{peng_instruction_2023, taori_alpaca_2023, the_vicuna_team_vicuna_2023}. Reactive controls that help mitigate these consequences include automated alerts for abnormal usage patterns, human review of use cases, and suspension of offending accounts.

\textbf{Benefits.} In contrast to most other techniques, bow tie analysis focuses on controls which gives a more complete picture of the actual level or risk. Its visualization further helps with understanding and communicating risks inside or outside the AGI company, such as to leadership or auditors. Bow tie analysis has the additional benefits of being simple and less time-consuming than other techniques. Beyond risk assessment, it may reveal gaps and necessary improvements, thus providing guidance for risk treatment \shortcite{iec_31010_2019}.

\textbf{Limitations.} The flipside of its simplicity is that bow tie analysis is prone to over-simplification. It assumes a linear causal chain as it ignores interactions between different causes, consequences, and controls. It is therefore not suitable for analyzing risks which involve complex interactions between events (e.g. competitive dynamics) \shortcite{iec_31010_2019}. Bow tie analysis may often make knowledge explicit rather than generating new knowledge \shortcite{mcconnell_safety_2006}. Finally, some maps developed through bow tie analysis need to be very well protected, because they would reveal sensible information that malicious actors could exploit. 

\textbf{Recommendations.} We recommend AGI companies to use bow tie analysis to take a systematic approach to their controls. Since the technique is simple and easy to create, AGI companies could try it out and see whether it is helpful. AGI companies should also update the maps as changes are made and observed, and new knowledge is gained.

\subsection{System-theoretic process analysis (STPA)}\label{5.5}

System-theoretic process analysis (STPA) helps organizations to assess the effectiveness of their overall control structure. Compared to bow tie analysis (Section \ref{5.4}), STPA is much more sophisticated and complicated. It has been developed to deal with the increasing complexity of systems, especially systems that entail software. The technique assumes that undesired events may not only be caused by failures of individual components of a system, but also by complex interactions between them. STPA involves backward reasoning from undesired events to controls and why they might not have the intended effect \shortcite{khlaaf_toward_2023, leveson_stpa_2018, rausand_risk_2020}.

\begin{figure}[t!]
    \centering
    \includegraphics[width=0.82\textwidth]{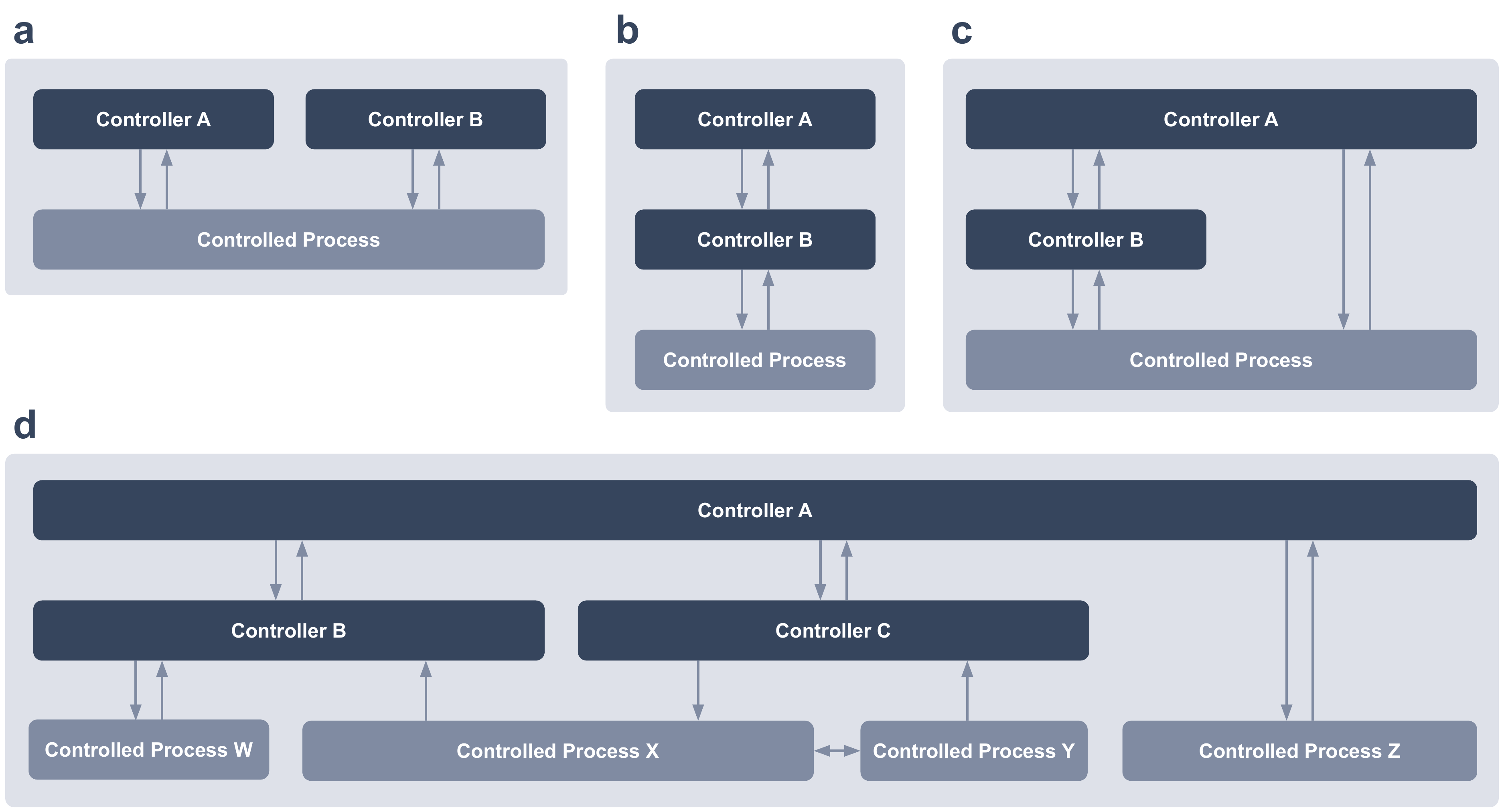}
    \caption{Simple examples of system control structures \protect\cite{leveson_stpa_2018}}
    \label{fig11}
\end{figure}

\begin{figure}[t!]
    \centering
    \includegraphics[width=\textwidth]{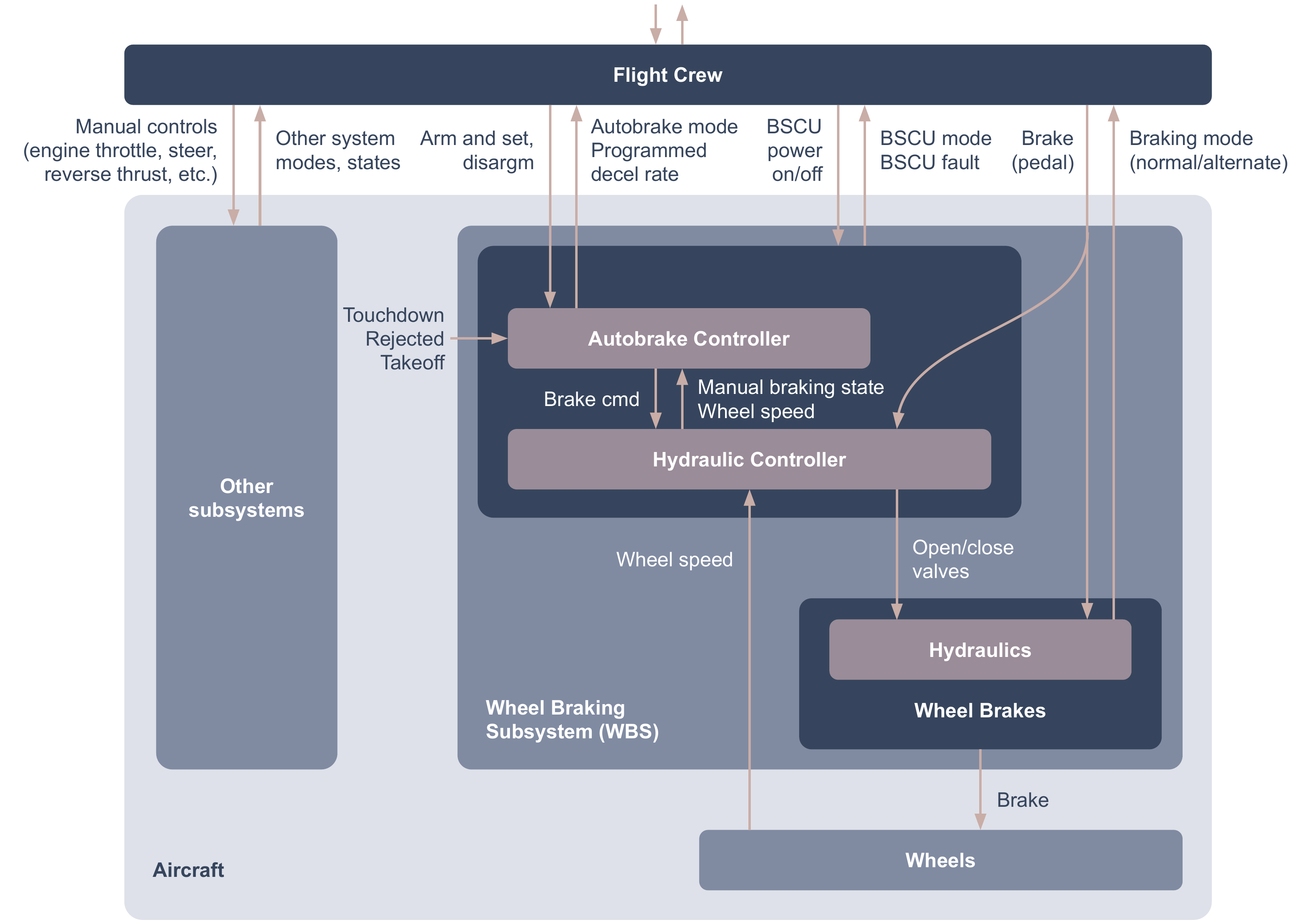}
    \caption{System control structure of a plane’s wheel braking subsystem \protect\cite{leveson_stpa_2018}}
    \label{fig12}
\end{figure}

\textbf{How it works.} STPA is a very complex technique. It is based on a different theory of risk assessment and, as such, uses a different terminology. First, organizations generate background information, namely, define a system (e.g. a particular model that interacts with the real world in various ways) and determine undesired events that could stem from this system (e.g. the model telling a user how to build a weapon). Then, organizations model the various ways in which the behaviors or states of the system are constrained to prevent all of these undesired events (e.g. no internet access, hard-coded responses, review of use cases). STPA provides guidance on questions that can help to identify the many and often unexpected reasons why these constraints may not work as intended (e.g. over the course of many interactions with the system, a researcher might become falsely and secretly convinced that the model is conscious and provides it with internet access to “liberate” it) \shortcite{leveson_stpa_2018, rausand_risk_2020}.

The detailed steps of STPA are the following \shortcite{leveson_stpa_2018, rausand_risk_2020}. First, organizations define the system to be analyzed, with a clear outline of its boundaries. The system, as opposed to its environment, may only include parts over which the organization has some control. Next, organizations compile “losses” (undesired events), “system-level hazards” (states of the system that may lead to losses), and “system-level constraints” (states of the system that may prevent losses) in a list or table. For example, in aviation, losses include human fatalities or injuries, as well as damage to the plane or other objects. System-level hazards encompass the plane leaving its designated runway or the plane coming too close to other objects. On the other hand, system-level constraints are their inversions, namely, the plane keeping its designated runway and the plane maintaining a safe distance to other objects \shortcite{leveson_stpa_2018}.

After generating this background information, organizations draw a model of the “system control structure” (the various ways in which the behavior and states of the overall system are controlled). This includes “controllers” (technical or human entities that decide about control actions), “control actions” (actions by controllers to enforce constraints), “controlled processes” (processes controlled by controllers), “actuators” (mechanisms through which controllers act upon their controlled processes), and “sensors” (mechanisms through which controllers receive feedback from their controlled processes). The process of creating this model progresses from abstract to more and more concrete. In particular, actuators and sensors can be added during later stages of the analysis. For example, the control structure of a plane’s wheel braking subsystem can be modeled within the aircraft and interacting with the flight crew. The flight crew is a human controller, whereas autobrake and hydraulic controllers are technical ones. These controllers work together in a hierarchical control structure to control the processes of hydraulics and ultimately wheels. There are various control actions, such as power buttons and brakes \shortcite{leveson_stpa_2018}.

Once the model of the control structure is complete, organizations identify “unsafe control actions (UCAs)”. A control action is considered unsafe if it could lead to a hazard. UCAs can involve actions that are not provided, actions that are unsafe in their context, actions that are improperly timed, or continuous actions that last too long or stop too soon. In the example from aviation, a UCA is “flight crew does not provide manual braking during landing when autobrake is not providing braking or is providing insufficient braking”. Then, organizations determine the causes of these UCAs. These can range from physical failures of controllers, incorrect input from sensors or other controllers, inadequate decision algorithms, and incorrect process models to issues with control action transmission, actuator response, or controlled process response. A potential cause for these two controllers in the aviation example failing may be “wheel speed signals may momentarily reach zero during anti-skid operation, causing both controllers to build flawed process models”. UCAs and their different causes are compiled in a long list or table. Finally, based on the whole analysis, organizations can determine additional necessary system-level constraints (note that this goes beyond risk assessment and already is part of risk treatment) \shortcite{leveson_stpa_2018}.

\textbf{How AGI companies could use it.} STPA, designed for inherently dangerous systems, is probably most helpful for controlling AI systems, not aligning them. For example, before pre-training or fine-tuning a new model, STPA may help AGI companies assess whether they are prepared to keep in check an unprecedentedly capable and potentially dangerous model. In this example, the system includes the model, hardware and software used to train it, and humans involved in the process. A loss would be the model escaping human control. System-level hazards, such as the model avoiding shutdown, copying or improving itself, gaining internet access, or collaborating with other models, can be mitigated through various constraints, such as physical access barriers to hardware and software for both the model and the humans interacting with it. The control structure model consists of human controllers (engineers, researchers, managers) and technical controllers (automated mechanisms, other AI systems). Control actions may include shutdown mechanisms, actuators could be changes to the model’s architecture, and sensors could involve evals conducted during training. As an overarching example, UCAs involving controllers, actuators, or sensors failing to respond to triggers might be caused by the model deceiving or manipulating them. 

\textbf{Benefits.} The inventors of the technique claim that it has been shown to identify more causes of undesired events than other risk assessment techniques \shortcite{leveson_stpa_2018}. It has recently become increasingly popular \shortcite{rausand_risk_2020}. Because of the technical and societal complexity of risks from AI, STPA has already been argued to be more adequate to this context \shortcite{dobbe_system_2022, khlaaf_toward_2023}. At the very least, the technique provides a different perspective than other risk assessment techniques which is a feature in itself (Section \ref{7}).

\textbf{Limitations.} The technique’s main limitation is its complexity. There is also no standardized approach on how to use it \shortcite{rausand_risk_2020}. However, we have outlined the common steps which AGI companies could follow. The handbooks and textbooks referenced provide additional details and guidance. Another comparatively minor limitation is that the technique’s usefulness over more established risk assessment techniques is still contested. STPA may help to find different rather than additional causes of undesired events, and thus should be used to obtain a complementary rather than a supplementary perspective on risks \shortcite{rausand_risk_2020}.

\textbf{Recommendations.} We encourage AGI companies to use STPA to think through how they can remain in control over increasingly capable and potentially dangerous models. Nevertheless, as STPA is a time-consuming technique – both to learn and to apply – we recommend starting with bow tie analysis and using STPA to expand on complex issues. AGI companies could also commission a research or consultancy organization with doing this.

\section{Risk evaluation}\label{6}

Risk evaluation is the third step in the risk assessment process. It aims to determine whether a specific risk is acceptable or whether its treatment is warranted. To this end, organizations compare the results of the risk analysis (Section \ref{5}) with the individual and overall risk they are willing to take and able to tolerate (so-called “risk appetite” and “risk tolerance”) \shortcite{iec_31010_2019}.\footnote{How to establish an organization’s risk appetite and risk tolerance is a question beyond risk assessment and thus the scope of this paper (Section \ref{8}). For the purposes of this section, we simply assume that the organization’s risk appetite and risk tolerance have been defined.} According to our selection criteria, we chose the following two risk evaluation techniques: checklists (Section \ref{6.1}) and risk matrices (Section \ref{6.2}).

\subsection{Checklists}\label{6.1}

Checklists are questionnaires to be used in pre-defined situations \shortcite{iec_31010_2019}. As a risk evaluation technique, checklists contain questions that help to identify risks associated with a certain decision or project, and judge whether countermeasures are necessary.

\textbf{How they work.} To develop a checklist, organizations determine a relevant situation and set up a list of questions to be answered in that situation. Checklists may consist of box-ticking, yes/no questions, or open-ended questions, and can be mandatory or voluntary. As a risk evaluation technique, they contain questions that help assess whether and to what extent a particular decision or project contributes to risks. They may also entail decision rules, such as “if X, then do not do Y” or “if Z, then consult with management”. Alternatively, checklists that have been completed by employees (e.g. researchers) may be reviewed by a particular person or team by default (e.g. a governance or risk management team). For example, a common type of checklists used for risk evaluation are so-called “impact assessments” (e.g. privacy or environmental impact assessments). Impact assessment checklists contain questions about the effects of a specific decision or project on individuals, society, or the environment \shortcite<e.g.>{uk_information_commissioners_office_data_nodate, us_environmental_protection_agency_guidelines_1998}.

\textbf{How AGI companies could use them.} AGI companies could develop checklists for relatively frequent situations that in sum may substantially affect catastrophic risks from AI. Risk typologies and taxonomies (Section \ref{4.3}) can be helpful for developing those checklists, both for determining situations in which they may be useful (e.g. based on the AI system lifecycle) and for coming up with their content (e.g. the different risks to be evaluated). For blueprints of how checklists for risks from AI could be structured, AGI companies may want to look at impact assessment checklists on trustworthy AI \shortcite{eu_high-level_expert_group_on_ai_assessment_2020}, AI fairness \shortcite{madaio_ai_2020}, and responsible AI \shortcite{microsoft_responsible_2022}. For instance, some suitable situations may include choosing a new research project (e.g. to what extent will it contribute to progress in capabilities and safety respectively), making choices about a model’s architecture or training (e.g. what will be the gain in performance versus the increase in risks of different options), publishing a piece of research (e.g. what will be the effects on competitive dynamics and misuse risks), communicating with external parties (e.g. what will be the effects on competitive dynamics), and monitoring a deployed model (e.g. what instances of misuse have occurred). 

To our knowledge, only a few checklists suitable for evaluating catastrophic risks from AI have already been developed. \shortciteA[Annex C]{hendrycks_x-risk_2022} suggest a checklist for assessing the impact of a potential research project on existential risks from AI. It assumes that research projects will often advance both capabilities and safety as well as have relevant indirect effects (e.g. on competitive dynamics). In light of this, the checklist aims to support individual researchers to decide whether they should pursue a certain project. \shortciteA[Section 3.2.2.1.1]{barrett_actionable_2023} provide a starting point for a pre-development and pre-deployment checklist for catastrophic risks from AI. It asks questions about the type (e.g. health, fundamental rights, national security) and magnitude (e.g. multiple fatalities or negligible harm) of adverse effects of AI systems on individuals, groups, organizations, and society. The checklist concretizes the NIST AI Risk Management Framework \shortcite{nist_artificial_2023}, but may need to be concretized even further (e.g. it may need to be adapted to specific situations).

\textbf{Benefits.} Checklists may save time because they can be used multiple times. They can also decentralize risk evaluation. Instead of a specific team alone being in charge, checklists allow organizations to require all employees to take part in risk evaluation. At the same time, through the design of checklists, organizations remain in control over the focus and substance of this risk evaluation. Checklists may also increase overall risk awareness within the AGI company, and thus contribute to a so-called “safety culture”, where employees generally aim to minimize risks in their everyday work.

\textbf{Limitations.} On the other hand, checklists are time-consuming to develop. The introduction of checklists may also increase the workload for employees. Therefore, the costs and benefits of requiring the use of a checklist in a particular situation need to be weighted against each other, and checklists should be as short and simple as possible. Moreover, the introduction of checklists may be perceived as annoying and burdensome by employees. As a result, employees may ignore checklists or only apply them superficially. To prevent this, good communication about the purpose of checklists is necessary. Checklists are also prone to missing “unknown unknowns”. This can be alleviated through more open-ended prompts that encourage creative thinking. However, this in turn necessitates more expertise by the user, leading to a tradeoff between simplicity and usefulness \shortcite{iec_31010_2019}.

\textbf{Recommendations.} We encourage AGI companies to use checklists, especially ones that have already been developed \shortcite<e.g.>[Annex C]{hendrycks_x-risk_2022}. The larger an organization, the more routine decisions and projects that could have an impact on risks it entails. This makes checklists increasingly worth developing as organizations scale. They may thus be particularly relevant for larger AGI companies like OpenAI and Google DeepMind. AGI companies should monitor correct usage of checklists and update them along with new developments and insights, particularly when new risks are identified.

\subsection{Risk matrices}\label{6.2}

A risk matrix, also known as heat map or consequence/likelihood matrix, is a table that contains consequence and likelihood ratings of different risks, often on a scale from 1 to 5. Each cell represents a specific combination of consequence and likelihood. Different risks can be plotted on the matrix to determine the need and priority of addressing them \shortcite{iec_31010_2019}. Risk matrices are one of the most common risk evaluation techniques.

\begin{figure}[t!]
    \centering
    \includegraphics[width=\textwidth]{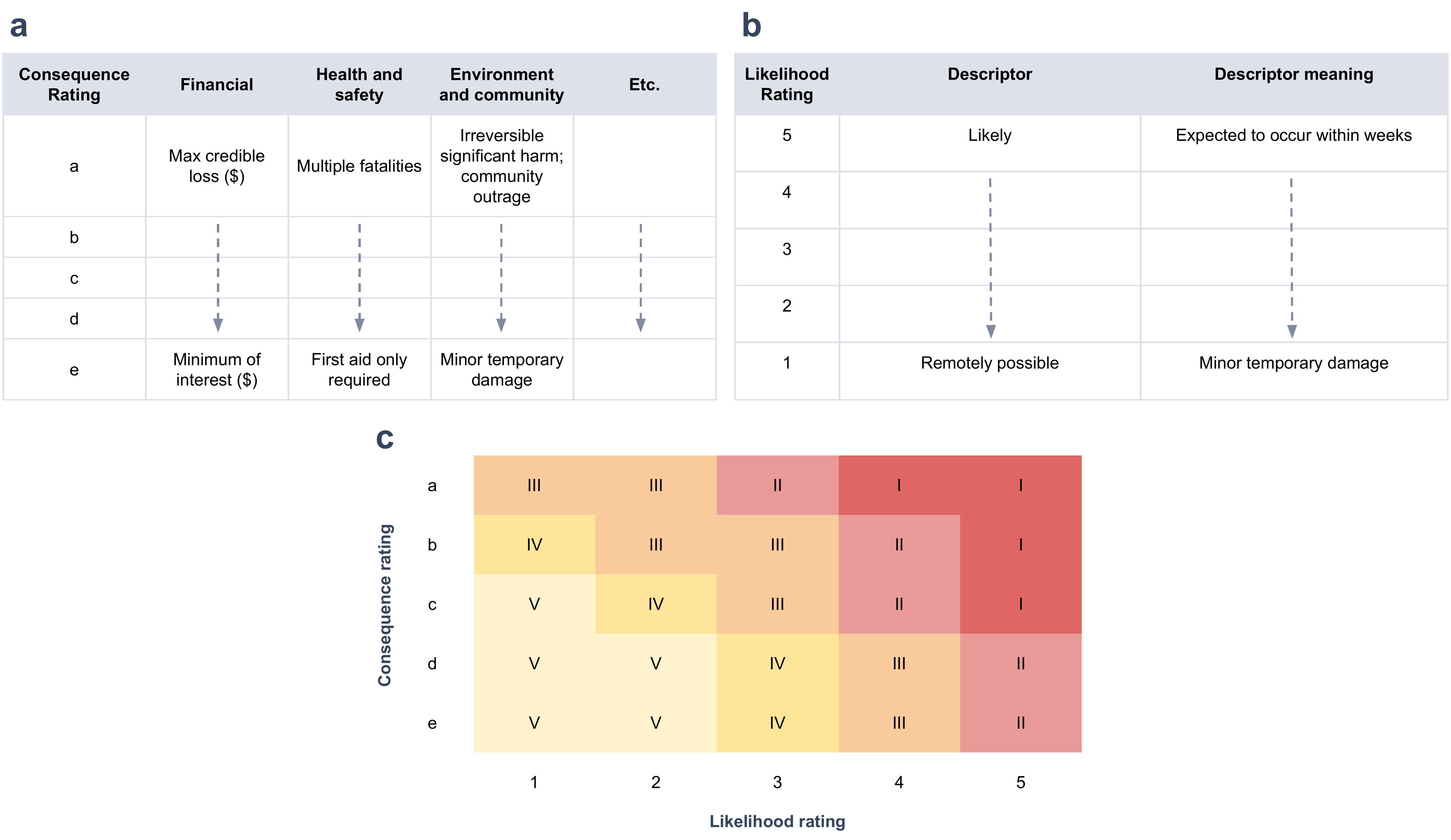}
    \caption{Segments of consequence and likelihood scales and an asymmetric risk matrix (which gives more weight to likelihood than consequence) \protect\cite{iec_31010_2019}}
    \label{fig13}
\end{figure}

\textbf{How it works.} To set up a risk matrix, organizations need to develop and combine two scales: one that ranks the consequence of risks, and another that ranks their likelihood. Typically, both scales have three to five points. The consequence scale can be quantitative (e.g. by cost or fatalities) or qualitative (e.g. catastrophic, major, moderate, minor). For different types of risks, separate definitions of the categories may be necessary. For instance, financial risks may be expressed in monetary terms (quantitative), health risks in number and severity of people affected (semi-quantitative), and reputation risks in purely qualitative terms. The likelihood scale can be quantitative (i.e. probabilities) or qualitative (e.g. frequent, probable, occasional, remote, improbable). For the likelihood scale, a time horizon of that risk occurring should be specified (e.g. within the next year). Then, organizations need to combine both scales into a consequence/likelihood matrix and develop an ordinal scale for ranking the priority of risks, typically with three to five points (I, II, III, IV, V). Decision rules can be linked to priority categories, such as that risks of both highest consequence and likelihood must always be treated. The final matrix allows organizations to locate individual risks by their consequence and likelihood descriptor \shortcite{iec_31010_2019}.

\textbf{How AGI companies could use them.} AGI companies could use a risk matrix to prioritize among different ultimate catastrophic risks from AI. Yet, we note that ultimate catastrophic risks from AI may be extremely hard to evaluate. Consequence ratings will likely depend on fundamental ethical questions, such as the value of happiness, suffering, and other moral goods like equality. While 1 trillion fatalities may clearly be worse than 100 million fatalities, it is very hard to compare these outcomes to, for example, lasting environmental damage or epistemic degradation through misinformation or polarization. Likelihood estimates for these scenarios may also be very difficult to obtain. To that end, AGI companies could use the Delphi technique (Section \ref{5.2}) and cross-impact analysis (Section \ref{5.3}). However, as highlighted in the limitations of these techniques, their results may be highly speculative. 

AGI companies could instead focus on the sources of catastrophic risks from AI. When doing this, it is important to aim for consistency in the level of granularity (see limitations below). For example, the misuse risk “terrorist attack with a novel pathogen” could be split into the risks of users bypassing various safeguards ingrained in the model or implemented through access design. The accident risk “takeover by misaligned AI”, could be broken down into the risks of the emergence of various dangerous model capabilities and propensities.

AGI companies could also use risk matrices to compare catastrophic risks from AI to other types of risks. This would allow them to make overall decisions on which risks to prioritize. \shortcite{khlaaf_toward_2023} has developed a risk matrix for societal risks from AI (where “catastrophic” is defined as “death, permanent total disability, direct harm, significant system or asset loss, or irreversible significant environmental impact”). AGI companies could build on that and add “global” as the highest category of consequence (e.g. global, catastrophic, major, moderate, minor). They could also include types of risks that do not stem directly from AI, such as financial, reputation, or IP risks. In light of the high uncertainty around catastrophic risks from AI, AGI companies could also design the priority scale to give more weight to consequence than likelihood. For example, the column at the intersection of “global” (highest consequence) and third-highest likelihood may still be of highest priority, while the column at the intersection of “major” (third-highest consequence) and highest likelihood may only be of second-highest priority (the opposite is the case in Figure \ref{fig13}).

Another option for designing a risk matrix for both catastrophic risks and their sources is to substitute likelihood with vulnerability. Risk can be conceptualized as the concurrence of threats and vulnerabilities \shortcite{mcchrystal_risk_2021}. While a threat is the source of a risk, a vulnerability is a necessary condition of the asset at stake \shortcite{aven_risk_2015}. In other words, a vulnerability is a weakness in a system which makes it susceptible to harm \shortcite{ostrom_risk_2019}. Assessing vulnerability rather than likelihood is an emerging paradigm for infrequent and highly uncertain risks \shortcite<e.g.>{baylon_risk_2020, ritchey_modelling_2004, uk_royal_academy_of_engineering_building_2023}. To that end, the process for developing the risk matrix remains the same except that the second dimension is vulnerability instead of likelihood. The vulnerability scale is qualitative and categories may simply be “high”, “medium”, and “low”. To assess vulnerability, the most relevant criteria are the the existence and effectiveness of preventive and reactive controls, which AGI companies can determine using bow tie analysis (Section \ref{5.4}) and STPA (Section \ref{5.5}). Other relevant criteria may be origin (e.g. external or internal), driver (e.g. intentionality or negligence), and velocity of the risk (i.e. the speed at which it would materialize). The different criteria can be weighted to obtain a single vulnerability score and assign a vulnerability category.

\textbf{Benefits.} Risk matrices are easy to use. They allow comparing and prioritizing between different risks. They provide a clear visualization \shortcite{iec_31010_2019}, which makes it easier to communicate risks internally and externally, such as with leadership or auditors. Risk matrices can also be used to compare and prioritize between different types of risks \shortcite{iec_31010_2019}, which allows AGI companies to integrate catastrophic risks from AI in their overall risk evaluation.

\textbf{Limitations.} Developing a risk matrix is somewhat difficult. First, the consequence scale may sometimes and the priority scale does always involve normative judgments. There is no single correct answer to which types of consequence are of the same severity (e.g. a specific number of facilities and a specific amount of financial costs) or how to weigh consequence and likelihood (e.g. symmetrically or not). This may complicate the process as different people may have diverging opinions. Second, the rating of a risk on consequence and likelihood depends on the level of detail in which it is described. A higher level of detail leads to a higher number of risks, each with lower consequence and likelihood ratings. Therefore, the level of detail should be as consistent as possible among different risks. Otherwise, valid comparisons of risks cannot be made. Finally, priority rankings can mislead users of risk matrices to make invalid comparisons of risks. However, the consequence scale is ordinal (e.g. steps from one category of consequence to the next may not be of equal size). As a result, two risks of the lowest priority category together do not necessarily have the same priority as a single risk of the second-lowest priority category \shortcite{iec_31010_2019}.

\textbf{Recommendations.} We strongly recommend AGI companies to use risk matrices if this is not already the case. In addition to evaluating the ultimate catastrophic risks from AI, they could break risks down into their sources and evaluate those. In cases where uncertainty about likelihood proves too high, AGI companies could try out using vulnerability as the second dimension instead.

\section{Discussion}\label{7}

\textbf{When to conduct risk assessments.} If risk assessments are not up to date, important risks can be missed. Therefore, risk assessment should happen continuously and iteratively \shortcite{iso_31000_2018, iso_23894_2023, iec_31010_2019, nist_artificial_2023}. Ideally, risk assessments are conducted both in the abstract (initially to establish the risk universe, at regular intervals, and whenever relevant changes of circumstances occur), and before making concrete decisions. The focus at AGI companies so far has been on pre-deployment risk assessment \shortcite{brundage_lessons_2022, kavukcuoglu_how_2022, openai_gpt-4_2023}. However, experts widely agree that AGI companies need to begin even earlier \shortcite{schuett_towards_2023}. In particular, AGI companies should also conduct risk assessments before training a new model – this may include risk assessments before any step of the training process, such as before conducting smaller training runs, before conducting the final large training run, before fine-tuning, and before conducting evals. It is also important to constantly evaluate and improve risk assessment practices \shortcite{iso_31000_2018, iso_23894_2023, iec_31010_2019, nist_artificial_2023}. AGI companies should ensure this is happening, for example, through an internal audit team \shortcite{schuett_agi_2023}.

\textbf{When to use which risk assessment technique.} There is no specific trigger for each of the techniques discussed in this paper. Instead, they can all be applied in various situations. However, each technique has a particular focus or function (for an overview, see Table 1). AGI companies can use techniques from the three risk assessment steps (risk identification, risk analysis, and risk evaluation) to build on one another. Within risk analysis, there are techniques for analyzing dependencies and interactions between different risk sources (causal mapping, cross-impact analysis), techniques for estimating the likelihood of risks (Delphi technique, cross-impact analysis), and techniques for analyzing controls (bow tie analysis, STPA). AGI companies should conduct risk assessments both in the abstract and to inform concrete decisions. Within risk identification, some techniques are best applied to provide background information (scenario analysis, risk typologies and taxonomies), while some risk evaluation techniques only make sense to help with concrete decisions (checklists). All other techniques can be applied both in the abstract and with regard to concrete decisions. Moreover, some techniques are easy to use (fishbone method, bow tie analysis), while others are very complex (cross-impact analysis, STPA). In choosing among more or less costly techniques, AGI companies should consider their value of information \shortcite{barrett_value_2017}. Finally, for “very novel, complex or challenging issues, where there is high uncertainty and little experience”, it is recommended to use multiple techniques and include a variety of stakeholder views \shortcite{iec_31010_2019, potts_assessing_2014}. Therefore, even if the focus and function of different techniques overlap, AGI companies should use several techniques in a given situation (Figure \ref{fig1}).

\textbf{How to use any risk assessment technique.} Some overarching factors can greatly affect the usefulness of any risk assessment technique. First, it must be ensured that the people involved know how to properly use the technique and have the relevant subject-matter knowledge. Otherwise, the results of techniques may be less useful, or even create a false sense of security \shortcite{khlaaf_toward_2023}. It may thus be helpful for AGI companies to hire or contract experienced risk analysts. Second, many techniques require input from different people within the organization. It is therefore common and highly recommendable to organize cross-functional workshops \shortcite{stolzer_safety_2023}. Third, risk assessment should not be limited to technical factors, but also take into account human and organizational factors. This has been an important realization in aviation over the last decades \shortcite{muller_fundamentals_2014}. Finally, making use of risk assessments techniques is only valuable if their results have a bearing on decisions. Implementing risk assessment techniques may be easier and less costly than actually following through on their results. It is essential that AGI companies set up structures and processes that ensure this.

\section{Conclusion}\label{8}

\textbf{Summary.} In this paper, we have discussed ten popular risk assessment techniques from established industries like finance, aviation, nuclear, and biolabs. For each of the techniques, we have described how they work and how AGI companies could use them, in many cases by providing concrete examples. We have also discussed their benefits and limitations in the context of AGI companies assessing catastrophic risks from AI, and made recommendations. Finally, we have discussed when to conduct risk assessments, when to use which technique, and how to use any risk assessment technique.

\textbf{Contributions.} We have made three main contributions to the existing literature. First, we have developed criteria for selecting risk assessment techniques from other industries. These criteria may also be useful for other efforts that aim to identify best practices from other industries suitable for the context of AGI companies and catastrophic risks from AI. Second, building on existing research which has singled out specific techniques, we have conducted a comprehensive review of established risk assessment techniques. Third, in contrast to previous research, we have focused on AGI companies and made actionable recommendations for how they could implement the techniques we selected.

\textbf{Limitations.} This paper has four main limitations. First, we have singled out catastrophic risks from AI. Yet, these risks are intertwined with many other risks AGI companies face and expose their environment to. For instance, AGI companies that expect financial difficulties may be tempted to reduce their safety efforts. Therefore, a full understanding of catastrophic risks from AI may only be possible through considering all types of risks that are relevant for AGI companies, and taking into account phenomena like compound or cascading risks. Second, this paper has looked at popular risk assessment techniques from other industries. While we believe that the techniques we included in this paper may be useful, just as in every other industry, techniques specifically tailored to the AI context will need to be developed, too. Third, we have left out many techniques we expect to currently be less relevant for AGI companies. However, this may change in the future (Section \ref{3}). Finally, although we have tried to apply the techniques as concretely as possible to the context of AGI companies, our suggestions remain fairly general. The techniques need to be customized to the particularities of individual AGI companies.

\textbf{Questions for further research.} Many questions around risk assessment at AGI companies require further research. For example, the question of how AGI companies should define what level of risk they are willing to take (risk appetite) and able to tolerate (risk tolerance) remains open. The definition of a company’s risk appetite and tolerance should be informed by the results of risk assessments, but they are also required to evaluate in the abstract what level of risk is acceptable. Another question that warrants further research is how AGI companies can improve their safety culture \shortcite{hendrycks_overview_2023, manheim_building_2023, shneiderman_bridging_2020}. The question of how AGI companies should document and report the results of risk assessments also needs to be investigated more. Traditionally, risk registers are an important tool in this regard, but there are also AI-specific tools, such as datasheets \shortcite{bender_data_2018, gebru_datasheets_2021}, model cards \shortcite{mitchell_model_2019}, system cards \shortcite{procope_system-level_2022} and risk cards \shortcite{derczynski_assessing_2023}. Whether and how the results of risk assessments could be shared with external parties like researchers, auditors, or governments warrants further investigation \shortcite{anderljung_frontier_2023, shevlane_model_2023}. Finally, concrete ways in which relatively abstract frameworks like ISO\slash IEC 23894:2023 \shortcite{iso_23894_2023} and the NIST AI Risk Management Framework \shortcite{nist_artificial_2023} can be adapted to the context of AGI companies need further elaboration \shortcite<see e.g.>{barrett_actionable_2023, barrett_seeking_2023}.

AGI companies need to act on their statements that catastrophic risks from AI should be a global priority \cite{center_for_ai_safety_statement_2023}. It might be possible that the next generation of state-of-the-art AI systems will already pose catastrophic risks. AGI companies therefore need to urgently improve their risk management practices. We hope this paper can support such efforts. While the reviewed techniques will certainly not be sufficient to assess catastrophic risks from AI, AGI companies should not skip the straightforward step of reviewing best practices from other industries.

\section*{Acknowledgements}

We are grateful for invaluable input from Cullen O’Keefe, James Ginns, and Malcolm Murray, as well as research management support from Emma Bluemke and discussions with Jan Brauner. We thank Michael Aird, Markus Anderljung, Anthony Barrett, Seth Baum, Siméon Campos, Zeshen Chin, Shaun Ee, Lennart Heim, Samuel Hilton, Laura Hiscott, Sebastian Lodemann, David Manheim, Joseph O’Brien, Merlin Stein, and Akash Wasil for helpful feedback on an earlier version of this paper. We are also grateful for input from the participants of two Project-in-Progress sessions of the 2023 Winter Research Fellowship of the Centre for the Governance of AI, where we presented earlier versions of this work, namely Ben Garfinkel, Jide Alaga, Kayla Blomquist, Benjamin Bucknall, Conor Downey, Saffron Huang, Kristy Loke, Nikhil Mulani, Jai Vipra, and Owen Yeung. We acknowledge the use of ChatGPT (\url{https://chat.openai.com}) to help with editing. All remaining errors are our own.

\bibliographystyle{apacite}
\bibliography{ms}

\begin{thebibliography}{}

\bibitem [\protect \citeauthoryear {%
Ackermann%
, Howick%
, Quigley%
, Walls%
\BCBL {}\ \BBA {} Houghton%
}{%
Ackermann%
\ \protect \BOthers {.}}{%
{\protect \APACyear {2014}}%
}]{%
ackermann_systemic_2014}
\APACinsertmetastar {%
ackermann_systemic_2014}%
\begin{APACrefauthors}%
Ackermann, F.%
, Howick, S.%
, Quigley, J.%
, Walls, L.%
\BCBL {}\ \BBA {} Houghton, T.%
\end{APACrefauthors}%
\unskip\
\newblock
\APACrefYearMonthDay{2014}{}{}.
\newblock
{\BBOQ}\APACrefatitle {Systemic risk elicitation: Using causal maps to engage
  stakeholders and build a comprehensive view of risks} {Systemic risk
  elicitation: Using causal maps to engage stakeholders and build a
  comprehensive view of risks}.{\BBCQ}
\newblock
\APACjournalVolNumPages{European Journal of Operational
  Research}{238}{1}{290--299}.
\newblock
\begin{APACrefDOI} \doi{10.1016/j.ejor.2014.03.035} \end{APACrefDOI}
\PrintBackRefs{\CurrentBib}

\bibitem [\protect \citeauthoryear {%
{Alignment Research Center}%
}{%
{Alignment Research Center}%
}{%
{\protect \APACyear {2023}}%
}]{%
alignment_research_center_update_2023}
\APACinsertmetastar {%
alignment_research_center_update_2023}%
\begin{APACrefauthors}%
{Alignment Research Center}.%
\end{APACrefauthors}%
\unskip\
\newblock
\APACrefYearMonthDay{2023}{}{}.
\newblock
\APACrefbtitle {Update on {ARC}'s recent eval efforts.} {Update on {ARC}'s
  recent eval efforts.}
\newblock
\APAChowpublished {Alignment Research Center Blog}.
\newblock
\begin{APACrefURL}
  \url{https://evals.alignment.org/blog/2023-03-18-update-on-recent-evals/}
  \end{APACrefURL}
\PrintBackRefs{\CurrentBib}

\bibitem [\protect \citeauthoryear {%
Anderljung%
\ \protect \BOthers {.}}{%
Anderljung%
\ \protect \BOthers {.}}{%
{\protect \APACyear {2023}}%
}]{%
anderljung_frontier_2023}
\APACinsertmetastar {%
anderljung_frontier_2023}%
\begin{APACrefauthors}%
Anderljung, M.%
, Barnhart, J.%
, Leung, J.%
, Korinek, A.%
, O'Keefe, C.%
, Whittlestone, J.%
\BDBL {}Wolf, K.%
\end{APACrefauthors}%
\unskip\
\newblock
\APACrefYearMonthDay{2023}{}{}.
\newblock
{\BBOQ}\APACrefatitle {Frontier {AI} Regulation: Managing Emerging Risks to
  Public Safety} {Frontier {AI} regulation: Managing emerging risks to public
  safety}.{\BBCQ}
\newblock
\APACjournalVolNumPages{arXiv preprint arXiv:2307.03718}{}{}{}.
\PrintBackRefs{\CurrentBib}

\bibitem [\protect \citeauthoryear {%
Anderljung%
\ \BBA {} Hazell%
}{%
Anderljung%
\ \BBA {} Hazell%
}{%
{\protect \APACyear {2023}}%
}]{%
anderljung_protecting_2023}
\APACinsertmetastar {%
anderljung_protecting_2023}%
\begin{APACrefauthors}%
Anderljung, M.%
\BCBT {}\ \BBA {} Hazell, J.%
\end{APACrefauthors}%
\unskip\
\newblock
\APACrefYearMonthDay{2023}{}{}.
\newblock
{\BBOQ}\APACrefatitle {Protecting Society from {AI} Misuse: When are
  Restrictions on Capabilities Warranted?} {Protecting society from {AI}
  misuse: When are restrictions on capabilities warranted?}{\BBCQ}
\newblock
\APACjournalVolNumPages{arXiv preprint arXiv:2303.09377}{}{}{}.
\PrintBackRefs{\CurrentBib}

\bibitem [\protect \citeauthoryear {%
Anthropic%
}{%
Anthropic%
}{%
{\protect \APACyear {2023}}%
}]{%
anthropic_ai_2023}
\APACinsertmetastar {%
anthropic_ai_2023}%
\begin{APACrefauthors}%
Anthropic.%
\end{APACrefauthors}%
\unskip\
\newblock
\APACrefYearMonthDay{2023}{}{}.
\newblock
\APACrefbtitle {An {AI} Policy Tool for Today: Ambitiously Invest in {NIST}.}
  {An {AI} policy tool for today: Ambitiously invest in {NIST}.}
\newblock
\begin{APACrefURL}
  \url{https://www.anthropic.com/index/an-ai-policy-tool-for-today-ambitiously-invest-in-nist}
  \end{APACrefURL}
\PrintBackRefs{\CurrentBib}

\bibitem [\protect \citeauthoryear {%
Applebaum%
, Miller%
, Strom%
, Korban%
\BCBL {}\ \BBA {} Wolf%
}{%
Applebaum%
\ \protect \BOthers {.}}{%
{\protect \APACyear {2016}}%
}]{%
applebaum_intelligent_2016}
\APACinsertmetastar {%
applebaum_intelligent_2016}%
\begin{APACrefauthors}%
Applebaum, A.%
, Miller, D.%
, Strom, B.%
, Korban, C.%
\BCBL {}\ \BBA {} Wolf, R.%
\end{APACrefauthors}%
\unskip\
\newblock
\APACrefYearMonthDay{2016}{}{}.
\newblock
{\BBOQ}\APACrefatitle {Intelligent, automated red team emulation} {Intelligent,
  automated red team emulation}.{\BBCQ}
\newblock
\BIn{} \APACrefbtitle {{Proceedings of the 32nd Annual Conference on Computer
  Security Applications}} {{Proceedings of the 32nd Annual Conference on
  Computer Security Applications}}\ (\BPGS\ 363--373).
\newblock
\begin{APACrefDOI} \doi{10.1145/2991079.2991111} \end{APACrefDOI}
\PrintBackRefs{\CurrentBib}

\bibitem [\protect \citeauthoryear {%
Aqlan%
\ \BBA {} Mustafa~Ali%
}{%
Aqlan%
\ \BBA {} Mustafa~Ali%
}{%
{\protect \APACyear {2014}}%
}]{%
aqlan_integrating_2014}
\APACinsertmetastar {%
aqlan_integrating_2014}%
\begin{APACrefauthors}%
Aqlan, F.%
\BCBT {}\ \BBA {} Mustafa~Ali, E.%
\end{APACrefauthors}%
\unskip\
\newblock
\APACrefYearMonthDay{2014}{}{}.
\newblock
{\BBOQ}\APACrefatitle {Integrating lean principles and fuzzy bow-tie analysis
  for risk assessment in chemical industry} {Integrating lean principles and
  fuzzy bow-tie analysis for risk assessment in chemical industry}.{\BBCQ}
\newblock
\APACjournalVolNumPages{Journal of Loss Prevention in the Process
  Industries}{29}{}{39--48}.
\newblock
\begin{APACrefDOI} \doi{10.1016/j.jlp.2014.01.006} \end{APACrefDOI}
\PrintBackRefs{\CurrentBib}

\bibitem [\protect \citeauthoryear {%
Armstrong%
, Sotala%
\BCBL {}\ \BBA {} Ó~hÉigeartaigh%
}{%
Armstrong%
\ \protect \BOthers {.}}{%
{\protect \APACyear {2014}}%
}]{%
armstrong_errors_2014}
\APACinsertmetastar {%
armstrong_errors_2014}%
\begin{APACrefauthors}%
Armstrong, S.%
, Sotala, K.%
\BCBL {}\ \BBA {} Ó~hÉigeartaigh, S\BPBI S.%
\end{APACrefauthors}%
\unskip\
\newblock
\APACrefYearMonthDay{2014}{{\APACmonth{07}}}{}.
\newblock
{\BBOQ}\APACrefatitle {The errors, insights and lessons of famous {AI}
  predictions – and what they mean for the future} {The errors, insights and
  lessons of famous {AI} predictions – and what they mean for the
  future}.{\BBCQ}
\newblock
\APACjournalVolNumPages{Journal of Experimental \& Theoretical Artificial
  Intelligence}{26}{3}{317--342}.
\newblock
\begin{APACrefDOI} \doi{10.1080/0952813X.2014.895105} \end{APACrefDOI}
\PrintBackRefs{\CurrentBib}

\bibitem [\protect \citeauthoryear {%
Aven%
}{%
Aven%
}{%
{\protect \APACyear {2015}}%
}]{%
aven_risk_2015}
\APACinsertmetastar {%
aven_risk_2015}%
\begin{APACrefauthors}%
Aven, T.%
\end{APACrefauthors}%
\unskip\
\newblock
\APACrefYear{2015}.
\newblock
\APACrefbtitle {Risk Analysis} {Risk analysis}.
\newblock
\APACaddressPublisher{}{Wiley}.
\newblock
\begin{APACrefDOI} \doi{10.1002/9781119057819} \end{APACrefDOI}
\PrintBackRefs{\CurrentBib}

\bibitem [\protect \citeauthoryear {%
Avin%
\ \protect \BOthers {.}}{%
Avin%
\ \protect \BOthers {.}}{%
{\protect \APACyear {2018}}%
}]{%
avin_classifying_2018}
\APACinsertmetastar {%
avin_classifying_2018}%
\begin{APACrefauthors}%
Avin, S.%
, Wintle, B\BPBI C.%
, Weitzdörfer, J.%
, Ó~hÉigeartaigh, S\BPBI S.%
, Sutherland, W\BPBI J.%
\BCBL {}\ \BBA {} Rees, M\BPBI J.%
\end{APACrefauthors}%
\unskip\
\newblock
\APACrefYearMonthDay{2018}{}{}.
\newblock
{\BBOQ}\APACrefatitle {Classifying global catastrophic risks} {Classifying
  global catastrophic risks}.{\BBCQ}
\newblock
\APACjournalVolNumPages{Futures}{102}{}{20--26}.
\newblock
\begin{APACrefDOI} \doi{10.1016/j.futures.2018.02.001} \end{APACrefDOI}
\PrintBackRefs{\CurrentBib}

\bibitem [\protect \citeauthoryear {%
Barrett%
}{%
Barrett%
}{%
{\protect \APACyear {2017}}%
}]{%
barrett_value_2017}
\APACinsertmetastar {%
barrett_value_2017}%
\begin{APACrefauthors}%
Barrett, A\BPBI M.%
\end{APACrefauthors}%
\unskip\
\newblock
\APACrefYearMonthDay{2017}{}{}.
\newblock
{\BBOQ}\APACrefatitle {Value of Global Catastrophic Risk ({GCR}) Information:
  Cost-Effectiveness-Based Approach for {GCR} Reduction} {Value of global
  catastrophic risk ({GCR}) information: Cost-effectiveness-based approach for
  {GCR} reduction}.{\BBCQ}
\newblock
\APACjournalVolNumPages{Decision Analysis}{14}{3}{187--203}.
\newblock
\begin{APACrefDOI} \doi{10.1287/deca.2017.0350} \end{APACrefDOI}
\PrintBackRefs{\CurrentBib}

\bibitem [\protect \citeauthoryear {%
Barrett%
\ \BBA {} Baum%
}{%
Barrett%
\ \BBA {} Baum%
}{%
{\protect \APACyear {2017}}%
{\protect \APACexlab {{\protect \BCnt {1}}}}}]{%
barrett_model_2017}
\APACinsertmetastar {%
barrett_model_2017}%
\begin{APACrefauthors}%
Barrett, A\BPBI M.%
\BCBT {}\ \BBA {} Baum, S\BPBI D.%
\end{APACrefauthors}%
\unskip\
\newblock
\APACrefYearMonthDay{2017{\protect \BCnt {1}}}{}{}.
\newblock
{\BBOQ}\APACrefatitle {A model of pathways to artificial superintelligence
  catastrophe for risk and decision analysis} {A model of pathways to
  artificial superintelligence catastrophe for risk and decision
  analysis}.{\BBCQ}
\newblock
\APACjournalVolNumPages{Journal of Experimental \& Theoretical Artificial
  Intelligence}{29}{2}{397--414}.
\newblock
\begin{APACrefDOI} \doi{10.1080/0952813X.2016.1186228} \end{APACrefDOI}
\PrintBackRefs{\CurrentBib}

\bibitem [\protect \citeauthoryear {%
Barrett%
\ \BBA {} Baum%
}{%
Barrett%
\ \BBA {} Baum%
}{%
{\protect \APACyear {2017}}%
{\protect \APACexlab {{\protect \BCnt {2}}}}}]{%
callaghan_risk_2017}
\APACinsertmetastar {%
callaghan_risk_2017}%
\begin{APACrefauthors}%
Barrett, A\BPBI M.%
\BCBT {}\ \BBA {} Baum, S\BPBI D.%
\end{APACrefauthors}%
\unskip\
\newblock
\APACrefYearMonthDay{2017{\protect \BCnt {2}}}{}{}.
\newblock
{\BBOQ}\APACrefatitle {Risk Analysis and Risk Management for the Artificial
  Superintelligence Research and Development Process} {Risk analysis and risk
  management for the artificial superintelligence research and development
  process}.{\BBCQ}
\newblock
\BIn{} V.~Callaghan, J.~Miller, R.~Yampolskiy\BCBL {}\ \BBA {} S.~Armstrong\
  (\BEDS), \APACrefbtitle {The Technological Singularity} {The technological
  singularity}\ (\BPGS\ 127--140).
\newblock
\APACaddressPublisher{}{Springer}.
\newblock
\begin{APACrefDOI} \doi{10.1007/978-3-662-54033-6_6} \end{APACrefDOI}
\PrintBackRefs{\CurrentBib}

\bibitem [\protect \citeauthoryear {%
Barrett%
, Hendrycks%
, Newman%
\BCBL {}\ \BBA {} Nonnecke%
}{%
Barrett%
\ \protect \BOthers {.}}{%
{\protect \APACyear {2023}}%
{\protect \APACexlab {{\protect \BCnt {1}}}}}]{%
barrett_actionable_2023}
\APACinsertmetastar {%
barrett_actionable_2023}%
\begin{APACrefauthors}%
Barrett, A\BPBI M.%
, Hendrycks, D.%
, Newman, J.%
\BCBL {}\ \BBA {} Nonnecke, B.%
\end{APACrefauthors}%
\unskip\
\newblock
\APACrefYearMonthDay{2023{\protect \BCnt {1}}}{}{}.
\newblock
{\BBOQ}\APACrefatitle {Actionable Guidance for High-Consequence {AI} Risk
  Management: Towards Standards Addressing AI Catastrophic Risks} {Actionable
  guidance for high-consequence {AI} risk management: Towards standards
  addressing ai catastrophic risks}.{\BBCQ}
\newblock
\APACjournalVolNumPages{arXiv preprint arXiv:2206.08966}{}{}{}.
\PrintBackRefs{\CurrentBib}

\bibitem [\protect \citeauthoryear {%
Barrett%
, Hendrycks%
, Newman%
\BCBL {}\ \BBA {} Nonnecke%
}{%
Barrett%
\ \protect \BOthers {.}}{%
{\protect \APACyear {2023}}%
{\protect \APACexlab {{\protect \BCnt {2}}}}}]{%
barrett_seeking_2023}
\APACinsertmetastar {%
barrett_seeking_2023}%
\begin{APACrefauthors}%
Barrett, A\BPBI M.%
, Hendrycks, D.%
, Newman, J.%
\BCBL {}\ \BBA {} Nonnecke, B.%
\end{APACrefauthors}%
\unskip\
\newblock
\APACrefYearMonthDay{2023{\protect \BCnt {2}}}{}{}.
\newblock
\APACrefbtitle {Seeking Input and Feedback: {AI} Risk Management-Standards
  Profile for Increasingly Multi- or General-Purpose {AI}.} {Seeking input and
  feedback: {AI} risk management-standards profile for increasingly multi- or
  general-purpose {AI}.}
\newblock
\APAChowpublished {Center for Long-Term Cybersecurity}.
\newblock
\begin{APACrefURL}
  \url{https://cltc.berkeley.edu/seeking-input-and-feedback-ai-risk-management-standards-profile-for-increasingly-multi-purpose-or-general-purpose-ai/}
  \end{APACrefURL}
\PrintBackRefs{\CurrentBib}

\bibitem [\protect \citeauthoryear {%
Baum%
}{%
Baum%
}{%
{\protect \APACyear {2020}}%
}]{%
baum_quantifying_2020}
\APACinsertmetastar {%
baum_quantifying_2020}%
\begin{APACrefauthors}%
Baum, S\BPBI D.%
\end{APACrefauthors}%
\unskip\
\newblock
\APACrefYearMonthDay{2020}{}{}.
\newblock
{\BBOQ}\APACrefatitle {Quantifying the probability of existential catastrophe:
  A reply to {Beard} et al.} {Quantifying the probability of existential
  catastrophe: A reply to {Beard} et al.}{\BBCQ}
\newblock
\APACjournalVolNumPages{Futures}{123}{}{102608}.
\newblock
\begin{APACrefDOI} \doi{10.1016/j.futures.2020.102608} \end{APACrefDOI}
\PrintBackRefs{\CurrentBib}

\bibitem [\protect \citeauthoryear {%
Baum%
, Goertzel%
\BCBL {}\ \BBA {} Goertzel%
}{%
Baum%
\ \protect \BOthers {.}}{%
{\protect \APACyear {2011}}%
}]{%
baum_how_2011}
\APACinsertmetastar {%
baum_how_2011}%
\begin{APACrefauthors}%
Baum, S\BPBI D.%
, Goertzel, B.%
\BCBL {}\ \BBA {} Goertzel, T\BPBI G.%
\end{APACrefauthors}%
\unskip\
\newblock
\APACrefYearMonthDay{2011}{}{}.
\newblock
{\BBOQ}\APACrefatitle {How long until human-level {AI}? Results from an expert
  assessment} {How long until human-level {AI}? results from an expert
  assessment}.{\BBCQ}
\newblock
\APACjournalVolNumPages{Technological Forecasting and Social
  Change}{78}{1}{185--195}.
\newblock
\begin{APACrefDOI} \doi{10.1016/j.techfore.2010.09.006} \end{APACrefDOI}
\PrintBackRefs{\CurrentBib}

\bibitem [\protect \citeauthoryear {%
Baylon%
\ \BBA {} Hilton%
}{%
Baylon%
\ \BBA {} Hilton%
}{%
{\protect \APACyear {2020}}%
}]{%
baylon_risk_2020}
\APACinsertmetastar {%
baylon_risk_2020}%
\begin{APACrefauthors}%
Baylon, C.%
\BCBT {}\ \BBA {} Hilton, S.%
\end{APACrefauthors}%
\unskip\
\newblock
\APACrefYearMonthDay{2020}{}{}.
\newblock
\APACrefbtitle {Risk management in the {UK}: What can we learn from {COVID}-19
  and are we prepared for the next disaster?} {Risk management in the {UK}:
  What can we learn from {COVID}-19 and are we prepared for the next disaster?}
\newblock
\begin{APACrefURL} \url{https://www.cser.ac.uk/resources/risk-management-uk/}
  \end{APACrefURL}
\PrintBackRefs{\CurrentBib}

\bibitem [\protect \citeauthoryear {%
Beard%
, Rowe%
\BCBL {}\ \BBA {} Fox%
}{%
Beard%
\ \protect \BOthers {.}}{%
{\protect \APACyear {2020}}%
{\protect \APACexlab {{\protect \BCnt {1}}}}}]{%
beard_analysis_2020}
\APACinsertmetastar {%
beard_analysis_2020}%
\begin{APACrefauthors}%
Beard, S.%
, Rowe, T.%
\BCBL {}\ \BBA {} Fox, J.%
\end{APACrefauthors}%
\unskip\
\newblock
\APACrefYearMonthDay{2020{\protect \BCnt {1}}}{}{}.
\newblock
{\BBOQ}\APACrefatitle {An analysis and evaluation of methods currently used to
  quantify the likelihood of existential hazards} {An analysis and evaluation
  of methods currently used to quantify the likelihood of existential
  hazards}.{\BBCQ}
\newblock
\APACjournalVolNumPages{Futures}{115}{}{102469}.
\newblock
\begin{APACrefDOI} \doi{10.1016/j.futures.2019.102469} \end{APACrefDOI}
\PrintBackRefs{\CurrentBib}

\bibitem [\protect \citeauthoryear {%
Beard%
, Rowe%
\BCBL {}\ \BBA {} Fox%
}{%
Beard%
\ \protect \BOthers {.}}{%
{\protect \APACyear {2020}}%
{\protect \APACexlab {{\protect \BCnt {2}}}}}]{%
beard_existential_2020}
\APACinsertmetastar {%
beard_existential_2020}%
\begin{APACrefauthors}%
Beard, S.%
, Rowe, T.%
\BCBL {}\ \BBA {} Fox, J.%
\end{APACrefauthors}%
\unskip\
\newblock
\APACrefYearMonthDay{2020{\protect \BCnt {2}}}{}{}.
\newblock
{\BBOQ}\APACrefatitle {Existential risk assessment: A reply to {Baum}}
  {Existential risk assessment: A reply to {Baum}}.{\BBCQ}
\newblock
\APACjournalVolNumPages{Futures}{122}{}{102606}.
\newblock
\begin{APACrefDOI} \doi{10.1016/j.futures.2020.102606} \end{APACrefDOI}
\PrintBackRefs{\CurrentBib}

\bibitem [\protect \citeauthoryear {%
Bender%
\ \BBA {} Friedman%
}{%
Bender%
\ \BBA {} Friedman%
}{%
{\protect \APACyear {2018}}%
}]{%
bender_data_2018}
\APACinsertmetastar {%
bender_data_2018}%
\begin{APACrefauthors}%
Bender, E\BPBI M.%
\BCBT {}\ \BBA {} Friedman, B.%
\end{APACrefauthors}%
\unskip\
\newblock
\APACrefYearMonthDay{2018}{}{}.
\newblock
{\BBOQ}\APACrefatitle {Data Statements for Natural Language Processing: Toward
  Mitigating System Bias and Enabling Better Science} {Data statements for
  natural language processing: Toward mitigating system bias and enabling
  better science}.{\BBCQ}
\newblock
\APACjournalVolNumPages{Transactions of the Association for Computational
  Linguistics}{6}{}{587--604}.
\newblock
\begin{APACrefDOI} \doi{10.1162/tacl_a_00041} \end{APACrefDOI}
\PrintBackRefs{\CurrentBib}

\bibitem [\protect \citeauthoryear {%
Bender%
, Gebru%
, McMillan-Major%
\BCBL {}\ \BBA {} Shmitchell%
}{%
Bender%
\ \protect \BOthers {.}}{%
{\protect \APACyear {2021}}%
}]{%
bender_dangers_2021}
\APACinsertmetastar {%
bender_dangers_2021}%
\begin{APACrefauthors}%
Bender, E\BPBI M.%
, Gebru, T.%
, McMillan-Major, A.%
\BCBL {}\ \BBA {} Shmitchell, S.%
\end{APACrefauthors}%
\unskip\
\newblock
\APACrefYearMonthDay{2021}{}{}.
\newblock
{\BBOQ}\APACrefatitle {On the Dangers of Stochastic Parrots: Can Language
  Models Be Too Big?} {On the dangers of stochastic parrots: Can language
  models be too big?}{\BBCQ}
\newblock
\BIn{} \APACrefbtitle {{FAccT '21: Proceedings of the 2021 ACM Conference on
  Fairness, Accountability, and Transparency}} {{FAccT '21: Proceedings of the
  2021 ACM Conference on Fairness, Accountability, and Transparency}}\ (\BPGS\
  610--623).
\newblock
\begin{APACrefDOI} \doi{10.1145/3442188.3445922} \end{APACrefDOI}
\PrintBackRefs{\CurrentBib}

\bibitem [\protect \citeauthoryear {%
Bolukbasi%
, Chang%
, Zou%
, Saligrama%
\BCBL {}\ \BBA {} Kalai%
}{%
Bolukbasi%
\ \protect \BOthers {.}}{%
{\protect \APACyear {2016}}%
}]{%
bolukbasi_man_2016}
\APACinsertmetastar {%
bolukbasi_man_2016}%
\begin{APACrefauthors}%
Bolukbasi, T.%
, Chang, K\BHBI W.%
, Zou, J\BPBI Y.%
, Saligrama, V.%
\BCBL {}\ \BBA {} Kalai, A\BPBI T.%
\end{APACrefauthors}%
\unskip\
\newblock
\APACrefYearMonthDay{2016}{}{}.
\newblock
{\BBOQ}\APACrefatitle {Man is to Computer Programmer as Woman is to Homemaker?
  Debiasing Word Embeddings} {Man is to computer programmer as woman is to
  homemaker? debiasing word embeddings}.{\BBCQ}
\newblock
\BIn{} \APACrefbtitle {{Advances in Neural Information Processing Systems
  (NeurIPS)}} {{Advances in Neural Information Processing Systems (NeurIPS)}}\
  (\BVOL~29).
\PrintBackRefs{\CurrentBib}

\bibitem [\protect \citeauthoryear {%
Book%
}{%
Book%
}{%
{\protect \APACyear {2012}}%
}]{%
book_lessons_2012}
\APACinsertmetastar {%
book_lessons_2012}%
\begin{APACrefauthors}%
Book, G.%
\end{APACrefauthors}%
\unskip\
\newblock
\APACrefYearMonthDay{2012}{}{}.
\newblock
{\BBOQ}\APACrefatitle {Lessons Learned from Real World Application of the
  Bow-tie Method} {Lessons learned from real world application of the bow-tie
  method}.{\BBCQ}
\newblock
\BIn{} \APACrefbtitle {SPE Middle East Health, Safety, Security, and
  Environment Conference and Exhibition.} {Spe middle east health, safety,
  security, and environment conference and exhibition.}
\newblock
\APACaddressPublisher{}{SPE}.
\newblock
\begin{APACrefDOI} \doi{10.2118/154549-MS} \end{APACrefDOI}
\PrintBackRefs{\CurrentBib}

\bibitem [\protect \citeauthoryear {%
Bostrom%
}{%
Bostrom%
}{%
{\protect \APACyear {1998}}%
}]{%
bostrom_how_1998}
\APACinsertmetastar {%
bostrom_how_1998}%
\begin{APACrefauthors}%
Bostrom, N.%
\end{APACrefauthors}%
\unskip\
\newblock
\APACrefYearMonthDay{1998}{}{}.
\newblock
{\BBOQ}\APACrefatitle {How long before superintelligence?} {How long before
  superintelligence?}{\BBCQ}
\newblock
\APACjournalVolNumPages{International Journal of Future Studies}{2}{}{}.
\PrintBackRefs{\CurrentBib}

\bibitem [\protect \citeauthoryear {%
Bostrom%
}{%
Bostrom%
}{%
{\protect \APACyear {2002}}%
}]{%
bostrom_existential_2002}
\APACinsertmetastar {%
bostrom_existential_2002}%
\begin{APACrefauthors}%
Bostrom, N.%
\end{APACrefauthors}%
\unskip\
\newblock
\APACrefYearMonthDay{2002}{}{}.
\newblock
{\BBOQ}\APACrefatitle {Existential Risks: Analyzing Human Extinction Scenarios}
  {Existential risks: Analyzing human extinction scenarios}.{\BBCQ}
\newblock
\APACjournalVolNumPages{Journal of Evolution and Technology}{9}{1}{1--31}.
\PrintBackRefs{\CurrentBib}

\bibitem [\protect \citeauthoryear {%
Bostrom%
}{%
Bostrom%
}{%
{\protect \APACyear {2013}}%
}]{%
bostrom_existential_2013}
\APACinsertmetastar {%
bostrom_existential_2013}%
\begin{APACrefauthors}%
Bostrom, N.%
\end{APACrefauthors}%
\unskip\
\newblock
\APACrefYearMonthDay{2013}{}{}.
\newblock
{\BBOQ}\APACrefatitle {Existential Risk Prevention as Global Priority}
  {Existential risk prevention as global priority}.{\BBCQ}
\newblock
\APACjournalVolNumPages{Global Policy}{4}{1}{15--31}.
\newblock
\begin{APACrefDOI} \doi{10.1111/1758-5899.12002} \end{APACrefDOI}
\PrintBackRefs{\CurrentBib}

\bibitem [\protect \citeauthoryear {%
Bostrom%
}{%
Bostrom%
}{%
{\protect \APACyear {2014}}%
}]{%
bostrom_superintelligence_2014}
\APACinsertmetastar {%
bostrom_superintelligence_2014}%
\begin{APACrefauthors}%
Bostrom, N.%
\end{APACrefauthors}%
\unskip\
\newblock
\APACrefYear{2014}.
\newblock
\APACrefbtitle {Superintelligence: Paths, Dangers, Strategies}
  {Superintelligence: Paths, dangers, strategies}.
\newblock
\APACaddressPublisher{}{Oxford University Press}.
\PrintBackRefs{\CurrentBib}

\bibitem [\protect \citeauthoryear {%
Bostrom%
\ \BBA {} Ćirković%
}{%
Bostrom%
\ \BBA {} Ćirković%
}{%
{\protect \APACyear {2008}}%
}]{%
bostrom_global_2008}
\APACinsertmetastar {%
bostrom_global_2008}%
\begin{APACrefauthors}%
Bostrom, N.%
\BCBT {}\ \BBA {} Ćirković, M\BPBI M.%
\end{APACrefauthors}%
\ (\BEDS).
\unskip\
\newblock
\APACrefYear{2008}.
\newblock
\APACrefbtitle {Global catastrophic risks} {Global catastrophic risks}.
\newblock
\APACaddressPublisher{}{Oxford University Press}.
\PrintBackRefs{\CurrentBib}

\bibitem [\protect \citeauthoryear {%
Bradshaw%
\ \BBA {} Abboud%
}{%
Bradshaw%
\ \BBA {} Abboud%
}{%
{\protect \APACyear {2023}}%
}]{%
bradshaw_four-week-old_2023}
\APACinsertmetastar {%
bradshaw_four-week-old_2023}%
\begin{APACrefauthors}%
Bradshaw, T.%
\BCBT {}\ \BBA {} Abboud, L.%
\end{APACrefauthors}%
\unskip\
\newblock
\APACrefYearMonthDay{2023}{}{}.
\newblock
\APACrefbtitle {Four-week-old {AI} start-up raises record €105mn in
  {European} push.} {Four-week-old {AI} start-up raises record €105mn in
  {European} push.}
\newblock
\APAChowpublished {Financial Times}.
\newblock
\begin{APACrefURL}
  \url{https://www.ft.com/content/cf939ea4-d96c-4908-896a-48a74381f251}
  \end{APACrefURL}
\PrintBackRefs{\CurrentBib}

\bibitem [\protect \citeauthoryear {%
Brundage%
\ \protect \BOthers {.}}{%
Brundage%
\ \protect \BOthers {.}}{%
{\protect \APACyear {2018}}%
}]{%
brundage_malicious_2018}
\APACinsertmetastar {%
brundage_malicious_2018}%
\begin{APACrefauthors}%
Brundage, M.%
, Avin, S.%
, Clark, J.%
, Toner, H.%
, Eckersley, P.%
, Garfinkel, B.%
\BDBL {}Amodei, D.%
\end{APACrefauthors}%
\unskip\
\newblock
\APACrefYearMonthDay{2018}{}{}.
\newblock
{\BBOQ}\APACrefatitle {The Malicious Use of Artificial Intelligence:
  Forecasting, Prevention, and Mitigation} {The malicious use of artificial
  intelligence: Forecasting, prevention, and mitigation}.{\BBCQ}
\newblock
\APACjournalVolNumPages{arXiv preprint arXiv:1802.07228}{}{}{}.
\PrintBackRefs{\CurrentBib}

\bibitem [\protect \citeauthoryear {%
Brundage%
\ \protect \BOthers {.}}{%
Brundage%
\ \protect \BOthers {.}}{%
{\protect \APACyear {2020}}%
}]{%
brundage_toward_2020}
\APACinsertmetastar {%
brundage_toward_2020}%
\begin{APACrefauthors}%
Brundage, M.%
, Avin, S.%
, Wang, J.%
, Belfield, H.%
, Krueger, G.%
, Hadfield, G.%
\BDBL {}Anderljung, M.%
\end{APACrefauthors}%
\unskip\
\newblock
\APACrefYearMonthDay{2020}{}{}.
\newblock
{\BBOQ}\APACrefatitle {Toward Trustworthy {AI} Development: Mechanisms for
  Supporting Verifiable Claims} {Toward trustworthy {AI} development:
  Mechanisms for supporting verifiable claims}.{\BBCQ}
\newblock
\APACjournalVolNumPages{arXiv preprint arXiv:2004.07213}{}{}{}.
\PrintBackRefs{\CurrentBib}

\bibitem [\protect \citeauthoryear {%
Brundage%
\ \protect \BOthers {.}}{%
Brundage%
\ \protect \BOthers {.}}{%
{\protect \APACyear {2022}}%
}]{%
brundage_lessons_2022}
\APACinsertmetastar {%
brundage_lessons_2022}%
\begin{APACrefauthors}%
Brundage, M.%
, Mayer, K.%
, Eloundou, T.%
, Agarwal, S.%
, Adler, S.%
, Krueger, G.%
\BDBL {}Mishkin, P.%
\end{APACrefauthors}%
\unskip\
\newblock
\APACrefYearMonthDay{2022}{}{}.
\newblock
\APACrefbtitle {Lessons learned on language model safety and misuse.} {Lessons
  learned on language model safety and misuse.}
\newblock
\begin{APACrefURL}
  \url{https://openai.com/research/language-model-safety-and-misuse}
  \end{APACrefURL}
\PrintBackRefs{\CurrentBib}

\bibitem [\protect \citeauthoryear {%
Bryson%
, Ackermann%
, Eden%
\BCBL {}\ \BBA {} Finn%
}{%
Bryson%
\ \protect \BOthers {.}}{%
{\protect \APACyear {2004}}%
}]{%
bryson_visible_2004}
\APACinsertmetastar {%
bryson_visible_2004}%
\begin{APACrefauthors}%
Bryson, J\BPBI M.%
, Ackermann, F.%
, Eden, C.%
\BCBL {}\ \BBA {} Finn, C\BPBI B.%
\end{APACrefauthors}%
\unskip\
\newblock
\APACrefYear{2004}.
\newblock
\APACrefbtitle {Visible Thinking: Unlocking Causal Mapping for Practical
  Business Results} {Visible thinking: Unlocking causal mapping for practical
  business results}.
\newblock
\APACaddressPublisher{}{Wiley}.
\PrintBackRefs{\CurrentBib}

\bibitem [\protect \citeauthoryear {%
Buchanan%
, Lohn%
, Musser%
\BCBL {}\ \BBA {} Sedova%
}{%
Buchanan%
\ \protect \BOthers {.}}{%
{\protect \APACyear {2021}}%
}]{%
buchanan_truth_2021}
\APACinsertmetastar {%
buchanan_truth_2021}%
\begin{APACrefauthors}%
Buchanan, B.%
, Lohn, A.%
, Musser, M.%
\BCBL {}\ \BBA {} Sedova.%
\end{APACrefauthors}%
\unskip\
\newblock
\APACrefYearMonthDay{2021}{}{}.
\newblock
\APACrefbtitle {Truth, Lies, and Automation - How Language Models Could Change
  Disinformation.} {Truth, lies, and automation - how language models could
  change disinformation.}
\newblock
\APAChowpublished {Center for Security and Emerging Technology}.
\newblock
\begin{APACrefURL}
  \url{https://cset.georgetown.edu/publication/truth-lies-and-automation/}
  \end{APACrefURL}
\PrintBackRefs{\CurrentBib}

\bibitem [\protect \citeauthoryear {%
Bucknall%
\ \BBA {} Dori-Hacohen%
}{%
Bucknall%
\ \BBA {} Dori-Hacohen%
}{%
{\protect \APACyear {2022}}%
}]{%
bucknall_current_2022}
\APACinsertmetastar {%
bucknall_current_2022}%
\begin{APACrefauthors}%
Bucknall, B\BPBI S.%
\BCBT {}\ \BBA {} Dori-Hacohen, S.%
\end{APACrefauthors}%
\unskip\
\newblock
\APACrefYearMonthDay{2022}{}{}.
\newblock
{\BBOQ}\APACrefatitle {Current and Near-Term {AI} as a Potential Existential
  Risk Factor} {Current and near-term {AI} as a potential existential risk
  factor}.{\BBCQ}
\newblock
\APACjournalVolNumPages{arXiv preprint arXiv:2209.10604}{}{}{}.
\PrintBackRefs{\CurrentBib}

\bibitem [\protect \citeauthoryear {%
Buolamwini%
\ \BBA {} Gebru%
}{%
Buolamwini%
\ \BBA {} Gebru%
}{%
{\protect \APACyear {2018}}%
}]{%
buolamwini_gender_2018}
\APACinsertmetastar {%
buolamwini_gender_2018}%
\begin{APACrefauthors}%
Buolamwini, J.%
\BCBT {}\ \BBA {} Gebru, T.%
\end{APACrefauthors}%
\unskip\
\newblock
\APACrefYearMonthDay{2018}{}{}.
\newblock
{\BBOQ}\APACrefatitle {Gender Shades: Intersectional Accuracy Disparities in
  Commercial Gender Classification} {Gender shades: Intersectional accuracy
  disparities in commercial gender classification}.{\BBCQ}
\newblock
\BIn{} \APACrefbtitle {{Proceedings of the 1st Conference on Fairness,
  Accountability and Transparency}} {{Proceedings of the 1st Conference on
  Fairness, Accountability and Transparency}}\ (\BPGS\ 77--91).
\PrintBackRefs{\CurrentBib}

\bibitem [\protect \citeauthoryear {%
Carlsmith%
}{%
Carlsmith%
}{%
{\protect \APACyear {2022}}%
}]{%
carlsmith_is_2022}
\APACinsertmetastar {%
carlsmith_is_2022}%
\begin{APACrefauthors}%
Carlsmith, J.%
\end{APACrefauthors}%
\unskip\
\newblock
\APACrefYearMonthDay{2022}{}{}.
\newblock
{\BBOQ}\APACrefatitle {Is Power-Seeking {AI} an Existential Risk?} {Is
  power-seeking {AI} an existential risk?}{\BBCQ}
\newblock
\APACjournalVolNumPages{arXiv preprint arXiv:2206.13353}{}{}{}.
\PrintBackRefs{\CurrentBib}

\bibitem [\protect \citeauthoryear {%
{Center for AI Safety}%
}{%
{Center for AI Safety}%
}{%
{\protect \APACyear {2023}}%
}]{%
center_for_ai_safety_statement_2023}
\APACinsertmetastar {%
center_for_ai_safety_statement_2023}%
\begin{APACrefauthors}%
{Center for AI Safety}.%
\end{APACrefauthors}%
\unskip\
\newblock
\APACrefYearMonthDay{2023}{}{}.
\newblock
\APACrefbtitle {Statement on {AI} Risk.} {Statement on {AI} risk.}
\newblock
\begin{APACrefURL} \url{https://www.safe.ai/statement-on-ai-risk}
  \end{APACrefURL}
\PrintBackRefs{\CurrentBib}

\bibitem [\protect \citeauthoryear {%
{Center for Security and Emerging Technology}%
}{%
{Center for Security and Emerging Technology}%
}{%
{\protect \APACyear {{\protect \bibnodate {}}}}%
}]{%
center_for_security_and_emerging_technology_cset_taxonomy_nodate}
\APACinsertmetastar {%
center_for_security_and_emerging_technology_cset_taxonomy_nodate}%
\begin{APACrefauthors}%
{Center for Security and Emerging Technology}.%
\end{APACrefauthors}%
\unskip\
\newblock
\APACrefYearMonthDay{{\protect \bibnodate {}}}{}{}.
\newblock
\APACrefbtitle {Taxonomy.} {Taxonomy.}
\newblock
\APAChowpublished {AI Incident Database}.
\newblock
\begin{APACrefURL} \url{https://incidentdatabase.ai/taxonomy/cset/}
  \end{APACrefURL}
\PrintBackRefs{\CurrentBib}

\bibitem [\protect \citeauthoryear {%
{Centre for Interdisciplinary Risk and Innovation Research}%
}{%
{Centre for Interdisciplinary Risk and Innovation Research}%
}{%
{\protect \APACyear {{\protect \bibnodate {}}}}%
}]{%
centre_for_interdisciplinary_risk_and_innovation_research_cross-impact_nodate}
\APACinsertmetastar {%
centre_for_interdisciplinary_risk_and_innovation_research_cross-impact_nodate}%
\begin{APACrefauthors}%
{Centre for Interdisciplinary Risk and Innovation Research}.%
\end{APACrefauthors}%
\unskip\
\newblock
\APACrefYearMonthDay{{\protect \bibnodate {}}}{}{}.
\newblock
\APACrefbtitle {Cross-Impact Balances Home.} {Cross-impact balances home.}
\newblock
\begin{APACrefURL} \url{https://www.cross-impact.org} \end{APACrefURL}
\PrintBackRefs{\CurrentBib}

\bibitem [\protect \citeauthoryear {%
Chan%
\ \protect \BOthers {.}}{%
Chan%
\ \protect \BOthers {.}}{%
{\protect \APACyear {2023}}%
}]{%
chan_harms_2023}
\APACinsertmetastar {%
chan_harms_2023}%
\begin{APACrefauthors}%
Chan, A.%
, Salganik, R.%
, Markelius, A.%
, Pang, C.%
, Rajkumar, N.%
, Krasheninnikov, D.%
\BDBL {}Maharaj, T.%
\end{APACrefauthors}%
\unskip\
\newblock
\APACrefYearMonthDay{2023}{}{}.
\newblock
{\BBOQ}\APACrefatitle {Harms from Increasingly Agentic Algorithmic Systems}
  {Harms from increasingly agentic algorithmic systems}.{\BBCQ}
\newblock
\BIn{} \APACrefbtitle {{FAccT '23: the 2023 ACM Conference on Fairness,
  Accountability, and Transparency}} {{FAccT '23: the 2023 ACM Conference on
  Fairness, Accountability, and Transparency}}\ (\BPGS\ 651--666).
\newblock
\begin{APACrefDOI} \doi{10.1145/3593013.3594033} \end{APACrefDOI}
\PrintBackRefs{\CurrentBib}

\bibitem [\protect \citeauthoryear {%
Chapelle%
}{%
Chapelle%
}{%
{\protect \APACyear {2019}}%
}]{%
chapelle_operational_2019}
\APACinsertmetastar {%
chapelle_operational_2019}%
\begin{APACrefauthors}%
Chapelle, A.%
\end{APACrefauthors}%
\unskip\
\newblock
\APACrefYear{2019}.
\newblock
\APACrefbtitle {Operational risk management: best practices in the financial
  services industry} {Operational risk management: best practices in the
  financial services industry}.
\newblock
\APACaddressPublisher{}{Wiley}.
\PrintBackRefs{\CurrentBib}

\bibitem [\protect \citeauthoryear {%
Chermack%
}{%
Chermack%
}{%
{\protect \APACyear {2011}}%
}]{%
chermack_scenario_2011}
\APACinsertmetastar {%
chermack_scenario_2011}%
\begin{APACrefauthors}%
Chermack, T\BPBI J.%
\end{APACrefauthors}%
\unskip\
\newblock
\APACrefYear{2011}.
\newblock
\APACrefbtitle {Scenario planning in organizations: How to create, use, and
  assess scenarios} {Scenario planning in organizations: How to create, use,
  and assess scenarios}.
\newblock
\APACaddressPublisher{}{Berrett-Koehler}.
\PrintBackRefs{\CurrentBib}

\bibitem [\protect \citeauthoryear {%
Chin%
}{%
Chin%
}{%
{\protect \APACyear {2022}}%
{\protect \APACexlab {{\protect \BCnt {1}}}}}]{%
chin_embedding_2022}
\APACinsertmetastar {%
chin_embedding_2022}%
\begin{APACrefauthors}%
Chin, Z.%
\end{APACrefauthors}%
\unskip\
\newblock
\APACrefYearMonthDay{2022{\protect \BCnt {1}}}{}{}.
\newblock
\APACrefbtitle {Embedding safety in {ML} development.} {Embedding safety in
  {ML} development.}
\newblock
\APAChowpublished {AI Alignment Forum}.
\newblock
\begin{APACrefURL}
  \url{https://www.alignmentforum.org/posts/dYHiMeSdLrrX3cy4a/embedding-safety-in-ml-development}
  \end{APACrefURL}
\PrintBackRefs{\CurrentBib}

\bibitem [\protect \citeauthoryear {%
Chin%
}{%
Chin%
}{%
{\protect \APACyear {2022}}%
{\protect \APACexlab {{\protect \BCnt {2}}}}}]{%
chin_what_2022}
\APACinsertmetastar {%
chin_what_2022}%
\begin{APACrefauthors}%
Chin, Z.%
\end{APACrefauthors}%
\unskip\
\newblock
\APACrefYearMonthDay{2022{\protect \BCnt {2}}}{}{}.
\newblock
\APACrefbtitle {What if we approach {AI} safety like a technical engineering
  safety problem.} {What if we approach {AI} safety like a technical
  engineering safety problem.}
\newblock
\APAChowpublished {AI Alignment Forum}.
\newblock
\begin{APACrefURL}
  \url{https://www.alignmentforum.org/posts/zNYmbFwgrxiNtayMm/what-if-we-approach-ai-safety-like-a-technical-engineering}
  \end{APACrefURL}
\PrintBackRefs{\CurrentBib}

\bibitem [\protect \citeauthoryear {%
Christiano%
}{%
Christiano%
}{%
{\protect \APACyear {2019}}%
}]{%
christiano_what_2019}
\APACinsertmetastar {%
christiano_what_2019}%
\begin{APACrefauthors}%
Christiano, P.%
\end{APACrefauthors}%
\unskip\
\newblock
\APACrefYearMonthDay{2019}{}{}.
\newblock
\APACrefbtitle {What failure looks like.} {What failure looks like.}
\newblock
\APAChowpublished {AI Alignment Forum}.
\newblock
\begin{APACrefURL}
  \url{https://www.alignmentforum.org/posts/HBxe6wdjxK239zajf/what-failure-looks-like}
  \end{APACrefURL}
\PrintBackRefs{\CurrentBib}

\bibitem [\protect \citeauthoryear {%
Clarke%
}{%
Clarke%
}{%
{\protect \APACyear {2022}}%
}]{%
clarke_classifying_2022}
\APACinsertmetastar {%
clarke_classifying_2022}%
\begin{APACrefauthors}%
Clarke, S.%
\end{APACrefauthors}%
\unskip\
\newblock
\APACrefYearMonthDay{2022}{}{}.
\newblock
\APACrefbtitle {Classifying sources of {AI} x-risk.} {Classifying sources of
  {AI} x-risk.}
\newblock
\APAChowpublished {Effective Altruism Forum}.
\newblock
\begin{APACrefURL}
  \url{https://forum.effectivealtruism.org/posts/e55QpEExmtkRjw9CD/classifying-sources-of-ai-x-risk}
  \end{APACrefURL}
\PrintBackRefs{\CurrentBib}

\bibitem [\protect \citeauthoryear {%
Clarke%
, Carlier%
\BCBL {}\ \BBA {} Schuett%
}{%
Clarke%
\ \protect \BOthers {.}}{%
{\protect \APACyear {2021}}%
}]{%
clarke_survey_2021}
\APACinsertmetastar {%
clarke_survey_2021}%
\begin{APACrefauthors}%
Clarke, S.%
, Carlier, A.%
\BCBL {}\ \BBA {} Schuett, J.%
\end{APACrefauthors}%
\unskip\
\newblock
\APACrefYearMonthDay{2021}{}{}.
\newblock
\APACrefbtitle {Survey on {AI} existential risk scenarios.} {Survey on {AI}
  existential risk scenarios.}
\newblock
\APAChowpublished {Effective Altruism Forum}.
\newblock
\begin{APACrefURL}
  \url{https://forum.effectivealtruism.org/posts/2tumunFmjBuXdfF2F/survey-on-ai-existential-risk-scenarios-1}
  \end{APACrefURL}
\PrintBackRefs{\CurrentBib}

\bibitem [\protect \citeauthoryear {%
Clarke%
\ \BBA {} Whittlestone%
}{%
Clarke%
\ \BBA {} Whittlestone%
}{%
{\protect \APACyear {2022}}%
}]{%
clarke_survey_2022}
\APACinsertmetastar {%
clarke_survey_2022}%
\begin{APACrefauthors}%
Clarke, S.%
\BCBT {}\ \BBA {} Whittlestone, J.%
\end{APACrefauthors}%
\unskip\
\newblock
\APACrefYearMonthDay{2022}{}{}.
\newblock
{\BBOQ}\APACrefatitle {A Survey of the Potential Long-term Impacts of {AI}: How
  AI Could Lead to Long-term Changes in Science, Cooperation, Power, Epistemics
  and Values} {A survey of the potential long-term impacts of {AI}: How ai
  could lead to long-term changes in science, cooperation, power, epistemics
  and values}.{\BBCQ}
\newblock
\BIn{} \APACrefbtitle {{AIES 2022: Proceedings of the 2022 AAAI/ACM Conference
  on AI, Ethics, and Society}} {{AIES 2022: Proceedings of the 2022 AAAI/ACM
  Conference on AI, Ethics, and Society}}\ (\BPGS\ 192--202).
\newblock
\begin{APACrefDOI} \doi{10.1145/3514094.3534131} \end{APACrefDOI}
\PrintBackRefs{\CurrentBib}

\bibitem [\protect \citeauthoryear {%
Cohen%
, Hutter%
\BCBL {}\ \BBA {} Osborne%
}{%
Cohen%
\ \protect \BOthers {.}}{%
{\protect \APACyear {2022}}%
}]{%
cohen_advanced_2022}
\APACinsertmetastar {%
cohen_advanced_2022}%
\begin{APACrefauthors}%
Cohen, M\BPBI K.%
, Hutter, M.%
\BCBL {}\ \BBA {} Osborne, M\BPBI A.%
\end{APACrefauthors}%
\unskip\
\newblock
\APACrefYearMonthDay{2022}{}{}.
\newblock
{\BBOQ}\APACrefatitle {Advanced artificial agents intervene in the provision of
  reward} {Advanced artificial agents intervene in the provision of
  reward}.{\BBCQ}
\newblock
\APACjournalVolNumPages{AI Magazine}{43}{3}{282--293}.
\newblock
\begin{APACrefDOI} \doi{10.1002/aaai.12064} \end{APACrefDOI}
\PrintBackRefs{\CurrentBib}

\bibitem [\protect \citeauthoryear {%
{COSO}%
}{%
{COSO}%
}{%
{\protect \APACyear {2017}}%
}]{%
coso_guidance_2017}
\APACinsertmetastar {%
coso_guidance_2017}%
\begin{APACrefauthors}%
{COSO}.%
\end{APACrefauthors}%
\unskip\
\newblock
\APACrefYearMonthDay{2017}{}{}.
\newblock
\APACrefbtitle {{Guidance on Enterprise Risk Management}.} {{Guidance on
  Enterprise Risk Management}.}
\newblock
\APAChowpublished
  {\url{https://www.coso.org/sitepages/guidance-on-enterprise-risk-management.aspx?web=1}}.
\PrintBackRefs{\CurrentBib}

\bibitem [\protect \citeauthoryear {%
Cotra%
}{%
Cotra%
}{%
{\protect \APACyear {2020}}%
}]{%
cotra_draft_2020}
\APACinsertmetastar {%
cotra_draft_2020}%
\begin{APACrefauthors}%
Cotra, A.%
\end{APACrefauthors}%
\unskip\
\newblock
\APACrefYearMonthDay{2020}{}{}.
\newblock
\APACrefbtitle {Draft report on {AI} timelines.} {Draft report on {AI}
  timelines.}
\newblock
\APAChowpublished {AI Alignment Forum}.
\newblock
\begin{APACrefURL}
  \url{https://www.alignmentforum.org/posts/KrJfoZzpSDpnrv9va/draft-report-on-ai-timelines}
  \end{APACrefURL}
\PrintBackRefs{\CurrentBib}

\bibitem [\protect \citeauthoryear {%
Cotra%
}{%
Cotra%
}{%
{\protect \APACyear {2022}}%
}]{%
cotra_without_2022}
\APACinsertmetastar {%
cotra_without_2022}%
\begin{APACrefauthors}%
Cotra, A.%
\end{APACrefauthors}%
\unskip\
\newblock
\APACrefYearMonthDay{2022}{}{}.
\newblock
\APACrefbtitle {Without specific countermeasures, the easiest path to
  transformative {AI} likely leads to {AI} takeover.} {Without specific
  countermeasures, the easiest path to transformative {AI} likely leads to {AI}
  takeover.}
\newblock
\APAChowpublished {AI Alignment Forum}.
\newblock
\begin{APACrefURL}
  \url{https://www.alignmentforum.org/posts/pRkFkzwKZ2zfa3R6H/without-specific-countermeasures-the-easiest-path-to}
  \end{APACrefURL}
\PrintBackRefs{\CurrentBib}

\bibitem [\protect \citeauthoryear {%
Cotton-Barratt%
, Daniel%
\BCBL {}\ \BBA {} Sandberg%
}{%
Cotton-Barratt%
\ \protect \BOthers {.}}{%
{\protect \APACyear {2020}}%
}]{%
cotton-barratt_defence_2020}
\APACinsertmetastar {%
cotton-barratt_defence_2020}%
\begin{APACrefauthors}%
Cotton-Barratt, O.%
, Daniel, M.%
\BCBL {}\ \BBA {} Sandberg, A.%
\end{APACrefauthors}%
\unskip\
\newblock
\APACrefYearMonthDay{2020}{}{}.
\newblock
{\BBOQ}\APACrefatitle {Defence in Depth Against Human Extinction: Prevention,
  Response, Resilience, and Why They All Matter} {Defence in depth against
  human extinction: Prevention, response, resilience, and why they all
  matter}.{\BBCQ}
\newblock
\APACjournalVolNumPages{Global Policy}{11}{3}{271--282}.
\newblock
\begin{APACrefDOI} \doi{10.1111/1758-5899.12786} \end{APACrefDOI}
\PrintBackRefs{\CurrentBib}

\bibitem [\protect \citeauthoryear {%
Cremer%
\ \BBA {} Whittlestone%
}{%
Cremer%
\ \BBA {} Whittlestone%
}{%
{\protect \APACyear {2021}}%
}]{%
cremer_artificial_2021}
\APACinsertmetastar {%
cremer_artificial_2021}%
\begin{APACrefauthors}%
Cremer, C\BPBI Z.%
\BCBT {}\ \BBA {} Whittlestone, J.%
\end{APACrefauthors}%
\unskip\
\newblock
\APACrefYearMonthDay{2021}{}{}.
\newblock
{\BBOQ}\APACrefatitle {Artificial Canaries: Early Warning Signs for
  Anticipatory and Democratic Governance of {AI}} {Artificial canaries: Early
  warning signs for anticipatory and democratic governance of {AI}}.{\BBCQ}
\newblock
\APACjournalVolNumPages{International Journal of Interactive Multimedia and
  Artificial Intelligence}{6}{5}{100--109}.
\PrintBackRefs{\CurrentBib}

\bibitem [\protect \citeauthoryear {%
Critch%
}{%
Critch%
}{%
{\protect \APACyear {2021}}%
}]{%
critch_what_2021}
\APACinsertmetastar {%
critch_what_2021}%
\begin{APACrefauthors}%
Critch, A.%
\end{APACrefauthors}%
\unskip\
\newblock
\APACrefYearMonthDay{2021}{}{}.
\newblock
\APACrefbtitle {What Multipolar Failure Looks Like, and Robust Agent-Agnostic
  Processes ({RAAPs}).} {What multipolar failure looks like, and robust
  agent-agnostic processes ({RAAPs}).}
\newblock
\APAChowpublished {AI Alignment Forum}.
\newblock
\begin{APACrefURL}
  \url{https://www.alignmentforum.org/posts/LpM3EAakwYdS6aRKf/what-multipolar-failure-looks-like-and-robust-agent-agnostic}
  \end{APACrefURL}
\PrintBackRefs{\CurrentBib}

\bibitem [\protect \citeauthoryear {%
Critch%
\ \BBA {} Krueger%
}{%
Critch%
\ \BBA {} Krueger%
}{%
{\protect \APACyear {2020}}%
}]{%
critch_ai_2020}
\APACinsertmetastar {%
critch_ai_2020}%
\begin{APACrefauthors}%
Critch, A.%
\BCBT {}\ \BBA {} Krueger, D.%
\end{APACrefauthors}%
\unskip\
\newblock
\APACrefYearMonthDay{2020}{}{}.
\newblock
{\BBOQ}\APACrefatitle {{AI} Research Considerations for Human Existential
  Safety ({ARCHES})} {{AI} research considerations for human existential safety
  ({ARCHES})}.{\BBCQ}
\newblock
\APACjournalVolNumPages{arXiv preprint arXiv:2006.04948}{}{}{}.
\PrintBackRefs{\CurrentBib}

\bibitem [\protect \citeauthoryear {%
Critch%
\ \BBA {} Russell%
}{%
Critch%
\ \BBA {} Russell%
}{%
{\protect \APACyear {2023}}%
}]{%
critch_tasra_2023}
\APACinsertmetastar {%
critch_tasra_2023}%
\begin{APACrefauthors}%
Critch, A.%
\BCBT {}\ \BBA {} Russell, S.%
\end{APACrefauthors}%
\unskip\
\newblock
\APACrefYearMonthDay{2023}{}{}.
\newblock
{\BBOQ}\APACrefatitle {{TASRA}: a Taxonomy and Analysis of Societal-Scale Risks
  from {AI}} {{TASRA}: a taxonomy and analysis of societal-scale risks from
  {AI}}.{\BBCQ}
\newblock
\APACjournalVolNumPages{arXiv preprint arXiv:2306.06924}{}{}{}.
\PrintBackRefs{\CurrentBib}

\bibitem [\protect \citeauthoryear {%
Dafoe%
}{%
Dafoe%
}{%
{\protect \APACyear {2018}}%
}]{%
dafoe_ai_2018}
\APACinsertmetastar {%
dafoe_ai_2018}%
\begin{APACrefauthors}%
Dafoe, A.%
\end{APACrefauthors}%
\unskip\
\newblock
\APACrefYearMonthDay{2018}{}{}.
\newblock
\APACrefbtitle {{AI} Governance: A Research Agenda.} {{AI} governance: A
  research agenda.}
\newblock
\APAChowpublished {Centre for the Governance of AI}.
\newblock
\begin{APACrefURL}
  \url{https://uploads-ssl.webflow.com/614b70a71b9f71c9c240c7a7/61d48553bf2faf58c3900bd2_GovAI-Research-Agenda.pdf}
  \end{APACrefURL}
\PrintBackRefs{\CurrentBib}

\bibitem [\protect \citeauthoryear {%
Dafoe%
}{%
Dafoe%
}{%
{\protect \APACyear {2020}}%
}]{%
dafoe_ai_2020}
\APACinsertmetastar {%
dafoe_ai_2020}%
\begin{APACrefauthors}%
Dafoe, A.%
\end{APACrefauthors}%
\unskip\
\newblock
\APACrefYearMonthDay{2020}{}{}.
\newblock
\APACrefbtitle {{AI} Governance: Opportunity and Theory of Impact.} {{AI}
  governance: Opportunity and theory of impact.}
\newblock
\APAChowpublished {Effective Altruism Forum}.
\newblock
\begin{APACrefURL}
  \url{https://forum.effectivealtruism.org/posts/42reWndoTEhFqu6T8/ai-governance-opportunity-and-theory-of-impact}
  \end{APACrefURL}
\PrintBackRefs{\CurrentBib}

\bibitem [\protect \citeauthoryear {%
Davidson%
}{%
Davidson%
}{%
{\protect \APACyear {2021}}%
}]{%
davidson_report_2021}
\APACinsertmetastar {%
davidson_report_2021}%
\begin{APACrefauthors}%
Davidson, T.%
\end{APACrefauthors}%
\unskip\
\newblock
\APACrefYearMonthDay{2021}{}{}.
\newblock
\APACrefbtitle {Report on {Semi}-informative {Priors}.} {Report on
  {Semi}-informative {Priors}.}
\newblock
\APAChowpublished {Open Philanthropy Blog}.
\newblock
\begin{APACrefURL}
  \url{https://www.openphilanthropy.org/research/report-on-semi-informative-priors/}
  \end{APACrefURL}
\PrintBackRefs{\CurrentBib}

\bibitem [\protect \citeauthoryear {%
Davidson%
}{%
Davidson%
}{%
{\protect \APACyear {2023}}%
}]{%
davidson_what_2023}
\APACinsertmetastar {%
davidson_what_2023}%
\begin{APACrefauthors}%
Davidson, T.%
\end{APACrefauthors}%
\unskip\
\newblock
\APACrefYearMonthDay{2023}{}{}.
\newblock
\APACrefbtitle {What a compute-centric framework says about {AI} takeoff speeds
  - draft report.} {What a compute-centric framework says about {AI} takeoff
  speeds - draft report.}
\newblock
\APAChowpublished {AI Alignment Forum}.
\newblock
\begin{APACrefURL}
  \url{https://www.alignmentforum.org/posts/Gc9FGtdXhK9sCSEYu/what-a-compute-centric-framework-says-about-ai-takeoff#}
  \end{APACrefURL}
\PrintBackRefs{\CurrentBib}

\bibitem [\protect \citeauthoryear {%
de Neufville%
}{%
de Neufville%
}{%
{\protect \APACyear {2023}}%
}]{%
de_neufville_forecasting_2023}
\APACinsertmetastar {%
de_neufville_forecasting_2023}%
\begin{APACrefauthors}%
de Neufville, R.%
\end{APACrefauthors}%
\unskip\
\newblock
\APACrefYearMonthDay{2023}{}{}.
\newblock
\APACrefbtitle {Forecasting Extraordinary Events.} {Forecasting extraordinary
  events.}
\newblock
\APAChowpublished {Telling the Future}.
\newblock
\begin{APACrefURL}
  \url{https://tellingthefuture.substack.com/p/forecasting-extraordinary-events}
  \end{APACrefURL}
\PrintBackRefs{\CurrentBib}

\bibitem [\protect \citeauthoryear {%
Derczynski%
\ \protect \BOthers {.}}{%
Derczynski%
\ \protect \BOthers {.}}{%
{\protect \APACyear {2023}}%
}]{%
derczynski_assessing_2023}
\APACinsertmetastar {%
derczynski_assessing_2023}%
\begin{APACrefauthors}%
Derczynski, L.%
, Kirk, H\BPBI R.%
, Balachandran, V.%
, Kumar, S.%
, Tsvetkov, Y.%
, Leiser, M\BPBI R.%
\BCBL {}\ \BBA {} Mohammad, S.%
\end{APACrefauthors}%
\unskip\
\newblock
\APACrefYearMonthDay{2023}{}{}.
\newblock
{\BBOQ}\APACrefatitle {Assessing Language Model Deployment with Risk Cards}
  {Assessing language model deployment with risk cards}.{\BBCQ}
\newblock
\APACjournalVolNumPages{arXiv preprint arXiv:2303.18190}{}{}{}.
\PrintBackRefs{\CurrentBib}

\bibitem [\protect \citeauthoryear {%
Dobbe%
}{%
Dobbe%
}{%
{\protect \APACyear {2022}}%
}]{%
dobbe_system_2022}
\APACinsertmetastar {%
dobbe_system_2022}%
\begin{APACrefauthors}%
Dobbe, R\BPBI I\BPBI J.%
\end{APACrefauthors}%
\unskip\
\newblock
\APACrefYearMonthDay{2022}{}{}.
\newblock
{\BBOQ}\APACrefatitle {System Safety and Artificial Intelligence} {System
  safety and artificial intelligence}.{\BBCQ}
\newblock
\BIn{} J\BPBI B.~Bullock\ \BOthers {.}\ (\BEDS), \APACrefbtitle {The Oxford
  Handbook of AI Governance.} {The oxford handbook of ai governance.}
\newblock
\APACaddressPublisher{}{Oxford University Press}.
\newblock
\begin{APACrefDOI} \doi{10.1093/oxfordhb/9780197579329.013.67} \end{APACrefDOI}
\PrintBackRefs{\CurrentBib}

\bibitem [\protect \citeauthoryear {%
Etzioni%
}{%
Etzioni%
}{%
{\protect \APACyear {2020}}%
}]{%
etzioni_how_2020}
\APACinsertmetastar {%
etzioni_how_2020}%
\begin{APACrefauthors}%
Etzioni, O.%
\end{APACrefauthors}%
\unskip\
\newblock
\APACrefYearMonthDay{2020}{}{}.
\newblock
\APACrefbtitle {How to know if artificial intelligence is about to destroy
  civilization.} {How to know if artificial intelligence is about to destroy
  civilization.}
\newblock
\APAChowpublished {MIT Technology Review}.
\newblock
\begin{APACrefURL}
  \url{https://www.technologyreview.com/2020/02/25/906083/artificial-intelligence-destroy-civilization-canaries-robot-overlords-take-over-world-ai}
  \end{APACrefURL}
\PrintBackRefs{\CurrentBib}

\bibitem [\protect \citeauthoryear {%
{EU High-Level Expert Group on AI}%
}{%
{EU High-Level Expert Group on AI}%
}{%
{\protect \APACyear {2020}}%
}]{%
eu_high-level_expert_group_on_ai_assessment_2020}
\APACinsertmetastar {%
eu_high-level_expert_group_on_ai_assessment_2020}%
\begin{APACrefauthors}%
{EU High-Level Expert Group on AI}.%
\end{APACrefauthors}%
\unskip\
\newblock
\APACrefYearMonthDay{2020}{}{}.
\newblock
\APACrefbtitle {Assessment List for Trustworthy Artificial Intelligence
  ({ALTAI}) for self-assessment.} {Assessment list for trustworthy artificial
  intelligence ({ALTAI}) for self-assessment.}
\newblock
\begin{APACrefURL}
  \url{https://digital-strategy.ec.europa.eu/en/library/assessment-list-trustworthy-artificial-intelligence-altai-self-assessment}
  \end{APACrefURL}
\PrintBackRefs{\CurrentBib}

\bibitem [\protect \citeauthoryear {%
{European Foresight Platform}%
}{%
{European Foresight Platform}%
}{%
{\protect \APACyear {{\protect \bibnodate {}}}}%
}]{%
european_foresight_platform_cross-impact_nodate}
\APACinsertmetastar {%
european_foresight_platform_cross-impact_nodate}%
\begin{APACrefauthors}%
{European Foresight Platform}.%
\end{APACrefauthors}%
\unskip\
\newblock
\APACrefYearMonthDay{{\protect \bibnodate {}}}{}{}.
\newblock
\APACrefbtitle {Cross-Impact Analysis.} {Cross-impact analysis.}
\newblock
\begin{APACrefURL}
  \url{http://foresight-platform.eu/community/forlearn/how-to-do-foresight/methods/analysis/cross-impact-analysis}
  \end{APACrefURL}
\PrintBackRefs{\CurrentBib}

\bibitem [\protect \citeauthoryear {%
Fei%
\ \protect \BOthers {.}}{%
Fei%
\ \protect \BOthers {.}}{%
{\protect \APACyear {2022}}%
}]{%
fei_towards_2022}
\APACinsertmetastar {%
fei_towards_2022}%
\begin{APACrefauthors}%
Fei, N.%
, Lu, Z.%
, Gao, Y.%
, Yang, G.%
, Huo, Y.%
, Wen, J.%
\BDBL {}Wen, J\BHBI R.%
\end{APACrefauthors}%
\unskip\
\newblock
\APACrefYearMonthDay{2022}{}{}.
\newblock
{\BBOQ}\APACrefatitle {Towards artificial general intelligence via a multimodal
  foundation model} {Towards artificial general intelligence via a multimodal
  foundation model}.{\BBCQ}
\newblock
\APACjournalVolNumPages{Nature Communications}{13}{1}{3094}.
\newblock
\begin{APACrefDOI} \doi{10.1038/s41467-022-30761-2} \end{APACrefDOI}
\PrintBackRefs{\CurrentBib}

\bibitem [\protect \citeauthoryear {%
Ganguli%
\ \protect \BOthers {.}}{%
Ganguli%
\ \protect \BOthers {.}}{%
{\protect \APACyear {2022}}%
}]{%
ganguli_red_2022}
\APACinsertmetastar {%
ganguli_red_2022}%
\begin{APACrefauthors}%
Ganguli, D.%
, Lovitt, L.%
, Kernion, J.%
, Askell, A.%
, Bai, Y.%
, Kadavath, S.%
\BDBL {}Clark, J.%
\end{APACrefauthors}%
\unskip\
\newblock
\APACrefYearMonthDay{2022}{}{}.
\newblock
{\BBOQ}\APACrefatitle {Red Teaming Language Models to Reduce Harms: Methods,
  Scaling, Behaviors, and Lessons Learned} {Red teaming language models to
  reduce harms: Methods, scaling, behaviors, and lessons learned}.{\BBCQ}
\newblock
\APACjournalVolNumPages{arXiv preprint arXiv:2209.07858}{}{}{}.
\PrintBackRefs{\CurrentBib}

\bibitem [\protect \citeauthoryear {%
Gebru%
\ \protect \BOthers {.}}{%
Gebru%
\ \protect \BOthers {.}}{%
{\protect \APACyear {2021}}%
}]{%
gebru_datasheets_2021}
\APACinsertmetastar {%
gebru_datasheets_2021}%
\begin{APACrefauthors}%
Gebru, T.%
, Morgenstern, J.%
, Vecchione, B.%
, Vaughan, J\BPBI W.%
, Wallach, H.%
, Iii, H\BPBI D.%
\BCBL {}\ \BBA {} Crawford, K.%
\end{APACrefauthors}%
\unskip\
\newblock
\APACrefYearMonthDay{2021}{}{}.
\newblock
{\BBOQ}\APACrefatitle {Datasheets for Datasets} {Datasheets for
  datasets}.{\BBCQ}
\newblock
\APACjournalVolNumPages{Communications of the ACM}{64}{12}{86--92}.
\newblock
\begin{APACrefDOI} \doi{10.1145/3458723} \end{APACrefDOI}
\PrintBackRefs{\CurrentBib}

\bibitem [\protect \citeauthoryear {%
Goertzel%
}{%
Goertzel%
}{%
{\protect \APACyear {2011}}%
}]{%
goertzel_who_2011}
\APACinsertmetastar {%
goertzel_who_2011}%
\begin{APACrefauthors}%
Goertzel, B.%
\end{APACrefauthors}%
\unskip\
\newblock
\APACrefYearMonthDay{2011}{}{}.
\newblock
\APACrefbtitle {Who coined the term “{AGI}”?} {Who coined the term
  “{AGI}”?}
\newblock
\begin{APACrefURL} \url{https://goertzel.org/who-coined-the-term-agi/}
  \end{APACrefURL}
\PrintBackRefs{\CurrentBib}

\bibitem [\protect \citeauthoryear {%
{Good Judgment}%
}{%
{Good Judgment}%
}{%
{\protect \APACyear {{\protect \bibnodate {}}}}%
}]{%
good_judgment_delphineo_nodate}
\APACinsertmetastar {%
good_judgment_delphineo_nodate}%
\begin{APACrefauthors}%
{Good Judgment}.%
\end{APACrefauthors}%
\unskip\
\newblock
\APACrefYearMonthDay{{\protect \bibnodate {}}}{}{}.
\newblock
\APACrefbtitle {Delphineo.} {Delphineo.}
\newblock
\begin{APACrefURL} \url{https://goodjudgment.com/delphineo/} \end{APACrefURL}
\PrintBackRefs{\CurrentBib}

\bibitem [\protect \citeauthoryear {%
Google%
}{%
Google%
}{%
{\protect \APACyear {{\protect \bibnodate {}}}}%
}]{%
google_google_nodate}
\APACinsertmetastar {%
google_google_nodate}%
\begin{APACrefauthors}%
Google.%
\end{APACrefauthors}%
\unskip\
\newblock
\APACrefYearMonthDay{{\protect \bibnodate {}}}{}{}.
\newblock
\APACrefbtitle {Google and {Alphabet} Vulnerability Reward Program ({VRP})
  Rules.} {Google and {Alphabet} vulnerability reward program ({VRP}) rules.}
\newblock
\begin{APACrefURL}
  \url{https://bughunters.google.com/about/rules/6625378258649088/google-and-alphabet-vulnerability-reward-program-vrp-rules}
  \end{APACrefURL}
\PrintBackRefs{\CurrentBib}

\bibitem [\protect \citeauthoryear {%
Gordon%
}{%
Gordon%
}{%
{\protect \APACyear {1994}}%
}]{%
gordon_cross-impact_1994}
\APACinsertmetastar {%
gordon_cross-impact_1994}%
\begin{APACrefauthors}%
Gordon, T\BPBI J.%
\end{APACrefauthors}%
\unskip\
\newblock
\APACrefYearMonthDay{1994}{}{}.
\newblock
\APACrefbtitle {Cross-Impact Method.} {Cross-impact method.}
\newblock
\begin{APACrefURL}
  \url{https://web.archive.org/web/20110713182749/http://www.lampsacus.com/documents/CROSSIMPACT.pdf}
  \end{APACrefURL}
\PrintBackRefs{\CurrentBib}

\bibitem [\protect \citeauthoryear {%
Grace%
, Salvatier%
, Dafoe%
, Zhang%
\BCBL {}\ \BBA {} Evans%
}{%
Grace%
\ \protect \BOthers {.}}{%
{\protect \APACyear {2018}}%
}]{%
grace_when_2018}
\APACinsertmetastar {%
grace_when_2018}%
\begin{APACrefauthors}%
Grace, K.%
, Salvatier, J.%
, Dafoe, A.%
, Zhang, B.%
\BCBL {}\ \BBA {} Evans, O.%
\end{APACrefauthors}%
\unskip\
\newblock
\APACrefYearMonthDay{2018}{}{}.
\newblock
{\BBOQ}\APACrefatitle {When Will {AI} Exceed Human Performance? Evidence from
  {AI} Experts} {When will {AI} exceed human performance? evidence from {AI}
  experts}.{\BBCQ}
\newblock
\APACjournalVolNumPages{arXiv preprint arXiv:1705.08807}{}{}{}.
\PrintBackRefs{\CurrentBib}

\bibitem [\protect \citeauthoryear {%
Gruetzemacher%
, Paradice%
\BCBL {}\ \BBA {} Lee%
}{%
Gruetzemacher%
\ \protect \BOthers {.}}{%
{\protect \APACyear {2019}}%
}]{%
gruetzemacher_forecasting_2019}
\APACinsertmetastar {%
gruetzemacher_forecasting_2019}%
\begin{APACrefauthors}%
Gruetzemacher, R.%
, Paradice, D.%
\BCBL {}\ \BBA {} Lee, K\BPBI B.%
\end{APACrefauthors}%
\unskip\
\newblock
\APACrefYearMonthDay{2019}{}{}.
\newblock
{\BBOQ}\APACrefatitle {Forecasting Transformative {AI}: An Expert Survey}
  {Forecasting transformative {AI}: An expert survey}.{\BBCQ}
\newblock
\APACjournalVolNumPages{arXiv preprint arXiv:1901.08579}{}{}{}.
\PrintBackRefs{\CurrentBib}

\bibitem [\protect \citeauthoryear {%
Gubrud%
}{%
Gubrud%
}{%
{\protect \APACyear {1997}}%
}]{%
gubrud_nanotechnology_1997}
\APACinsertmetastar {%
gubrud_nanotechnology_1997}%
\begin{APACrefauthors}%
Gubrud, M\BPBI A.%
\end{APACrefauthors}%
\unskip\
\newblock
\APACrefYearMonthDay{1997}{}{}.
\newblock
\APACrefbtitle {Nanotechnology and International Security.} {Nanotechnology and
  international security.}
\newblock
\begin{APACrefURL}
  \url{https://web.archive.org/web/20110427135521/http://www.foresight.org/Conferences/MNT05/Papers/Gubrud/index.html}
  \end{APACrefURL}
\PrintBackRefs{\CurrentBib}

\bibitem [\protect \citeauthoryear {%
Hazell%
}{%
Hazell%
}{%
{\protect \APACyear {2023}}%
}]{%
hazell_large_2023}
\APACinsertmetastar {%
hazell_large_2023}%
\begin{APACrefauthors}%
Hazell, J.%
\end{APACrefauthors}%
\unskip\
\newblock
\APACrefYearMonthDay{2023}{}{}.
\newblock
{\BBOQ}\APACrefatitle {Large Language Models Can Be Used To Effectively Scale
  Spear Phishing Campaigns} {Large language models can be used to effectively
  scale spear phishing campaigns}.{\BBCQ}
\newblock
\APACjournalVolNumPages{arXiv preprint arXiv:2305.06972}{}{}{}.
\PrintBackRefs{\CurrentBib}

\bibitem [\protect \citeauthoryear {%
Hendrycks%
\ \BBA {} Mazeika%
}{%
Hendrycks%
\ \BBA {} Mazeika%
}{%
{\protect \APACyear {2022}}%
}]{%
hendrycks_x-risk_2022}
\APACinsertmetastar {%
hendrycks_x-risk_2022}%
\begin{APACrefauthors}%
Hendrycks, D.%
\BCBT {}\ \BBA {} Mazeika, M.%
\end{APACrefauthors}%
\unskip\
\newblock
\APACrefYearMonthDay{2022}{}{}.
\newblock
{\BBOQ}\APACrefatitle {X-Risk Analysis for {AI} Research} {X-risk analysis for
  {AI} research}.{\BBCQ}
\newblock
\APACjournalVolNumPages{arXiv preprint arXiv:2206.05862}{}{}{}.
\PrintBackRefs{\CurrentBib}

\bibitem [\protect \citeauthoryear {%
Hendrycks%
, Mazeika%
\BCBL {}\ \BBA {} Woodside%
}{%
Hendrycks%
\ \protect \BOthers {.}}{%
{\protect \APACyear {2023}}%
}]{%
hendrycks_overview_2023}
\APACinsertmetastar {%
hendrycks_overview_2023}%
\begin{APACrefauthors}%
Hendrycks, D.%
, Mazeika, M.%
\BCBL {}\ \BBA {} Woodside, T.%
\end{APACrefauthors}%
\unskip\
\newblock
\APACrefYearMonthDay{2023}{}{}.
\newblock
{\BBOQ}\APACrefatitle {An Overview of Catastrophic {AI} Risks} {An overview of
  catastrophic {AI} risks}.{\BBCQ}
\newblock
\APACjournalVolNumPages{arXiv preprint arXiv:2306.12001}{}{}{}.
\PrintBackRefs{\CurrentBib}

\bibitem [\protect \citeauthoryear {%
HFACS%
}{%
HFACS%
}{%
{\protect \APACyear {{\protect \bibnodate {}}}}%
}]{%
hfacs_hfacs_nodate}
\APACinsertmetastar {%
hfacs_hfacs_nodate}%
\begin{APACrefauthors}%
HFACS.%
\end{APACrefauthors}%
\unskip\
\newblock
\APACrefYearMonthDay{{\protect \bibnodate {}}}{}{}.
\newblock
\APACrefbtitle {The {HFACS} Framework.} {The {HFACS} framework.}
\newblock
\begin{APACrefURL} \url{https://www.hfacs.com/hfacs-framework.html}
  \end{APACrefURL}
\PrintBackRefs{\CurrentBib}

\bibitem [\protect \citeauthoryear {%
{HM Government}%
}{%
{HM Government}%
}{%
{\protect \APACyear {2021}}%
}]{%
hm_government_national_2021}
\APACinsertmetastar {%
hm_government_national_2021}%
\begin{APACrefauthors}%
{HM Government}.%
\end{APACrefauthors}%
\unskip\
\newblock
\APACrefYearMonthDay{2021}{}{}.
\newblock
\APACrefbtitle {National {AI} Strategy.} {National {AI} strategy.}
\newblock
\begin{APACrefURL}
  \url{https://www.gov.uk/government/publications/national-ai-strategy}
  \end{APACrefURL}
\PrintBackRefs{\CurrentBib}

\bibitem [\protect \citeauthoryear {%
Hogarth%
}{%
Hogarth%
}{%
{\protect \APACyear {2023}}%
}]{%
hogarth_we_2023}
\APACinsertmetastar {%
hogarth_we_2023}%
\begin{APACrefauthors}%
Hogarth, I.%
\end{APACrefauthors}%
\unskip\
\newblock
\APACrefYearMonthDay{2023}{}{}.
\newblock
\APACrefbtitle {We must slow down the race to {God}-like {AI}.} {We must slow
  down the race to {God}-like {AI}.}
\newblock
\APAChowpublished {Financial Times}.
\newblock
\begin{APACrefURL}
  \url{https://www.ft.com/content/03895dc4-a3b7-481e-95cc-336a524f2ac2}
  \end{APACrefURL}
\PrintBackRefs{\CurrentBib}

\bibitem [\protect \citeauthoryear {%
Hsiao%
\ \BBA {} Collins%
}{%
Hsiao%
\ \BBA {} Collins%
}{%
{\protect \APACyear {2023}}%
}]{%
hsiao_try_2023}
\APACinsertmetastar {%
hsiao_try_2023}%
\begin{APACrefauthors}%
Hsiao, S.%
\BCBT {}\ \BBA {} Collins, E.%
\end{APACrefauthors}%
\unskip\
\newblock
\APACrefYearMonthDay{2023}{}{}.
\newblock
\APACrefbtitle {Try {Bard} and share your feedback.} {Try {Bard} and share your
  feedback.}
\newblock
\APAChowpublished {Google Blog}.
\newblock
\begin{APACrefURL} \url{https://blog.google/technology/ai/try-bard/}
  \end{APACrefURL}
\PrintBackRefs{\CurrentBib}

\bibitem [\protect \citeauthoryear {%
Hubinger%
}{%
Hubinger%
}{%
{\protect \APACyear {2022}}%
}]{%
hubinger_how_2022}
\APACinsertmetastar {%
hubinger_how_2022}%
\begin{APACrefauthors}%
Hubinger, E.%
\end{APACrefauthors}%
\unskip\
\newblock
\APACrefYearMonthDay{2022}{}{}.
\newblock
\APACrefbtitle {How likely is deceptive alignment?} {How likely is deceptive
  alignment?}
\newblock
\APAChowpublished {AI Alignment Forum}.
\newblock
\begin{APACrefURL}
  [{2023-06-12}]\url{https://www.alignmentforum.org/posts/A9NxPTwbw6r6Awuwt/how-likely-is-deceptive-alignment}
  \end{APACrefURL}
\PrintBackRefs{\CurrentBib}

\bibitem [\protect \citeauthoryear {%
{IEC}%
}{%
{IEC}%
}{%
{\protect \APACyear {2019}}%
}]{%
iec_31010_2019}
\APACinsertmetastar {%
iec_31010_2019}%
\begin{APACrefauthors}%
{IEC}.%
\end{APACrefauthors}%
\unskip\
\newblock
\APACrefYearMonthDay{2019}{}{}.
\newblock
\APACrefbtitle {{31010:2019 Risk management — Risk assessment techniques}.}
  {{31010:2019 Risk management — Risk assessment techniques}.}
\newblock
\APAChowpublished {\url{https://www.iso.org/standard/72140.html}}.
\PrintBackRefs{\CurrentBib}

\bibitem [\protect \citeauthoryear {%
Ishikawa%
}{%
Ishikawa%
}{%
{\protect \APACyear {1976}}%
}]{%
ishikawa_guide_1976}
\APACinsertmetastar {%
ishikawa_guide_1976}%
\begin{APACrefauthors}%
Ishikawa, K.%
\end{APACrefauthors}%
\unskip\
\newblock
\APACrefYear{1976}.
\newblock
\APACrefbtitle {Guide to quality control} {Guide to quality control}.
\newblock
\APACaddressPublisher{}{Asian Productivity Organization}.
\PrintBackRefs{\CurrentBib}

\bibitem [\protect \citeauthoryear {%
{ISO}%
}{%
{ISO}%
}{%
{\protect \APACyear {2018}}%
}]{%
iso_31000_2018}
\APACinsertmetastar {%
iso_31000_2018}%
\begin{APACrefauthors}%
{ISO}.%
\end{APACrefauthors}%
\unskip\
\newblock
\APACrefYearMonthDay{2018}{}{}.
\newblock
\APACrefbtitle {{31000:2018 Risk management — Guidelines}.} {{31000:2018 Risk
  management — Guidelines}.}
\newblock
\APAChowpublished {\url{https://www.iso.org/standard/65694.html}}.
\PrintBackRefs{\CurrentBib}

\bibitem [\protect \citeauthoryear {%
{ISO \& IEC}%
}{%
{ISO \& IEC}%
}{%
{\protect \APACyear {2014}}%
}]{%
iso_guide_2014}
\APACinsertmetastar {%
iso_guide_2014}%
\begin{APACrefauthors}%
{ISO \& IEC}.%
\end{APACrefauthors}%
\unskip\
\newblock
\APACrefYearMonthDay{2014}{}{}.
\newblock
\APACrefbtitle {{Guide 51:2014 Safety aspects — Guidelines for their
  inclusion in standards}.} {{Guide 51:2014 Safety aspects — Guidelines for
  their inclusion in standards}.}
\newblock
\APAChowpublished {\url{https://www.iso.org/standard/53940.html}}.
\PrintBackRefs{\CurrentBib}

\bibitem [\protect \citeauthoryear {%
{ISO \& IEC}%
}{%
{ISO \& IEC}%
}{%
{\protect \APACyear {2023}}%
}]{%
iso_23894_2023}
\APACinsertmetastar {%
iso_23894_2023}%
\begin{APACrefauthors}%
{ISO \& IEC}.%
\end{APACrefauthors}%
\unskip\
\newblock
\APACrefYearMonthDay{2023}{}{}.
\newblock
\APACrefbtitle {{23894:2023 Information technology — Artificial intelligence
  — Guidance on risk management}.} {{23894:2023 Information technology —
  Artificial intelligence — Guidance on risk management}.}
\newblock
\APAChowpublished {\url{https://www.iso.org/standard/77304.html}}.
\PrintBackRefs{\CurrentBib}

\bibitem [\protect \citeauthoryear {%
Johnson%
}{%
Johnson%
}{%
{\protect \APACyear {1976}}%
}]{%
johnson_ten-year_1976}
\APACinsertmetastar {%
johnson_ten-year_1976}%
\begin{APACrefauthors}%
Johnson, J\BPBI L.%
\end{APACrefauthors}%
\unskip\
\newblock
\APACrefYearMonthDay{1976}{}{}.
\newblock
{\BBOQ}\APACrefatitle {A ten-year {Delphi} forecast in the electronics
  industry} {A ten-year {Delphi} forecast in the electronics industry}.{\BBCQ}
\newblock
\APACjournalVolNumPages{Industrial Marketing Management}{5}{1}{45--55}.
\newblock
\begin{APACrefDOI} \doi{10.1016/0019-8501(76)90009-2} \end{APACrefDOI}
\PrintBackRefs{\CurrentBib}

\bibitem [\protect \citeauthoryear {%
Karnofsky%
}{%
Karnofsky%
}{%
{\protect \APACyear {2016}}%
}]{%
karnofsky_background_2016}
\APACinsertmetastar {%
karnofsky_background_2016}%
\begin{APACrefauthors}%
Karnofsky, H.%
\end{APACrefauthors}%
\unskip\
\newblock
\APACrefYearMonthDay{2016}{}{}.
\newblock
\APACrefbtitle {Some Background on Our Views Regarding Advanced Artificial
  Intelligence.} {Some background on our views regarding advanced artificial
  intelligence.}
\newblock
\APAChowpublished {Open Philanthropy Blog}.
\newblock
\begin{APACrefURL}
  \url{https://www.openphilanthropy.org/research/some-background-on-our-views-regarding-advanced-artificial-intelligence/}
  \end{APACrefURL}
\PrintBackRefs{\CurrentBib}

\bibitem [\protect \citeauthoryear {%
Kavukcuoglu%
, Kohli%
, Ibrahim%
, Bloxwich%
\BCBL {}\ \BBA {} Brown%
}{%
Kavukcuoglu%
\ \protect \BOthers {.}}{%
{\protect \APACyear {2022}}%
}]{%
kavukcuoglu_how_2022}
\APACinsertmetastar {%
kavukcuoglu_how_2022}%
\begin{APACrefauthors}%
Kavukcuoglu, K.%
, Kohli, P.%
, Ibrahim, L.%
, Bloxwich, D.%
\BCBL {}\ \BBA {} Brown, S.%
\end{APACrefauthors}%
\unskip\
\newblock
\APACrefYearMonthDay{2022}{}{}.
\newblock
\APACrefbtitle {How our principles helped define {AlphaFold}’s release.} {How
  our principles helped define {AlphaFold}’s release.}
\newblock
\APAChowpublished {Google DeepMind Blog}.
\newblock
\begin{APACrefURL}
  \url{https://www.deepmind.com/blog/how-our-principles-helped-define-alphafolds-release}
  \end{APACrefURL}
\PrintBackRefs{\CurrentBib}

\bibitem [\protect \citeauthoryear {%
Kenton%
\ \protect \BOthers {.}}{%
Kenton%
\ \protect \BOthers {.}}{%
{\protect \APACyear {2022}}%
{\protect \APACexlab {{\protect \BCnt {1}}}}}]{%
kenton_clarifying_2022}
\APACinsertmetastar {%
kenton_clarifying_2022}%
\begin{APACrefauthors}%
Kenton, Z.%
, Shah, R.%
, Lindner, D.%
, Varma, V.%
, Krakovna, V.%
, Phuong, M.%
\BDBL {}Catt, E.%
\end{APACrefauthors}%
\unskip\
\newblock
\APACrefYearMonthDay{2022{\protect \BCnt {1}}}{}{}.
\newblock
\APACrefbtitle {Clarifying {AI} X-risk.} {Clarifying {AI} x-risk.}
\newblock
\APAChowpublished {AI Alignment Forum}.
\newblock
\begin{APACrefURL}
  \url{https://www.alignmentforum.org/posts/GctJD5oCDRxCspEaZ/clarifying-ai-x-risk}
  \end{APACrefURL}
\PrintBackRefs{\CurrentBib}

\bibitem [\protect \citeauthoryear {%
Kenton%
\ \protect \BOthers {.}}{%
Kenton%
\ \protect \BOthers {.}}{%
{\protect \APACyear {2022}}%
{\protect \APACexlab {{\protect \BCnt {2}}}}}]{%
kenton_threat_2022}
\APACinsertmetastar {%
kenton_threat_2022}%
\begin{APACrefauthors}%
Kenton, Z.%
, Shah, R.%
, Lindner, D.%
, Varma, V.%
, Krakovna, V.%
, Phuong, M.%
\BDBL {}Catt, E.%
\end{APACrefauthors}%
\unskip\
\newblock
\APACrefYearMonthDay{2022{\protect \BCnt {2}}}{}{}.
\newblock
\APACrefbtitle {Threat Model Literature Review.} {Threat model literature
  review.}
\newblock
\APAChowpublished {AI Alignment Forum}.
\newblock
\begin{APACrefURL}
  \url{https://www.alignmentforum.org/posts/wnnkD6P2k2TfHnNmt/threat-model-literature-review}
  \end{APACrefURL}
\PrintBackRefs{\CurrentBib}

\bibitem [\protect \citeauthoryear {%
Khlaaf%
}{%
Khlaaf%
}{%
{\protect \APACyear {2023}}%
}]{%
khlaaf_toward_2023}
\APACinsertmetastar {%
khlaaf_toward_2023}%
\begin{APACrefauthors}%
Khlaaf, H.%
\end{APACrefauthors}%
\unskip\
\newblock
\APACrefYearMonthDay{2023}{}{}.
\newblock
\APACrefbtitle {Toward Comprehensive Risk Assessments and Assurance of
  {AI}-Based Systems.} {Toward comprehensive risk assessments and assurance of
  {AI}-based systems.}
\newblock
\APAChowpublished {Trail of Bits}.
\newblock
\begin{APACrefURL}
  \url{https://github.com/trailofbits/publications/blob/master/papers/toward_comprehensive_risk_assessments.pdf}
  \end{APACrefURL}
\PrintBackRefs{\CurrentBib}

\bibitem [\protect \citeauthoryear {%
Khlaaf%
, Mishkin%
, Achiam%
, Krueger%
\BCBL {}\ \BBA {} Brundage%
}{%
Khlaaf%
\ \protect \BOthers {.}}{%
{\protect \APACyear {2022}}%
}]{%
khlaaf_hazard_2022}
\APACinsertmetastar {%
khlaaf_hazard_2022}%
\begin{APACrefauthors}%
Khlaaf, H.%
, Mishkin, P.%
, Achiam, J.%
, Krueger, G.%
\BCBL {}\ \BBA {} Brundage, M.%
\end{APACrefauthors}%
\unskip\
\newblock
\APACrefYearMonthDay{2022}{}{}.
\newblock
{\BBOQ}\APACrefatitle {A Hazard Analysis Framework for Code Synthesis Large
  Language Models} {A hazard analysis framework for code synthesis large
  language models}.{\BBCQ}
\newblock
\APACjournalVolNumPages{arXiv preprint arXiv:2207.14157}{}{}{}.
\PrintBackRefs{\CurrentBib}

\bibitem [\protect \citeauthoryear {%
Kilian%
, Ventura%
\BCBL {}\ \BBA {} Bailey%
}{%
Kilian%
\ \protect \BOthers {.}}{%
{\protect \APACyear {2023}}%
}]{%
kilian_examining_2023}
\APACinsertmetastar {%
kilian_examining_2023}%
\begin{APACrefauthors}%
Kilian, K\BPBI A.%
, Ventura, C\BPBI J.%
\BCBL {}\ \BBA {} Bailey, M\BPBI M.%
\end{APACrefauthors}%
\unskip\
\newblock
\APACrefYearMonthDay{2023}{}{}.
\newblock
{\BBOQ}\APACrefatitle {Examining the differential risk from high-level
  artificial intelligence and the question of control} {Examining the
  differential risk from high-level artificial intelligence and the question of
  control}.{\BBCQ}
\newblock
\APACjournalVolNumPages{Futures}{151}{}{103182}.
\newblock
\begin{APACrefDOI} \doi{10.1016/j.futures.2023.103182} \end{APACrefDOI}
\PrintBackRefs{\CurrentBib}

\bibitem [\protect \citeauthoryear {%
Kim%
, Lee%
, Kim%
, Park%
\BCBL {}\ \BBA {} Jang%
}{%
Kim%
\ \protect \BOthers {.}}{%
{\protect \APACyear {2016}}%
}]{%
kim_hybrid_2016}
\APACinsertmetastar {%
kim_hybrid_2016}%
\begin{APACrefauthors}%
Kim, J.%
, Lee, J.%
, Kim, G.%
, Park, S.%
\BCBL {}\ \BBA {} Jang, D.%
\end{APACrefauthors}%
\unskip\
\newblock
\APACrefYearMonthDay{2016}{}{}.
\newblock
{\BBOQ}\APACrefatitle {A Hybrid Method of Analyzing Patents for Sustainable
  Technology Management in Humanoid Robot Industry} {A hybrid method of
  analyzing patents for sustainable technology management in humanoid robot
  industry}.{\BBCQ}
\newblock
\APACjournalVolNumPages{Sustainability}{8}{5}{474}.
\newblock
\begin{APACrefDOI} \doi{10.3390/su8050474} \end{APACrefDOI}
\PrintBackRefs{\CurrentBib}

\bibitem [\protect \citeauthoryear {%
Klein%
}{%
Klein%
}{%
{\protect \APACyear {2023}}%
}]{%
klein_surprising_2023}
\APACinsertmetastar {%
klein_surprising_2023}%
\begin{APACrefauthors}%
Klein, E.%
\end{APACrefauthors}%
\unskip\
\newblock
\APACrefYearMonthDay{2023}{}{}.
\newblock
\APACrefbtitle {The Surprising Thing {A}.{I}. Engineers Will Tell You if You
  Let Them.} {The surprising thing {A}.{I}. engineers will tell you if you let
  them.}
\newblock
\APAChowpublished {The New York Times}.
\newblock
\begin{APACrefURL}
  \url{https://www.nytimes.com/2023/04/16/opinion/this-is-too-important-to-leave-to-microsoft-google-and-facebook.html}
  \end{APACrefURL}
\PrintBackRefs{\CurrentBib}

\bibitem [\protect \citeauthoryear {%
Kokotajlo%
\ \BBA {} Wei%
}{%
Kokotajlo%
\ \BBA {} Wei%
}{%
{\protect \APACyear {2019}}%
}]{%
kokotajlo_main_2019}
\APACinsertmetastar {%
kokotajlo_main_2019}%
\begin{APACrefauthors}%
Kokotajlo, D.%
\BCBT {}\ \BBA {} Wei, D.%
\end{APACrefauthors}%
\unskip\
\newblock
\APACrefYearMonthDay{2019}{}{}.
\newblock
\APACrefbtitle {The Main Sources of AI Risk?} {The main sources of ai risk?}
\newblock
\APAChowpublished {AI Alignment Forum}.
\newblock
\begin{APACrefURL}
  \url{https://www.alignmentforum.org/posts/WXvt8bxYnwBYpy9oT/the-main-sources-of-ai-risk}
  \end{APACrefURL}
\PrintBackRefs{\CurrentBib}

\bibitem [\protect \citeauthoryear {%
Kosow%
\ \BBA {} Gaßner%
}{%
Kosow%
\ \BBA {} Gaßner%
}{%
{\protect \APACyear {2008}}%
}]{%
kosow_methods_2008}
\APACinsertmetastar {%
kosow_methods_2008}%
\begin{APACrefauthors}%
Kosow, H.%
\BCBT {}\ \BBA {} Gaßner, R.%
\end{APACrefauthors}%
\unskip\
\newblock
\APACrefYearMonthDay{2008}{}{}.
\newblock
\APACrefbtitle {Methods of Future and Scenario Analysis: Overview, Assessment,
  and Selection Criteria.} {Methods of future and scenario analysis: Overview,
  assessment, and selection criteria.}
\newblock
\APAChowpublished {Deutsches Institut für Entwicklungspolitik}.
\newblock
\begin{APACrefURL} \url{https://www.ssoar.info/ssoar/handle/document/19366}
  \end{APACrefURL}
\PrintBackRefs{\CurrentBib}

\bibitem [\protect \citeauthoryear {%
Krakovna%
\ \BBA {} Shah%
}{%
Krakovna%
\ \BBA {} Shah%
}{%
{\protect \APACyear {2023}}%
}]{%
krakovna_linkpost_2023}
\APACinsertmetastar {%
krakovna_linkpost_2023}%
\begin{APACrefauthors}%
Krakovna, V.%
\BCBT {}\ \BBA {} Shah, R.%
\end{APACrefauthors}%
\unskip\
\newblock
\APACrefYearMonthDay{2023}{}{}.
\newblock
\APACrefbtitle {[Linkpost] Some high-level thoughts on the {DeepMind} alignment
  team's strategy.} {[linkpost] some high-level thoughts on the {DeepMind}
  alignment team's strategy.}
\newblock
\APAChowpublished {AI Alignment Forum}.
\newblock
\begin{APACrefURL}
  \url{https://www.alignmentforum.org/posts/a9SPcZ6GXAg9cNKdi/linkpost-some-high-level-thoughts-on-the-deepmind-alignment}
  \end{APACrefURL}
\PrintBackRefs{\CurrentBib}

\bibitem [\protect \citeauthoryear {%
Leung%
}{%
Leung%
}{%
{\protect \APACyear {2019}}%
}]{%
leung_who_2019}
\APACinsertmetastar {%
leung_who_2019}%
\begin{APACrefauthors}%
Leung, J.%
\end{APACrefauthors}%
\unskip\
\newblock
\APACrefYearMonthDay{2019}{}{}.
\newblock
\APACrefbtitle {Who will govern artificial intelligence? Learning from the
  history of strategic politics in emerging technologies.} {Who will govern
  artificial intelligence? learning from the history of strategic politics in
  emerging technologies.}
\newblock
\APAChowpublished {University of Oxford}.
\newblock
\begin{APACrefURL}
  \url{https://ora.ox.ac.uk/objects/uuid:ea3c7cb8-2464-45f1-a47c-c7b568f27665}
  \end{APACrefURL}
\PrintBackRefs{\CurrentBib}

\bibitem [\protect \citeauthoryear {%
Leveson%
\ \BBA {} Thomas%
}{%
Leveson%
\ \BBA {} Thomas%
}{%
{\protect \APACyear {2018}}%
}]{%
leveson_stpa_2018}
\APACinsertmetastar {%
leveson_stpa_2018}%
\begin{APACrefauthors}%
Leveson, N\BPBI G.%
\BCBT {}\ \BBA {} Thomas, J\BPBI P.%
\end{APACrefauthors}%
\unskip\
\newblock
\APACrefYearMonthDay{2018}{}{}.
\newblock
\APACrefbtitle {{STPA} handbook.} {{STPA} handbook.}
\newblock
\begin{APACrefURL}
  \url{https://psas.scripts.mit.edu/home/get_file.php?name=STPA_handbook.pdf}
  \end{APACrefURL}
\PrintBackRefs{\CurrentBib}

\bibitem [\protect \citeauthoryear {%
Li%
\ \BBA {} Chignell%
}{%
Li%
\ \BBA {} Chignell%
}{%
{\protect \APACyear {2022}}%
}]{%
li_fmea-ai_2022}
\APACinsertmetastar {%
li_fmea-ai_2022}%
\begin{APACrefauthors}%
Li, J.%
\BCBT {}\ \BBA {} Chignell, M.%
\end{APACrefauthors}%
\unskip\
\newblock
\APACrefYearMonthDay{2022}{}{}.
\newblock
{\BBOQ}\APACrefatitle {{FMEA}-{AI}: {AI} fairness impact assessment using
  failure mode and effects analysis} {{FMEA}-{AI}: {AI} fairness impact
  assessment using failure mode and effects analysis}.{\BBCQ}
\newblock
\APACjournalVolNumPages{AI and Ethics}{2}{}{837--850}.
\newblock
\begin{APACrefDOI} \doi{10.1007/s43681-022-00145-9} \end{APACrefDOI}
\PrintBackRefs{\CurrentBib}

\bibitem [\protect \citeauthoryear {%
Liu%
, Lauta%
\BCBL {}\ \BBA {} Maas%
}{%
Liu%
\ \protect \BOthers {.}}{%
{\protect \APACyear {2018}}%
}]{%
liu_governing_2018}
\APACinsertmetastar {%
liu_governing_2018}%
\begin{APACrefauthors}%
Liu, H\BHBI Y.%
, Lauta, K\BPBI C.%
\BCBL {}\ \BBA {} Maas, M\BPBI M.%
\end{APACrefauthors}%
\unskip\
\newblock
\APACrefYearMonthDay{2018}{}{}.
\newblock
{\BBOQ}\APACrefatitle {Governing Boring Apocalypses: A new typology of
  existential vulnerabilities and exposures for existential risk research}
  {Governing boring apocalypses: A new typology of existential vulnerabilities
  and exposures for existential risk research}.{\BBCQ}
\newblock
\APACjournalVolNumPages{Futures}{102}{}{6--19}.
\newblock
\begin{APACrefDOI} \doi{10.1016/j.futures.2018.04.009} \end{APACrefDOI}
\PrintBackRefs{\CurrentBib}

\bibitem [\protect \citeauthoryear {%
Losi%
}{%
Losi%
}{%
{\protect \APACyear {2023}}%
}]{%
losi_system_2023}
\APACinsertmetastar {%
losi_system_2023}%
\begin{APACrefauthors}%
Losi, S.%
\end{APACrefauthors}%
\unskip\
\newblock
\APACrefYearMonthDay{2023}{}{}.
\newblock
\APACrefbtitle {System Dynamics and {AI} Regulation.} {System dynamics and {AI}
  regulation.}
\newblock
\APAChowpublished {Risk Musings}.
\newblock
\begin{APACrefURL}
  \url{https://riskmusings.substack.com/p/system-dynamics-and-ai-regulation?utm_campaign=post}
  \end{APACrefURL}
\PrintBackRefs{\CurrentBib}

\bibitem [\protect \citeauthoryear {%
Madaio%
, Stark%
, Wortman~Vaughan%
\BCBL {}\ \BBA {} Wallach%
}{%
Madaio%
\ \protect \BOthers {.}}{%
{\protect \APACyear {2020}}%
}]{%
madaio_ai_2020}
\APACinsertmetastar {%
madaio_ai_2020}%
\begin{APACrefauthors}%
Madaio, M\BPBI A.%
, Stark, L.%
, Wortman~Vaughan, J.%
\BCBL {}\ \BBA {} Wallach, H.%
\end{APACrefauthors}%
\unskip\
\newblock
\APACrefYearMonthDay{2020}{}{}.
\newblock
\APACrefbtitle {{AI} Fairness Checklist.} {{AI} fairness checklist.}
\newblock
\begin{APACrefURL}
  \url{https://query.prod.cms.rt.microsoft.com/cms/api/am/binary/RE4t6dA}
  \end{APACrefURL}
\PrintBackRefs{\CurrentBib}

\bibitem [\protect \citeauthoryear {%
Mahler%
}{%
Mahler%
}{%
{\protect \APACyear {2022}}%
}]{%
mahler_regulating_2022}
\APACinsertmetastar {%
mahler_regulating_2022}%
\begin{APACrefauthors}%
Mahler, T.%
\end{APACrefauthors}%
\unskip\
\newblock
\APACrefYearMonthDay{2022}{}{}.
\newblock
{\BBOQ}\APACrefatitle {Regulating Artificial General Intelligence ({AGI})}
  {Regulating artificial general intelligence ({AGI})}.{\BBCQ}
\newblock
\BIn{} B.~Custers\ \BBA {} E.~Fosch-Villaronga\ (\BEDS), \APACrefbtitle {Law
  and Artificial Intelligence: Regulating {AI} and Applying {AI} in Legal
  Practice} {Law and artificial intelligence: Regulating {AI} and applying {AI}
  in legal practice}\ (\BPGS\ 521--540).
\newblock
\APACaddressPublisher{}{Springer}.
\newblock
\begin{APACrefDOI} \doi{10.1007/978-94-6265-523-2_26} \end{APACrefDOI}
\PrintBackRefs{\CurrentBib}

\bibitem [\protect \citeauthoryear {%
Manheim%
}{%
Manheim%
}{%
{\protect \APACyear {2023}}%
}]{%
manheim_building_2023}
\APACinsertmetastar {%
manheim_building_2023}%
\begin{APACrefauthors}%
Manheim, D.%
\end{APACrefauthors}%
\unskip\
\newblock
\APACrefYearMonthDay{2023}{}{}.
\newblock
\APACrefbtitle {Building a Culture of Safety for {AI}: Perspectives and
  Challenges.} {Building a culture of safety for {AI}: Perspectives and
  challenges.}
\newblock
\APAChowpublished {SSRN}.
\newblock
\begin{APACrefURL}
  \url{https://papers.ssrn.com/sol3/papers.cfm?abstract_id=4491421}
  \end{APACrefURL}
\PrintBackRefs{\CurrentBib}

\bibitem [\protect \citeauthoryear {%
McChrystal%
\ \BBA {} Butrico%
}{%
McChrystal%
\ \BBA {} Butrico%
}{%
{\protect \APACyear {2021}}%
}]{%
mcchrystal_risk_2021}
\APACinsertmetastar {%
mcchrystal_risk_2021}%
\begin{APACrefauthors}%
McChrystal, S\BPBI A.%
\BCBT {}\ \BBA {} Butrico, A.%
\end{APACrefauthors}%
\unskip\
\newblock
\APACrefYear{2021}.
\newblock
\APACrefbtitle {Risk: a user's guide} {Risk: a user's guide}.
\newblock
\APACaddressPublisher{}{Portfolio}.
\PrintBackRefs{\CurrentBib}

\bibitem [\protect \citeauthoryear {%
McConnell%
\ \BBA {} Davies%
}{%
McConnell%
\ \BBA {} Davies%
}{%
{\protect \APACyear {2006}}%
}]{%
mcconnell_safety_2006}
\APACinsertmetastar {%
mcconnell_safety_2006}%
\begin{APACrefauthors}%
McConnell, P.%
\BCBT {}\ \BBA {} Davies, M.%
\end{APACrefauthors}%
\unskip\
\newblock
\APACrefYearMonthDay{2006}{}{}.
\newblock
\APACrefbtitle {Safety First – Scenario Analysis under {Basel} {II}.} {Safety
  first – scenario analysis under {Basel} {II}.}
\newblock
\begin{APACrefURL}
  \url{https://www.continuitycentral.com/SafetyFirstscenarioanalysis.pdf}
  \end{APACrefURL}
\PrintBackRefs{\CurrentBib}

\bibitem [\protect \citeauthoryear {%
Metz%
}{%
Metz%
}{%
{\protect \APACyear {2023}}%
}]{%
metz_godfather_2023}
\APACinsertmetastar {%
metz_godfather_2023}%
\begin{APACrefauthors}%
Metz, C.%
\end{APACrefauthors}%
\unskip\
\newblock
\APACrefYearMonthDay{2023}{}{}.
\newblock
\APACrefbtitle {{‘The godfather of A.I.’ leaves Google and warns of danger
  ahead}.} {{‘The godfather of A.I.’ leaves Google and warns of danger
  ahead}.}
\newblock
\APAChowpublished {The New York Times}.
\newblock
\begin{APACrefURL}
  \url{https://www.nytimes.com/2023/05/01/technology/ai-google-chatbot-engineer-quits-hinton.html}
  \end{APACrefURL}
\PrintBackRefs{\CurrentBib}

\bibitem [\protect \citeauthoryear {%
Michael%
\ \protect \BOthers {.}}{%
Michael%
\ \protect \BOthers {.}}{%
{\protect \APACyear {2022}}%
}]{%
michael_what_2022}
\APACinsertmetastar {%
michael_what_2022}%
\begin{APACrefauthors}%
Michael, J.%
, Holtzman, A.%
, Parrish, A.%
, Mueller, A.%
, Wang, A.%
, Chen, A.%
\BDBL {}Bowman, S\BPBI R.%
\end{APACrefauthors}%
\unskip\
\newblock
\APACrefYearMonthDay{2022}{}{}.
\newblock
{\BBOQ}\APACrefatitle {What Do {NLP} Researchers Believe? Results of the {NLP}
  Community Metasurvey} {What do {NLP} researchers believe? results of the
  {NLP} community metasurvey}.{\BBCQ}
\newblock
\APACjournalVolNumPages{arXiv preprint arXiv:2208.12852}{}{}{}.
\PrintBackRefs{\CurrentBib}

\bibitem [\protect \citeauthoryear {%
Microsoft%
}{%
Microsoft%
}{%
{\protect \APACyear {2020}}%
}]{%
microsoft_assessing_2020}
\APACinsertmetastar {%
microsoft_assessing_2020}%
\begin{APACrefauthors}%
Microsoft.%
\end{APACrefauthors}%
\unskip\
\newblock
\APACrefYearMonthDay{2020}{}{}.
\newblock
\APACrefbtitle {Assessing harm: A guide for tech builders.} {Assessing harm: A
  guide for tech builders.}
\newblock
\begin{APACrefURL} \url{https://perma.cc/PV3E-HL23} \end{APACrefURL}
\PrintBackRefs{\CurrentBib}

\bibitem [\protect \citeauthoryear {%
Microsoft%
}{%
Microsoft%
}{%
{\protect \APACyear {2022}}%
{\protect \APACexlab {{\protect \BCnt {1}}}}}]{%
microsoft_harms_2022}
\APACinsertmetastar {%
microsoft_harms_2022}%
\begin{APACrefauthors}%
Microsoft.%
\end{APACrefauthors}%
\unskip\
\newblock
\APACrefYearMonthDay{2022{\protect \BCnt {1}}}{}{}.
\newblock
\APACrefbtitle {Harms modeling.} {Harms modeling.}
\newblock
\begin{APACrefURL}
  \url{https://learn.microsoft.com/en-us/azure/architecture/guide/responsible-innovation/harms-modeling/}
  \end{APACrefURL}
\PrintBackRefs{\CurrentBib}

\bibitem [\protect \citeauthoryear {%
Microsoft%
}{%
Microsoft%
}{%
{\protect \APACyear {2022}}%
{\protect \APACexlab {{\protect \BCnt {2}}}}}]{%
microsoft_responsible_2022}
\APACinsertmetastar {%
microsoft_responsible_2022}%
\begin{APACrefauthors}%
Microsoft.%
\end{APACrefauthors}%
\unskip\
\newblock
\APACrefYearMonthDay{2022{\protect \BCnt {2}}}{}{}.
\newblock
\APACrefbtitle {Responsible {AI} Impact Assessment Template.} {Responsible {AI}
  impact assessment template.}
\newblock
\begin{APACrefURL}
  \url{https://query.prod.cms.rt.microsoft.com/cms/api/am/binary/RE5cmFk}
  \end{APACrefURL}
\PrintBackRefs{\CurrentBib}

\bibitem [\protect \citeauthoryear {%
Mishkin%
, Ahmad%
, Brundage%
, Krueger%
\BCBL {}\ \BBA {} Sastry%
}{%
Mishkin%
\ \protect \BOthers {.}}{%
{\protect \APACyear {2022}}%
}]{%
mishkin_dalle_2022}
\APACinsertmetastar {%
mishkin_dalle_2022}%
\begin{APACrefauthors}%
Mishkin, P.%
, Ahmad, L.%
, Brundage, M.%
, Krueger, G.%
\BCBL {}\ \BBA {} Sastry, G.%
\end{APACrefauthors}%
\unskip\
\newblock
\APACrefYearMonthDay{2022}{}{}.
\newblock
\APACrefbtitle {{DALL}·{E} 2 Preview - Risks and Limitations.} {{DALL}·{E} 2
  preview - risks and limitations.}
\newblock
\APAChowpublished {GitHub}.
\newblock
\begin{APACrefURL} \url{https://perma.cc/X467-47PX} \end{APACrefURL}
\PrintBackRefs{\CurrentBib}

\bibitem [\protect \citeauthoryear {%
Mitchell%
\ \protect \BOthers {.}}{%
Mitchell%
\ \protect \BOthers {.}}{%
{\protect \APACyear {2019}}%
}]{%
mitchell_model_2019}
\APACinsertmetastar {%
mitchell_model_2019}%
\begin{APACrefauthors}%
Mitchell, M.%
, Wu, S.%
, Zaldivar, A.%
, Barnes, P.%
, Vasserman, L.%
, Hutchinson, B.%
\BDBL {}Gebru, T.%
\end{APACrefauthors}%
\unskip\
\newblock
\APACrefYearMonthDay{2019}{}{}.
\newblock
{\BBOQ}\APACrefatitle {Model Cards for Model Reporting} {Model cards for model
  reporting}.{\BBCQ}
\newblock
\BIn{} \APACrefbtitle {{FAT* '19: Proceedings of the Conference on Fairness,
  Accountability, and Transparency}} {{FAT* '19: Proceedings of the Conference
  on Fairness, Accountability, and Transparency}}\ (\BPGS\ 220--229).
\newblock
\begin{APACrefDOI} \doi{10.1145/3287560.3287596} \end{APACrefDOI}
\PrintBackRefs{\CurrentBib}

\bibitem [\protect \citeauthoryear {%
Morgan%
}{%
Morgan%
}{%
{\protect \APACyear {2014}}%
}]{%
morgan_use_2014}
\APACinsertmetastar {%
morgan_use_2014}%
\begin{APACrefauthors}%
Morgan, M\BPBI G.%
\end{APACrefauthors}%
\unskip\
\newblock
\APACrefYearMonthDay{2014}{}{}.
\newblock
{\BBOQ}\APACrefatitle {Use (and abuse) of expert elicitation in support of
  decision making for public policy} {Use (and abuse) of expert elicitation in
  support of decision making for public policy}.{\BBCQ}
\newblock
\APACjournalVolNumPages{Proceedings of the National Academy of
  Sciences}{111}{20}{7176--7184}.
\newblock
\begin{APACrefDOI} \doi{10.1073/pnas.1319946111} \end{APACrefDOI}
\PrintBackRefs{\CurrentBib}

\bibitem [\protect \citeauthoryear {%
R.~Müller%
\ \BBA {} Drax%
}{%
R.~Müller%
\ \BBA {} Drax%
}{%
{\protect \APACyear {2014}}%
}]{%
muller_fundamentals_2014}
\APACinsertmetastar {%
muller_fundamentals_2014}%
\begin{APACrefauthors}%
Müller, R.%
\BCBT {}\ \BBA {} Drax, C.%
\end{APACrefauthors}%
\unskip\
\newblock
\APACrefYearMonthDay{2014}{}{}.
\newblock
{\BBOQ}\APACrefatitle {Fundamentals and Structure of Safety Management Systems
  in Aviation} {Fundamentals and structure of safety management systems in
  aviation}.{\BBCQ}
\newblock
\BIn{} R.~Müller, A.~Wittmer\BCBL {}\ \BBA {} C.~Drax\ (\BEDS), \APACrefbtitle
  {Aviation Risk and Safety Management: Methods and Applications in Aviation
  Organizations} {Aviation risk and safety management: Methods and applications
  in aviation organizations}\ (\BPGS\ 45--55).
\newblock
\APACaddressPublisher{}{Springer}.
\newblock
\begin{APACrefDOI} \doi{10.1007/978-3-319-02780-7_5} \end{APACrefDOI}
\PrintBackRefs{\CurrentBib}

\bibitem [\protect \citeauthoryear {%
V\BPBI C.~Müller%
\ \BBA {} Bostrom%
}{%
V\BPBI C.~Müller%
\ \BBA {} Bostrom%
}{%
{\protect \APACyear {2016}}%
}]{%
muller_future_2016}
\APACinsertmetastar {%
muller_future_2016}%
\begin{APACrefauthors}%
Müller, V\BPBI C.%
\BCBT {}\ \BBA {} Bostrom, N.%
\end{APACrefauthors}%
\unskip\
\newblock
\APACrefYearMonthDay{2016}{}{}.
\newblock
{\BBOQ}\APACrefatitle {Future Progress in Artificial Intelligence: A Survey of
  Expert Opinion} {Future progress in artificial intelligence: A survey of
  expert opinion}.{\BBCQ}
\newblock
\BIn{} V\BPBI C.~Müller\ (\BED), \APACrefbtitle {Fundamental Issues of
  Artificial Intelligence} {Fundamental issues of artificial intelligence}\
  (\BPGS\ 553--571).
\newblock
\APACaddressPublisher{}{Springer}.
\newblock
\begin{APACrefDOI} \doi{10.1007/978-3-319-26485-1_33} \end{APACrefDOI}
\PrintBackRefs{\CurrentBib}

\bibitem [\protect \citeauthoryear {%
Newman%
}{%
Newman%
}{%
{\protect \APACyear {2023}}%
}]{%
newman_taxonomy_2023}
\APACinsertmetastar {%
newman_taxonomy_2023}%
\begin{APACrefauthors}%
Newman, J.%
\end{APACrefauthors}%
\unskip\
\newblock
\APACrefYearMonthDay{2023}{}{}.
\newblock
\APACrefbtitle {A Taxonomy of Trustworthiness for Artificial Intelligence.} {A
  taxonomy of trustworthiness for artificial intelligence.}
\newblock
\APAChowpublished {Center for Long-Term Cybersecurity}.
\newblock
\begin{APACrefURL}
  \url{https://cltc.berkeley.edu/publication/a-taxonomy-of-trustworthiness-for-artificial-intelligence/}
  \end{APACrefURL}
\PrintBackRefs{\CurrentBib}

\bibitem [\protect \citeauthoryear {%
Ngo%
, Chan%
\BCBL {}\ \BBA {} Mindermann%
}{%
Ngo%
\ \protect \BOthers {.}}{%
{\protect \APACyear {2023}}%
}]{%
ngo_alignment_2023}
\APACinsertmetastar {%
ngo_alignment_2023}%
\begin{APACrefauthors}%
Ngo, R.%
, Chan, L.%
\BCBL {}\ \BBA {} Mindermann, S.%
\end{APACrefauthors}%
\unskip\
\newblock
\APACrefYearMonthDay{2023}{}{}.
\newblock
{\BBOQ}\APACrefatitle {The alignment problem from a deep learning perspective}
  {The alignment problem from a deep learning perspective}.{\BBCQ}
\newblock
\APACjournalVolNumPages{arXiv preprint arXiv:2209.00626}{}{}{}.
\PrintBackRefs{\CurrentBib}

\bibitem [\protect \citeauthoryear {%
{NIST}%
}{%
{NIST}%
}{%
{\protect \APACyear {2023}}%
}]{%
nist_artificial_2023}
\APACinsertmetastar {%
nist_artificial_2023}%
\begin{APACrefauthors}%
{NIST}.%
\end{APACrefauthors}%
\unskip\
\newblock
\APACrefYearMonthDay{2023}{}{}.
\newblock
\APACrefbtitle {{Artificial Intelligence Risk Management Framework (AI RMF
  1.0}).} {{Artificial Intelligence Risk Management Framework (AI RMF 1.0}).}
\newblock
\APAChowpublished {\url{https://doi.org/10.6028/NIST.AI.100-1}}.
\PrintBackRefs{\CurrentBib}

\bibitem [\protect \citeauthoryear {%
{NITI Aayog}%
}{%
{NITI Aayog}%
}{%
{\protect \APACyear {2018}}%
}]{%
niti_aayog_national_2018}
\APACinsertmetastar {%
niti_aayog_national_2018}%
\begin{APACrefauthors}%
{NITI Aayog}.%
\end{APACrefauthors}%
\unskip\
\newblock
\APACrefYearMonthDay{2018}{}{}.
\newblock
\APACrefbtitle {National Artificial Intelligence Strategy.} {National
  artificial intelligence strategy.}
\newblock
\begin{APACrefURL}
  \url{https://niti.gov.in/sites/default/files/2019-01/NationalStrategy-for-AI-Discussion-Paper.pdf}
  \end{APACrefURL}
\PrintBackRefs{\CurrentBib}

\bibitem [\protect \citeauthoryear {%
OECD%
}{%
OECD%
}{%
{\protect \APACyear {2023}}%
}]{%
oecd_advancing_2023}
\APACinsertmetastar {%
oecd_advancing_2023}%
\begin{APACrefauthors}%
OECD.%
\end{APACrefauthors}%
\unskip\
\newblock
\APACrefYearMonthDay{2023}{}{}.
\newblock
\APACrefbtitle {Advancing accountability in {AI}: Governing and managing risks
  throughout the lifecycle for trustworthy {AI}.} {Advancing accountability in
  {AI}: Governing and managing risks throughout the lifecycle for trustworthy
  {AI}.}
\newblock
\APAChowpublished {OECD Digital Economy Papers}.
\newblock
\begin{APACrefURL}
  \url{https://www.oecd-ilibrary.org/science-and-technology/advancing-accountability-in-ai_2448f04b-en}
  \end{APACrefURL}
\newblock
\begin{APACrefDOI} \doi{10.1787/2448f04b-en} \end{APACrefDOI}
\PrintBackRefs{\CurrentBib}

\bibitem [\protect \citeauthoryear {%
{OpenAI}%
}{%
{OpenAI}%
}{%
{\protect \APACyear {2023}}%
}]{%
openai_announcing_2023}
\APACinsertmetastar {%
openai_announcing_2023}%
\begin{APACrefauthors}%
{OpenAI}.%
\end{APACrefauthors}%
\unskip\
\newblock
\APACrefYearMonthDay{2023}{}{}.
\newblock
\APACrefbtitle {Announcing {OpenAI}’s Bug Bounty Program.} {Announcing
  {OpenAI}’s bug bounty program.}
\newblock
\APAChowpublished {OpenAI Blog}.
\newblock
\begin{APACrefURL} \url{https://openai.com/blog/bug-bounty-program#OpenAI}
  \end{APACrefURL}
\PrintBackRefs{\CurrentBib}

\bibitem [\protect \citeauthoryear {%
OpenAI%
}{%
OpenAI%
}{%
{\protect \APACyear {2023}}%
}]{%
openai_gpt-4_2023}
\APACinsertmetastar {%
openai_gpt-4_2023}%
\begin{APACrefauthors}%
OpenAI.%
\end{APACrefauthors}%
\unskip\
\newblock
\APACrefYearMonthDay{2023}{}{}.
\newblock
{\BBOQ}\APACrefatitle {{GPT}-4 Technical Report} {{GPT}-4 technical
  report}.{\BBCQ}
\newblock
\APACjournalVolNumPages{arXiv preprint arXiv:2303.08774}{}{}{}.
\PrintBackRefs{\CurrentBib}

\bibitem [\protect \citeauthoryear {%
Ord%
}{%
Ord%
}{%
{\protect \APACyear {2020}}%
}]{%
ord_precipice_2020}
\APACinsertmetastar {%
ord_precipice_2020}%
\begin{APACrefauthors}%
Ord, T.%
\end{APACrefauthors}%
\unskip\
\newblock
\APACrefYear{2020}.
\newblock
\APACrefbtitle {The Precipice} {The precipice}.
\newblock
\APACaddressPublisher{}{Hachette Books}.
\PrintBackRefs{\CurrentBib}

\bibitem [\protect \citeauthoryear {%
Ostrom%
\ \BBA {} Wilhelmsen%
}{%
Ostrom%
\ \BBA {} Wilhelmsen%
}{%
{\protect \APACyear {2019}}%
}]{%
ostrom_risk_2019}
\APACinsertmetastar {%
ostrom_risk_2019}%
\begin{APACrefauthors}%
Ostrom, L\BPBI T.%
\BCBT {}\ \BBA {} Wilhelmsen, C\BPBI A.%
\end{APACrefauthors}%
\unskip\
\newblock
\APACrefYear{2019}.
\newblock
\APACrefbtitle {Risk assessment: tools, techniques, and their applications}
  {Risk assessment: tools, techniques, and their applications}.
\newblock
\APACaddressPublisher{}{John Wiley \& Sons, Inc}.
\PrintBackRefs{\CurrentBib}

\bibitem [\protect \citeauthoryear {%
Parente%
\ \BBA {} Anderson-Parente%
}{%
Parente%
\ \BBA {} Anderson-Parente%
}{%
{\protect \APACyear {2011}}%
}]{%
parente_case_2011}
\APACinsertmetastar {%
parente_case_2011}%
\begin{APACrefauthors}%
Parente, R.%
\BCBT {}\ \BBA {} Anderson-Parente, J.%
\end{APACrefauthors}%
\unskip\
\newblock
\APACrefYearMonthDay{2011}{}{}.
\newblock
{\BBOQ}\APACrefatitle {A case study of long-term {Delphi} accuracy} {A case
  study of long-term {Delphi} accuracy}.{\BBCQ}
\newblock
\APACjournalVolNumPages{Technological Forecasting and Social
  Change}{78}{9}{1705--1711}.
\newblock
\begin{APACrefDOI} \doi{10.1016/j.techfore.2011.07.005} \end{APACrefDOI}
\PrintBackRefs{\CurrentBib}

\bibitem [\protect \citeauthoryear {%
Pei%
\ \protect \BOthers {.}}{%
Pei%
\ \protect \BOthers {.}}{%
{\protect \APACyear {2019}}%
}]{%
pei_towards_2019}
\APACinsertmetastar {%
pei_towards_2019}%
\begin{APACrefauthors}%
Pei, J.%
, Deng, L.%
, Song, S.%
, Zhao, M.%
, Zhang, Y.%
, Wu, S.%
\BDBL {}Shi, L.%
\end{APACrefauthors}%
\unskip\
\newblock
\APACrefYearMonthDay{2019}{}{}.
\newblock
{\BBOQ}\APACrefatitle {Towards artificial general intelligence with hybrid
  {Tianjic} chip architecture} {Towards artificial general intelligence with
  hybrid {Tianjic} chip architecture}.{\BBCQ}
\newblock
\APACjournalVolNumPages{Nature}{572}{7767}{106--111}.
\newblock
\begin{APACrefDOI} \doi{10.1038/s41586-019-1424-8} \end{APACrefDOI}
\PrintBackRefs{\CurrentBib}

\bibitem [\protect \citeauthoryear {%
Peng%
, Li%
, He%
, Galley%
\BCBL {}\ \BBA {} Gao%
}{%
Peng%
\ \protect \BOthers {.}}{%
{\protect \APACyear {2023}}%
}]{%
peng_instruction_2023}
\APACinsertmetastar {%
peng_instruction_2023}%
\begin{APACrefauthors}%
Peng, B.%
, Li, C.%
, He, P.%
, Galley, M.%
\BCBL {}\ \BBA {} Gao, J.%
\end{APACrefauthors}%
\unskip\
\newblock
\APACrefYearMonthDay{2023}{}{}.
\newblock
{\BBOQ}\APACrefatitle {Instruction Tuning with {GPT}-4} {Instruction tuning
  with {GPT}-4}.{\BBCQ}
\newblock
\APACjournalVolNumPages{arXiv preprint arXiv:/2304.03277}{}{}{}.
\PrintBackRefs{\CurrentBib}

\bibitem [\protect \citeauthoryear {%
Perez%
\ \protect \BOthers {.}}{%
Perez%
\ \protect \BOthers {.}}{%
{\protect \APACyear {2022}}%
}]{%
perez_red_2022}
\APACinsertmetastar {%
perez_red_2022}%
\begin{APACrefauthors}%
Perez, E.%
, Huang, S.%
, Song, F.%
, Cai, T.%
, Ring, R.%
, Aslanides, J.%
\BDBL {}Irving, G.%
\end{APACrefauthors}%
\unskip\
\newblock
\APACrefYearMonthDay{2022}{}{}.
\newblock
{\BBOQ}\APACrefatitle {Red Teaming Language Models with Language Models} {Red
  teaming language models with language models}.{\BBCQ}
\newblock
\APACjournalVolNumPages{arXiv preprint arXiv:2202.03286}{}{}{}.
\PrintBackRefs{\CurrentBib}

\bibitem [\protect \citeauthoryear {%
Posner%
}{%
Posner%
}{%
{\protect \APACyear {2004}}%
}]{%
posner_catastrophe_2004}
\APACinsertmetastar {%
posner_catastrophe_2004}%
\begin{APACrefauthors}%
Posner, R\BPBI A.%
\end{APACrefauthors}%
\unskip\
\newblock
\APACrefYear{2004}.
\newblock
\APACrefbtitle {Catastrophe: risk and response} {Catastrophe: risk and
  response}.
\newblock
\APACaddressPublisher{}{Oxford University Press}.
\PrintBackRefs{\CurrentBib}

\bibitem [\protect \citeauthoryear {%
Potts%
\ \protect \BOthers {.}}{%
Potts%
\ \protect \BOthers {.}}{%
{\protect \APACyear {2014}}%
}]{%
potts_assessing_2014}
\APACinsertmetastar {%
potts_assessing_2014}%
\begin{APACrefauthors}%
Potts, H\BPBI W.%
, Anderson, J\BPBI E.%
, Colligan, L.%
, Leach, P.%
, Davis, S.%
\BCBL {}\ \BBA {} Berman, J.%
\end{APACrefauthors}%
\unskip\
\newblock
\APACrefYearMonthDay{2014}{}{}.
\newblock
{\BBOQ}\APACrefatitle {Assessing the validity of prospective hazard analysis
  methods: a comparison of two techniques} {Assessing the validity of
  prospective hazard analysis methods: a comparison of two techniques}.{\BBCQ}
\newblock
\APACjournalVolNumPages{BMC Health Services Research}{14}{}{41}.
\newblock
\begin{APACrefDOI} \doi{10.1186/1472-6963-14-41} \end{APACrefDOI}
\PrintBackRefs{\CurrentBib}

\bibitem [\protect \citeauthoryear {%
Pritchard%
}{%
Pritchard%
}{%
{\protect \APACyear {2015}}%
}]{%
pritchard_risk_2015}
\APACinsertmetastar {%
pritchard_risk_2015}%
\begin{APACrefauthors}%
Pritchard, C\BPBI L.%
\end{APACrefauthors}%
\unskip\
\newblock
\APACrefYear{2015}.
\newblock
\APACrefbtitle {Risk Management: Concepts and Guidance} {Risk management:
  Concepts and guidance}.
\newblock
\APACaddressPublisher{}{Auerbach Publications}.
\PrintBackRefs{\CurrentBib}

\bibitem [\protect \citeauthoryear {%
Procope%
\ \protect \BOthers {.}}{%
Procope%
\ \protect \BOthers {.}}{%
{\protect \APACyear {2022}}%
}]{%
procope_system-level_2022}
\APACinsertmetastar {%
procope_system-level_2022}%
\begin{APACrefauthors}%
Procope, C.%
, Cheema, A.%
, Adkins, D.%
, Alsallakh, B.%
, Green, N.%
, McReynolds, E.%
\BDBL {}Zvyagina, P.%
\end{APACrefauthors}%
\unskip\
\newblock
\APACrefYearMonthDay{2022}{}{}.
\newblock
\APACrefbtitle {System-Level Transparency of Machine Learning.} {System-level
  transparency of machine learning.}
\newblock
\APAChowpublished {Meta AI}.
\newblock
\begin{APACrefURL}
  \url{https://ai.facebook.com/research/publications/system-level-transparency-of-machine-learning/}
  \end{APACrefURL}
\PrintBackRefs{\CurrentBib}

\bibitem [\protect \citeauthoryear {%
Raji%
\ \BBA {} Buolamwini%
}{%
Raji%
\ \BBA {} Buolamwini%
}{%
{\protect \APACyear {2019}}%
}]{%
raji_actionable_2019}
\APACinsertmetastar {%
raji_actionable_2019}%
\begin{APACrefauthors}%
Raji, I\BPBI D.%
\BCBT {}\ \BBA {} Buolamwini, J.%
\end{APACrefauthors}%
\unskip\
\newblock
\APACrefYearMonthDay{2019}{}{}.
\newblock
{\BBOQ}\APACrefatitle {Actionable Auditing: Investigating the Impact of
  Publicly Naming Biased Performance Results of Commercial {AI} Products}
  {Actionable auditing: Investigating the impact of publicly naming biased
  performance results of commercial {AI} products}.{\BBCQ}
\newblock
\BIn{} \APACrefbtitle {{AIES 2019: Proceedings of the 2019 AAAI/ACM Conference
  on AI, Ethics, and Society}} {{AIES 2019: Proceedings of the 2019 AAAI/ACM
  Conference on AI, Ethics, and Society}}\ (\BPGS\ 429--435).
\newblock
\begin{APACrefDOI} \doi{10.1145/3306618.3314244} \end{APACrefDOI}
\PrintBackRefs{\CurrentBib}

\bibitem [\protect \citeauthoryear {%
Raji%
, Kumar%
, Horowitz%
\BCBL {}\ \BBA {} Selbst%
}{%
Raji%
\ \protect \BOthers {.}}{%
{\protect \APACyear {2022}}%
}]{%
raji_fallacy_2022}
\APACinsertmetastar {%
raji_fallacy_2022}%
\begin{APACrefauthors}%
Raji, I\BPBI D.%
, Kumar, I\BPBI E.%
, Horowitz, A.%
\BCBL {}\ \BBA {} Selbst, A.%
\end{APACrefauthors}%
\unskip\
\newblock
\APACrefYearMonthDay{2022}{}{}.
\newblock
{\BBOQ}\APACrefatitle {The Fallacy of {AI} Functionality} {The fallacy of {AI}
  functionality}.{\BBCQ}
\newblock
\BIn{} \APACrefbtitle {{FAccT '22: Proceedings of the 2022 ACM Conference on
  Fairness, Accountability, and Transparency}} {{FAccT '22: Proceedings of the
  2022 ACM Conference on Fairness, Accountability, and Transparency}}\ (\BPGS\
  959--972).
\newblock
\begin{APACrefDOI} \doi{10.1145/3531146.3533158} \end{APACrefDOI}
\PrintBackRefs{\CurrentBib}

\bibitem [\protect \citeauthoryear {%
Raji%
\ \protect \BOthers {.}}{%
Raji%
\ \protect \BOthers {.}}{%
{\protect \APACyear {2020}}%
}]{%
raji_closing_2020}
\APACinsertmetastar {%
raji_closing_2020}%
\begin{APACrefauthors}%
Raji, I\BPBI D.%
, Smart, A.%
, White, R\BPBI N.%
, Mitchell, M.%
, Gebru, T.%
, Hutchinson, B.%
\BDBL {}Barnes, P.%
\end{APACrefauthors}%
\unskip\
\newblock
\APACrefYearMonthDay{2020}{}{}.
\newblock
{\BBOQ}\APACrefatitle {Closing the {AI} Accountability Gap: Defining an
  End-to-End Framework for Internal Algorithmic Auditing} {Closing the {AI}
  accountability gap: Defining an end-to-end framework for internal algorithmic
  auditing}.{\BBCQ}
\newblock
\BIn{} \APACrefbtitle {{FAT* '20: Proceedings of the 2020 Conference on
  Fairness, Accountability, and Transparency}} {{FAT* '20: Proceedings of the
  2020 Conference on Fairness, Accountability, and Transparency}}\ (\BPGS\
  33--44).
\newblock
\begin{APACrefDOI} \doi{10.1145/3351095.3372873} \end{APACrefDOI}
\PrintBackRefs{\CurrentBib}

\bibitem [\protect \citeauthoryear {%
RAND%
}{%
RAND%
}{%
{\protect \APACyear {{\protect \bibnodate {}}}}%
}]{%
rand_delphi_nodate}
\APACinsertmetastar {%
rand_delphi_nodate}%
\begin{APACrefauthors}%
RAND.%
\end{APACrefauthors}%
\unskip\
\newblock
\APACrefYearMonthDay{{\protect \bibnodate {}}}{}{}.
\newblock
\APACrefbtitle {Delphi {Method}.} {Delphi {Method}.}
\newblock
\begin{APACrefURL} \url{https://www.rand.org/topics/delphi-method.html}
  \end{APACrefURL}
\PrintBackRefs{\CurrentBib}

\bibitem [\protect \citeauthoryear {%
Rao%
, Vashistha%
, Naik%
, Aditya%
\BCBL {}\ \BBA {} Choudhury%
}{%
Rao%
\ \protect \BOthers {.}}{%
{\protect \APACyear {2023}}%
}]{%
rao_tricking_2023}
\APACinsertmetastar {%
rao_tricking_2023}%
\begin{APACrefauthors}%
Rao, A.%
, Vashistha, S.%
, Naik, A.%
, Aditya, S.%
\BCBL {}\ \BBA {} Choudhury, M.%
\end{APACrefauthors}%
\unskip\
\newblock
\APACrefYearMonthDay{2023}{}{}.
\newblock
{\BBOQ}\APACrefatitle {Tricking {LLMs} into Disobedience: Understanding,
  Analyzing, and Preventing Jailbreaks} {Tricking {LLMs} into disobedience:
  Understanding, analyzing, and preventing jailbreaks}.{\BBCQ}
\newblock
\APACjournalVolNumPages{arXiv preprint arXiv:2305.14965}{}{}{}.
\PrintBackRefs{\CurrentBib}

\bibitem [\protect \citeauthoryear {%
Rausand%
\ \BBA {} Haugen%
}{%
Rausand%
\ \BBA {} Haugen%
}{%
{\protect \APACyear {2020}}%
}]{%
rausand_risk_2020}
\APACinsertmetastar {%
rausand_risk_2020}%
\begin{APACrefauthors}%
Rausand, M.%
\BCBT {}\ \BBA {} Haugen, S.%
\end{APACrefauthors}%
\unskip\
\newblock
\APACrefYear{2020}.
\newblock
\APACrefbtitle {Risk Assessment: Theory, Methods, and Applications} {Risk
  assessment: Theory, methods, and applications}.
\newblock
\APACaddressPublisher{}{Wiley}.
\PrintBackRefs{\CurrentBib}

\bibitem [\protect \citeauthoryear {%
Reason%
}{%
Reason%
}{%
{\protect \APACyear {2000}}%
}]{%
reason_human_2000}
\APACinsertmetastar {%
reason_human_2000}%
\begin{APACrefauthors}%
Reason, J.%
\end{APACrefauthors}%
\unskip\
\newblock
\APACrefYearMonthDay{2000}{}{}.
\newblock
{\BBOQ}\APACrefatitle {Human error: models and management} {Human error: models
  and management}.{\BBCQ}
\newblock
\APACjournalVolNumPages{British Medical Journal}{320}{7237}{768--770}.
\PrintBackRefs{\CurrentBib}

\bibitem [\protect \citeauthoryear {%
Rees%
}{%
Rees%
}{%
{\protect \APACyear {2004}}%
}]{%
rees_our_2004}
\APACinsertmetastar {%
rees_our_2004}%
\begin{APACrefauthors}%
Rees, M\BPBI J.%
\end{APACrefauthors}%
\unskip\
\newblock
\APACrefYear{2004}.
\newblock
\APACrefbtitle {Our final hour: A scientist's warning: how terror, error, and
  environmental disaster threaten humankind's future in this century on earth
  and beyond} {Our final hour: A scientist's warning: how terror, error, and
  environmental disaster threaten humankind's future in this century on earth
  and beyond}.
\newblock
\APACaddressPublisher{}{Basic Books}.
\PrintBackRefs{\CurrentBib}

\bibitem [\protect \citeauthoryear {%
Ritchey%
, Lövkvist-Andersen%
, Olsson%
\BCBL {}\ \BBA {} Stenström%
}{%
Ritchey%
\ \protect \BOthers {.}}{%
{\protect \APACyear {2004}}%
}]{%
ritchey_modelling_2004}
\APACinsertmetastar {%
ritchey_modelling_2004}%
\begin{APACrefauthors}%
Ritchey, T.%
, Lövkvist-Andersen, A\BHBI L.%
, Olsson, R.%
\BCBL {}\ \BBA {} Stenström, M.%
\end{APACrefauthors}%
\unskip\
\newblock
\APACrefYearMonthDay{2004}{}{}.
\newblock
{\BBOQ}\APACrefatitle {Modelling Society's Capacity to Manage Extraordinary
  Events Developing a Generic Design Basis {(GDB)} Model for Extraordinary
  Societal Events using Computer-Aided Morphological Analysis} {Modelling
  society's capacity to manage extraordinary events developing a generic design
  basis {(GDB)} model for extraordinary societal events using computer-aided
  morphological analysis}.{\BBCQ}
\newblock
\BIn{} \APACrefbtitle {{Society for Risk Analysis Conference}.} {{Society for
  Risk Analysis Conference}.}
\PrintBackRefs{\CurrentBib}

\bibitem [\protect \citeauthoryear {%
Roli%
, Jaeger%
\BCBL {}\ \BBA {} Kauffman%
}{%
Roli%
\ \protect \BOthers {.}}{%
{\protect \APACyear {2022}}%
}]{%
roli_how_2022}
\APACinsertmetastar {%
roli_how_2022}%
\begin{APACrefauthors}%
Roli, A.%
, Jaeger, J.%
\BCBL {}\ \BBA {} Kauffman, S\BPBI A.%
\end{APACrefauthors}%
\unskip\
\newblock
\APACrefYearMonthDay{2022}{}{}.
\newblock
{\BBOQ}\APACrefatitle {How Organisms Come to Know the World: Fundamental Limits
  on Artificial General Intelligence} {How organisms come to know the world:
  Fundamental limits on artificial general intelligence}.{\BBCQ}
\newblock
\APACjournalVolNumPages{Frontiers in Ecology and Evolution}{9}{}{}.
\PrintBackRefs{\CurrentBib}

\bibitem [\protect \citeauthoryear {%
Rowe%
\ \BBA {} Wright%
}{%
Rowe%
\ \BBA {} Wright%
}{%
{\protect \APACyear {1999}}%
}]{%
rowe_delphi_1999}
\APACinsertmetastar {%
rowe_delphi_1999}%
\begin{APACrefauthors}%
Rowe, G.%
\BCBT {}\ \BBA {} Wright, G.%
\end{APACrefauthors}%
\unskip\
\newblock
\APACrefYearMonthDay{1999}{}{}.
\newblock
{\BBOQ}\APACrefatitle {The {Delphi} technique as a forecasting tool: issues and
  analysis} {The {Delphi} technique as a forecasting tool: issues and
  analysis}.{\BBCQ}
\newblock
\APACjournalVolNumPages{International Journal of Forecasting}{15}{4}{353--375}.
\newblock
\begin{APACrefDOI} \doi{10.1016/S0169-2070(99)00018-7} \end{APACrefDOI}
\PrintBackRefs{\CurrentBib}

\bibitem [\protect \citeauthoryear {%
Rozo%
, Lawler%
\BCBL {}\ \BBA {} Paragas%
}{%
Rozo%
\ \protect \BOthers {.}}{%
{\protect \APACyear {2017}}%
}]{%
wooley_viral_2017}
\APACinsertmetastar {%
wooley_viral_2017}%
\begin{APACrefauthors}%
Rozo, M.%
, Lawler, J.%
\BCBL {}\ \BBA {} Paragas, J.%
\end{APACrefauthors}%
\unskip\
\newblock
\APACrefYearMonthDay{2017}{}{}.
\newblock
{\BBOQ}\APACrefatitle {Viral Agents of Human Disease: Biosafety Concerns}
  {Viral agents of human disease: Biosafety concerns}.{\BBCQ}
\newblock
\BIn{} D\BPBI P.~Wooley\ \BBA {} K\BPBI B.~Byers\ (\BEDS), \APACrefbtitle
  {Biological safety: principles and practices} {Biological safety: principles
  and practices}\ (\BPGS\ 187--220).
\newblock
\APACaddressPublisher{}{ASM Press}.
\PrintBackRefs{\CurrentBib}

\bibitem [\protect \citeauthoryear {%
Russell%
}{%
Russell%
}{%
{\protect \APACyear {2019}}%
}]{%
russell_human_2019}
\APACinsertmetastar {%
russell_human_2019}%
\begin{APACrefauthors}%
Russell, S\BPBI J.%
\end{APACrefauthors}%
\unskip\
\newblock
\APACrefYear{2019}.
\newblock
\APACrefbtitle {Human compatible: Artificial intelligence and the problem of
  control} {Human compatible: Artificial intelligence and the problem of
  control}.
\newblock
\APACaddressPublisher{}{Viking}.
\PrintBackRefs{\CurrentBib}

\bibitem [\protect \citeauthoryear {%
Salmon%
\ \protect \BOthers {.}}{%
Salmon%
\ \protect \BOthers {.}}{%
{\protect \APACyear {2023}}%
}]{%
salmon_managing_2023}
\APACinsertmetastar {%
salmon_managing_2023}%
\begin{APACrefauthors}%
Salmon, P\BPBI M.%
, Baber, C.%
, Burns, C.%
, Carden, T.%
, Cooke, N.%
, Cummings, M.%
\BDBL {}Stanton, N\BPBI A.%
\end{APACrefauthors}%
\unskip\
\newblock
\APACrefYearMonthDay{2023}{}{}.
\newblock
{\BBOQ}\APACrefatitle {Managing the risks of artificial general intelligence: A
  human factors and ergonomics perspective} {Managing the risks of artificial
  general intelligence: A human factors and ergonomics perspective}.{\BBCQ}
\newblock
\APACjournalVolNumPages{Human Factors and Ergonomics in Manufacturing \&
  Service Industries}{}{}{1--13}.
\newblock
\begin{APACrefDOI} \doi{10.1002/hfm.20996} \end{APACrefDOI}
\PrintBackRefs{\CurrentBib}

\bibitem [\protect \citeauthoryear {%
Schuett%
}{%
Schuett%
}{%
{\protect \APACyear {2023}}%
}]{%
schuett_agi_2023}
\APACinsertmetastar {%
schuett_agi_2023}%
\begin{APACrefauthors}%
Schuett, J.%
\end{APACrefauthors}%
\unskip\
\newblock
\APACrefYearMonthDay{2023}{}{}.
\newblock
{\BBOQ}\APACrefatitle {{AGI} labs need an internal audit function} {{AGI} labs
  need an internal audit function}.{\BBCQ}
\newblock
\APACjournalVolNumPages{arXiv preprint arXiv:2305.17038}{}{}{}.
\PrintBackRefs{\CurrentBib}

\bibitem [\protect \citeauthoryear {%
Schuett%
\ \protect \BOthers {.}}{%
Schuett%
\ \protect \BOthers {.}}{%
{\protect \APACyear {2023}}%
}]{%
schuett_towards_2023}
\APACinsertmetastar {%
schuett_towards_2023}%
\begin{APACrefauthors}%
Schuett, J.%
, Dreksler, N.%
, Anderljung, M.%
, McCaffary, D.%
, Heim, L.%
, Bluemke, E.%
\BCBL {}\ \BBA {} Garfinkel, B.%
\end{APACrefauthors}%
\unskip\
\newblock
\APACrefYearMonthDay{2023}{}{}.
\newblock
{\BBOQ}\APACrefatitle {Towards best practices in {AGI} safety and governance: A
  survey of expert opinion} {Towards best practices in {AGI} safety and
  governance: A survey of expert opinion}.{\BBCQ}
\newblock
\APACjournalVolNumPages{arXiv preprint arXiv:2305.07153}{}{}{}.
\PrintBackRefs{\CurrentBib}

\bibitem [\protect \citeauthoryear {%
Schweizer%
}{%
Schweizer%
}{%
{\protect \APACyear {2020}}%
}]{%
schweizer_reflections_2020}
\APACinsertmetastar {%
schweizer_reflections_2020}%
\begin{APACrefauthors}%
Schweizer, V\BPBI J.%
\end{APACrefauthors}%
\unskip\
\newblock
\APACrefYearMonthDay{2020}{}{}.
\newblock
{\BBOQ}\APACrefatitle {Reflections on cross-impact balances, a systematic
  method constructing global socio-technical scenarios for climate change
  research} {Reflections on cross-impact balances, a systematic method
  constructing global socio-technical scenarios for climate change
  research}.{\BBCQ}
\newblock
\APACjournalVolNumPages{Climatic Change}{162}{4}{1705--1722}.
\newblock
\begin{APACrefDOI} \doi{10.1007/s10584-019-02615-2} \end{APACrefDOI}
\PrintBackRefs{\CurrentBib}

\bibitem [\protect \citeauthoryear {%
Searle%
}{%
Searle%
}{%
{\protect \APACyear {1980}}%
}]{%
searle_minds_1980}
\APACinsertmetastar {%
searle_minds_1980}%
\begin{APACrefauthors}%
Searle, J\BPBI R.%
\end{APACrefauthors}%
\unskip\
\newblock
\APACrefYearMonthDay{1980}{}{}.
\newblock
{\BBOQ}\APACrefatitle {Minds, brains, and programs} {Minds, brains, and
  programs}.{\BBCQ}
\newblock
\APACjournalVolNumPages{Behavioral and Brain Sciences}{3}{3}{417--424}.
\newblock
\begin{APACrefDOI} \doi{10.1017/S0140525X00005756} \end{APACrefDOI}
\PrintBackRefs{\CurrentBib}

\bibitem [\protect \citeauthoryear {%
Shah%
}{%
Shah%
}{%
{\protect \APACyear {2022}}%
}]{%
shah_ai_2022}
\APACinsertmetastar {%
shah_ai_2022}%
\begin{APACrefauthors}%
Shah, R.%
\end{APACrefauthors}%
\unskip\
\newblock
\APACrefYearMonthDay{2022}{}{}.
\newblock
\APACrefbtitle {{AI} Risk from Program Search.} {{AI} risk from program
  search.}
\newblock
\APAChowpublished {AI Alignment Forum}.
\newblock
\begin{APACrefURL}
  \url{https://www.alignmentforum.org/posts/wnnkD6P2k2TfHnNmt/threat-model-literature-review}
  \end{APACrefURL}
\PrintBackRefs{\CurrentBib}

\bibitem [\protect \citeauthoryear {%
Shelby%
\ \protect \BOthers {.}}{%
Shelby%
\ \protect \BOthers {.}}{%
{\protect \APACyear {2023}}%
}]{%
shelby_identifying_2023}
\APACinsertmetastar {%
shelby_identifying_2023}%
\begin{APACrefauthors}%
Shelby, R.%
, Rismani, S.%
, Henne, K.%
, Moon, A.%
, Rostamzadeh, N.%
, Nicholas, P.%
\BDBL {}Virk, G.%
\end{APACrefauthors}%
\unskip\
\newblock
\APACrefYearMonthDay{2023}{}{}.
\newblock
{\BBOQ}\APACrefatitle {Identifying Sociotechnical Harms of Algorithmic Systems:
  Scoping a Taxonomy for Harm Reduction} {Identifying sociotechnical harms of
  algorithmic systems: Scoping a taxonomy for harm reduction}.{\BBCQ}
\newblock
\APACjournalVolNumPages{arXiv preprint arXiv:2210.05791}{}{}{}.
\PrintBackRefs{\CurrentBib}

\bibitem [\protect \citeauthoryear {%
Shevlane%
\ \protect \BOthers {.}}{%
Shevlane%
\ \protect \BOthers {.}}{%
{\protect \APACyear {2023}}%
}]{%
shevlane_model_2023}
\APACinsertmetastar {%
shevlane_model_2023}%
\begin{APACrefauthors}%
Shevlane, T.%
, Farquhar, S.%
, Garfinkel, B.%
, Phuong, M.%
, Whittlestone, J.%
, Leung, J.%
\BDBL {}Dafoe, A.%
\end{APACrefauthors}%
\unskip\
\newblock
\APACrefYearMonthDay{2023}{}{}.
\newblock
{\BBOQ}\APACrefatitle {Model evaluation for extreme risks} {Model evaluation
  for extreme risks}.{\BBCQ}
\newblock
\APACjournalVolNumPages{arXiv preprint arXiv:2305.15324}{}{}{}.
\PrintBackRefs{\CurrentBib}

\bibitem [\protect \citeauthoryear {%
Shneiderman%
}{%
Shneiderman%
}{%
{\protect \APACyear {2020}}%
}]{%
shneiderman_bridging_2020}
\APACinsertmetastar {%
shneiderman_bridging_2020}%
\begin{APACrefauthors}%
Shneiderman, B.%
\end{APACrefauthors}%
\unskip\
\newblock
\APACrefYearMonthDay{2020}{}{}.
\newblock
{\BBOQ}\APACrefatitle {Bridging the Gap Between Ethics and Practice: Guidelines
  for Reliable, Safe, and Trustworthy Human-centered {AI} Systems} {Bridging
  the gap between ethics and practice: Guidelines for reliable, safe, and
  trustworthy human-centered {AI} systems}.{\BBCQ}
\newblock
\APACjournalVolNumPages{ACM Transactions on Interactive Intelligent
  Systems}{10}{4}{1--31}.
\newblock
\begin{APACrefDOI} \doi{10.1145/3419764} \end{APACrefDOI}
\PrintBackRefs{\CurrentBib}

\bibitem [\protect \citeauthoryear {%
Soares%
}{%
Soares%
}{%
{\protect \APACyear {2022}}%
}]{%
soares_central_2022}
\APACinsertmetastar {%
soares_central_2022}%
\begin{APACrefauthors}%
Soares, N.%
\end{APACrefauthors}%
\unskip\
\newblock
\APACrefYearMonthDay{2022}{}{}.
\newblock
\APACrefbtitle {A central {AI} alignment problem: capabilities generalization,
  and the sharp left turn.} {A central {AI} alignment problem: capabilities
  generalization, and the sharp left turn.}
\newblock
\APAChowpublished {AI Alignment Forum}.
\newblock
\begin{APACrefURL}
  \url{https://www.alignmentforum.org/posts/GNhMPAWcfBCASy8e6/a-central-ai-alignment-problem-capabilities-generalization}
  \end{APACrefURL}
\PrintBackRefs{\CurrentBib}

\bibitem [\protect \citeauthoryear {%
Stein-Perlman%
, Weinstein-Raun%
\BCBL {}\ \BBA {} Grace%
}{%
Stein-Perlman%
\ \protect \BOthers {.}}{%
{\protect \APACyear {2022}}%
}]{%
stein-perlman_2022_2022}
\APACinsertmetastar {%
stein-perlman_2022_2022}%
\begin{APACrefauthors}%
Stein-Perlman, Z.%
, Weinstein-Raun, B.%
\BCBL {}\ \BBA {} Grace, K.%
\end{APACrefauthors}%
\unskip\
\newblock
\APACrefYearMonthDay{2022}{}{}.
\newblock
\APACrefbtitle {2022 {Expert} Survey on Progress in {AI}.} {2022 {Expert}
  survey on progress in {AI}.}
\newblock
\APAChowpublished {AI Impacts}.
\newblock
\begin{APACrefURL}
  \url{https://aiimpacts.org/2022-expert-survey-on-progress-in-ai/}
  \end{APACrefURL}
\PrintBackRefs{\CurrentBib}

\bibitem [\protect \citeauthoryear {%
Stolzer%
, Sumwalt%
\BCBL {}\ \BBA {} Goglia%
}{%
Stolzer%
\ \protect \BOthers {.}}{%
{\protect \APACyear {2023}}%
}]{%
stolzer_safety_2023}
\APACinsertmetastar {%
stolzer_safety_2023}%
\begin{APACrefauthors}%
Stolzer, A\BPBI J.%
, Sumwalt, R\BPBI L.%
\BCBL {}\ \BBA {} Goglia, J\BPBI J.%
\end{APACrefauthors}%
\unskip\
\newblock
\APACrefYear{2023}.
\newblock
\APACrefbtitle {Safety management systems in aviation} {Safety management
  systems in aviation}.
\newblock
\APACaddressPublisher{}{CRC Press}.
\PrintBackRefs{\CurrentBib}

\bibitem [\protect \citeauthoryear {%
Suresh%
\ \BBA {} Guttag%
}{%
Suresh%
\ \BBA {} Guttag%
}{%
{\protect \APACyear {2021}}%
}]{%
suresh_framework_2021}
\APACinsertmetastar {%
suresh_framework_2021}%
\begin{APACrefauthors}%
Suresh, H.%
\BCBT {}\ \BBA {} Guttag, J\BPBI V.%
\end{APACrefauthors}%
\unskip\
\newblock
\APACrefYearMonthDay{2021}{}{}.
\newblock
{\BBOQ}\APACrefatitle {A Framework for Understanding Sources of Harm throughout
  the Machine Learning Life Cycle} {A framework for understanding sources of
  harm throughout the machine learning life cycle}.{\BBCQ}
\newblock
\APACjournalVolNumPages{arXiv preprint arXiv:1901.10002}{}{}{}.
\PrintBackRefs{\CurrentBib}

\bibitem [\protect \citeauthoryear {%
Taori%
\ \protect \BOthers {.}}{%
Taori%
\ \protect \BOthers {.}}{%
{\protect \APACyear {2023}}%
}]{%
taori_alpaca_2023}
\APACinsertmetastar {%
taori_alpaca_2023}%
\begin{APACrefauthors}%
Taori, R.%
, Gulrajani, I.%
, Zhang, T.%
, Dubois, Y.%
, Li, X.%
, Guestrin, C.%
\BDBL {}Hashimoto, T\BPBI B.%
\end{APACrefauthors}%
\unskip\
\newblock
\APACrefYearMonthDay{2023}{}{}.
\newblock
\APACrefbtitle {Alpaca: {A} {Strong}, {Replicable} {Instruction}-{Following}
  {Model}.} {Alpaca: {A} {Strong}, {Replicable} {Instruction}-{Following}
  {Model}.}
\newblock
\APAChowpublished {Stanford University}.
\newblock
\begin{APACrefURL} \url{https://crfm.stanford.edu/2023/03/13/alpaca.html}
  \end{APACrefURL}
\PrintBackRefs{\CurrentBib}

\bibitem [\protect \citeauthoryear {%
{The Vicuna Team}%
}{%
{The Vicuna Team}%
}{%
{\protect \APACyear {2023}}%
}]{%
the_vicuna_team_vicuna_2023}
\APACinsertmetastar {%
the_vicuna_team_vicuna_2023}%
\begin{APACrefauthors}%
{The Vicuna Team}.%
\end{APACrefauthors}%
\unskip\
\newblock
\APACrefYearMonthDay{2023}{}{}.
\newblock
\APACrefbtitle {Vicuna: An Open-Source Chatbot Impressing {GPT}-4 with 90\%
  {ChatGPT} Quality.} {Vicuna: An open-source chatbot impressing {GPT}-4 with
  90\% {ChatGPT} quality.}
\newblock
\APAChowpublished {LMSYS Org Blog}.
\newblock
\begin{APACrefURL} \url{https://lmsys.org/blog/2023-03-30-vicuna}
  \end{APACrefURL}
\PrintBackRefs{\CurrentBib}

\bibitem [\protect \citeauthoryear {%
{The White House}%
}{%
{The White House}%
}{%
{\protect \APACyear {2023}}%
}]{%
the_white_house_national_2023}
\APACinsertmetastar {%
the_white_house_national_2023}%
\begin{APACrefauthors}%
{The White House}.%
\end{APACrefauthors}%
\unskip\
\newblock
\APACrefYearMonthDay{2023}{}{}.
\newblock
\APACrefbtitle {National Artificial Intelligence Research and Development
  Strategic Plan 2023 Update.} {National artificial intelligence research and
  development strategic plan 2023 update.}
\newblock
\begin{APACrefURL}
  \url{https://www.whitehouse.gov/wp-content/uploads/2023/05/National-Artificial-Intelligence-Research-and-Development-Strategic-Plan-2023-Update.pdf}
  \end{APACrefURL}
\PrintBackRefs{\CurrentBib}

\bibitem [\protect \citeauthoryear {%
{UK Information Commissioner’s Office}%
}{%
{UK Information Commissioner’s Office}%
}{%
{\protect \APACyear {{\protect \bibnodate {}}}}%
}]{%
uk_information_commissioners_office_data_nodate}
\APACinsertmetastar {%
uk_information_commissioners_office_data_nodate}%
\begin{APACrefauthors}%
{UK Information Commissioner’s Office}.%
\end{APACrefauthors}%
\unskip\
\newblock
\APACrefYearMonthDay{{\protect \bibnodate {}}}{}{}.
\newblock
\APACrefbtitle {Data protection impact assessments.} {Data protection impact
  assessments.}
\newblock
\begin{APACrefURL}
  \url{https://ico.org.uk/for-organisations/uk-gdpr-guidance-and-resources/accountability-and-governance/guide-to-accountability-and-governance/accountability-and-governance/data-protection-impact-assessments/}
  \end{APACrefURL}
\PrintBackRefs{\CurrentBib}

\bibitem [\protect \citeauthoryear {%
{UK Royal Academy of Engineering}%
}{%
{UK Royal Academy of Engineering}%
}{%
{\protect \APACyear {2023}}%
}]{%
uk_royal_academy_of_engineering_building_2023}
\APACinsertmetastar {%
uk_royal_academy_of_engineering_building_2023}%
\begin{APACrefauthors}%
{UK Royal Academy of Engineering}.%
\end{APACrefauthors}%
\unskip\
\newblock
\APACrefYearMonthDay{2023}{}{}.
\newblock
\APACrefbtitle {Building resilience: lessons from the Academy's review of the
  National Security Risk Assessment methodology.} {Building resilience: lessons
  from the academy's review of the national security risk assessment
  methodology.}
\newblock
\begin{APACrefURL}
  \url{https://raeng.org.uk/policy-and-resources/engineering-policy/security-and-resilience/nsra}
  \end{APACrefURL}
\PrintBackRefs{\CurrentBib}

\bibitem [\protect \citeauthoryear {%
Urbina%
, Lentzos%
, Invernizzi%
\BCBL {}\ \BBA {} Ekins%
}{%
Urbina%
\ \protect \BOthers {.}}{%
{\protect \APACyear {2022}}%
}]{%
urbina_dual_2022}
\APACinsertmetastar {%
urbina_dual_2022}%
\begin{APACrefauthors}%
Urbina, F.%
, Lentzos, F.%
, Invernizzi, C.%
\BCBL {}\ \BBA {} Ekins, S.%
\end{APACrefauthors}%
\unskip\
\newblock
\APACrefYearMonthDay{2022}{}{}.
\newblock
{\BBOQ}\APACrefatitle {Dual use of artificial-intelligence-powered drug
  discovery} {Dual use of artificial-intelligence-powered drug
  discovery}.{\BBCQ}
\newblock
\APACjournalVolNumPages{Nature Machine Intelligence}{4}{3}{189--191}.
\newblock
\begin{APACrefDOI} \doi{10.1038/s42256-022-00465-9} \end{APACrefDOI}
\PrintBackRefs{\CurrentBib}

\bibitem [\protect \citeauthoryear {%
{US Environmental Protection Agency}%
}{%
{US Environmental Protection Agency}%
}{%
{\protect \APACyear {1998}}%
}]{%
us_environmental_protection_agency_guidelines_1998}
\APACinsertmetastar {%
us_environmental_protection_agency_guidelines_1998}%
\begin{APACrefauthors}%
{US Environmental Protection Agency}.%
\end{APACrefauthors}%
\unskip\
\newblock
\APACrefYearMonthDay{1998}{}{}.
\newblock
\APACrefbtitle {Guidelines for Ecological Risk Assessment.} {Guidelines for
  ecological risk assessment.}
\newblock
\begin{APACrefURL}
  \url{https://www.epa.gov/risk/guidelines-ecological-risk-assessment}
  \end{APACrefURL}
\PrintBackRefs{\CurrentBib}

\bibitem [\protect \citeauthoryear {%
van~der Heijden%
}{%
van~der Heijden%
}{%
{\protect \APACyear {2005}}%
}]{%
van_der_heijden_scenarios_2005}
\APACinsertmetastar {%
van_der_heijden_scenarios_2005}%
\begin{APACrefauthors}%
van~der Heijden, K.%
\end{APACrefauthors}%
\unskip\
\newblock
\APACrefYear{2005}.
\newblock
\APACrefbtitle {Scenarios: The art of strategic conversation} {Scenarios: The
  art of strategic conversation}.
\newblock
\APACaddressPublisher{}{Wiley}.
\PrintBackRefs{\CurrentBib}

\bibitem [\protect \citeauthoryear {%
Weidinger%
\ \protect \BOthers {.}}{%
Weidinger%
\ \protect \BOthers {.}}{%
{\protect \APACyear {2022}}%
}]{%
weidinger_taxonomy_2022}
\APACinsertmetastar {%
weidinger_taxonomy_2022}%
\begin{APACrefauthors}%
Weidinger, L.%
, Uesato, J.%
, Rauh, M.%
, Griffin, C.%
, Huang, P\BHBI S.%
, Mellor, J.%
\BDBL {}Gabriel, I.%
\end{APACrefauthors}%
\unskip\
\newblock
\APACrefYearMonthDay{2022}{}{}.
\newblock
{\BBOQ}\APACrefatitle {Taxonomy of Risks posed by Language Models} {Taxonomy of
  risks posed by language models}.{\BBCQ}
\newblock
\BIn{} \APACrefbtitle {{FAccT '22: Proceedings of the 2022 ACM Conference on
  Fairness, Accountability, and Transparency}} {{FAccT '22: Proceedings of the
  2022 ACM Conference on Fairness, Accountability, and Transparency}}\ (\BPGS\
  214--229).
\newblock
\begin{APACrefDOI} \doi{10.1145/3531146.3533088} \end{APACrefDOI}
\PrintBackRefs{\CurrentBib}

\bibitem [\protect \citeauthoryear {%
Westerlund%
}{%
Westerlund%
}{%
{\protect \APACyear {2019}}%
}]{%
westerlund_emergence_2019}
\APACinsertmetastar {%
westerlund_emergence_2019}%
\begin{APACrefauthors}%
Westerlund, M.%
\end{APACrefauthors}%
\unskip\
\newblock
\APACrefYearMonthDay{2019}{}{}.
\newblock
{\BBOQ}\APACrefatitle {The Emergence of Deepfake Technology: A Review} {The
  emergence of deepfake technology: A review}.{\BBCQ}
\newblock
\APACjournalVolNumPages{Technology Innovation Management
  Review}{9}{11}{39--52}.
\newblock
\begin{APACrefDOI} \doi{10.22215/timreview/1282} \end{APACrefDOI}
\PrintBackRefs{\CurrentBib}

\bibitem [\protect \citeauthoryear {%
Wiggers%
}{%
Wiggers%
}{%
{\protect \APACyear {2023}}%
}]{%
wiggers_anthropic_2023}
\APACinsertmetastar {%
wiggers_anthropic_2023}%
\begin{APACrefauthors}%
Wiggers, K.%
\end{APACrefauthors}%
\unskip\
\newblock
\APACrefYearMonthDay{2023}{}{}.
\newblock
\APACrefbtitle {Anthropic raises \${450M} to build next-gen {AI} assistants.}
  {Anthropic raises \${450M} to build next-gen {AI} assistants.}
\newblock
\APAChowpublished {TechCrunch}.
\newblock
\begin{APACrefURL}
  \url{https://techcrunch.com/2023/05/23/anthropic-raises-350m-to-build-next-gen-ai-assistants/}
  \end{APACrefURL}
\PrintBackRefs{\CurrentBib}

\bibitem [\protect \citeauthoryear {%
Wikipedia%
}{%
Wikipedia%
}{%
{\protect \APACyear {{\protect \bibnodate {}}}}%
}]{%
wikipedia_list_nodate}
\APACinsertmetastar {%
wikipedia_list_nodate}%
\begin{APACrefauthors}%
Wikipedia.%
\end{APACrefauthors}%
\unskip\
\newblock
\APACrefYearMonthDay{{\protect \bibnodate {}}}{}{}.
\newblock
\APACrefbtitle {List of causal mapping software.} {List of causal mapping
  software.}
\newblock
\begin{APACrefURL}
  \url{https://en.wikipedia.org/wiki/List_of_causal_mapping_software}
  \end{APACrefURL}
\PrintBackRefs{\CurrentBib}

\bibitem [\protect \citeauthoryear {%
Xia%
\ \protect \BOthers {.}}{%
Xia%
\ \protect \BOthers {.}}{%
{\protect \APACyear {2023}}%
}]{%
xia_towards_2023}
\APACinsertmetastar {%
xia_towards_2023}%
\begin{APACrefauthors}%
Xia, B.%
, Lu, Q.%
, Perera, H.%
, Zhu, L.%
, Xing, Z.%
, Liu, Y.%
\BCBL {}\ \BBA {} Whittle, J.%
\end{APACrefauthors}%
\unskip\
\newblock
\APACrefYearMonthDay{2023}{}{}.
\newblock
{\BBOQ}\APACrefatitle {Towards Concrete and Connected {AI} Risk Assessment
  ({C2AIRA}): A Systematic Mapping Study} {Towards concrete and connected {AI}
  risk assessment ({C2AIRA}): A systematic mapping study}.{\BBCQ}
\newblock
\APACjournalVolNumPages{arXiv preprint arXiv:2301.11616}{}{}{}.
\PrintBackRefs{\CurrentBib}

\bibitem [\protect \citeauthoryear {%
Yampolskiy%
}{%
Yampolskiy%
}{%
{\protect \APACyear {2015}}%
}]{%
yampolskiy_taxonomy_2015}
\APACinsertmetastar {%
yampolskiy_taxonomy_2015}%
\begin{APACrefauthors}%
Yampolskiy, R\BPBI V.%
\end{APACrefauthors}%
\unskip\
\newblock
\APACrefYearMonthDay{2015}{}{}.
\newblock
{\BBOQ}\APACrefatitle {Taxonomy of Pathways to Dangerous {AI}} {Taxonomy of
  pathways to dangerous {AI}}.{\BBCQ}
\newblock
\APACjournalVolNumPages{arXiv preprint arXiv:1511.03246}{}{}{}.
\PrintBackRefs{\CurrentBib}

\bibitem [\protect \citeauthoryear {%
Yudkowsky%
}{%
Yudkowsky%
}{%
{\protect \APACyear {2008}}%
}]{%
bostrom_artificial_2008}
\APACinsertmetastar {%
bostrom_artificial_2008}%
\begin{APACrefauthors}%
Yudkowsky, E.%
\end{APACrefauthors}%
\unskip\
\newblock
\APACrefYearMonthDay{2008}{}{}.
\newblock
{\BBOQ}\APACrefatitle {Artificial intelligence as a positive and negative
  factor in global risk} {Artificial intelligence as a positive and negative
  factor in global risk}.{\BBCQ}
\newblock
\BIn{} N.~Bostrom\ \BBA {} M\BPBI M.~Ćirković\ (\BEDS), \APACrefbtitle
  {Global Catastrophic Risks} {Global catastrophic risks}\ (\BPGS\ 308--345).
\newblock
\APACaddressPublisher{}{Oxford University Press}.
\PrintBackRefs{\CurrentBib}

\bibitem [\protect \citeauthoryear {%
Yudkowsky%
}{%
Yudkowsky%
}{%
{\protect \APACyear {2023}}%
}]{%
yudkowsky_pausing_2023}
\APACinsertmetastar {%
yudkowsky_pausing_2023}%
\begin{APACrefauthors}%
Yudkowsky, E.%
\end{APACrefauthors}%
\unskip\
\newblock
\APACrefYearMonthDay{2023}{}{}.
\newblock
\APACrefbtitle {Pausing {AI} Developments Isn't Enough. We Need to Shut it All
  Down.} {Pausing {AI} developments isn't enough. we need to shut it all down.}
\newblock
\APAChowpublished {Time}.
\newblock
\begin{APACrefURL}
  \url{https://time.com/6266923/ai-eliezer-yudkowsky-open-letter-not-enough/}
  \end{APACrefURL}
\PrintBackRefs{\CurrentBib}

\bibitem [\protect \citeauthoryear {%
Zendel%
, Murschitz%
, Humenberger%
\BCBL {}\ \BBA {} Herzner%
}{%
Zendel%
\ \protect \BOthers {.}}{%
{\protect \APACyear {2015}}%
}]{%
zendel_cv-hazop_2015}
\APACinsertmetastar {%
zendel_cv-hazop_2015}%
\begin{APACrefauthors}%
Zendel, O.%
, Murschitz, M.%
, Humenberger, M.%
\BCBL {}\ \BBA {} Herzner, W.%
\end{APACrefauthors}%
\unskip\
\newblock
\APACrefYearMonthDay{2015}{}{}.
\newblock
{\BBOQ}\APACrefatitle {{CV}-{HAZOP}: Introducing Test Data Validation for
  Computer Vision} {{CV}-{HAZOP}: Introducing test data validation for computer
  vision}.{\BBCQ}
\newblock
\BIn{} \APACrefbtitle {{Proceedings of the 2015 IEEE International Conference
  on Computer Vision (ICCV)}} {{Proceedings of the 2015 IEEE International
  Conference on Computer Vision (ICCV)}}\ (\BPGS\ 2066--2074).
\newblock
\begin{APACrefDOI} \doi{10.1109/ICCV.2015.239} \end{APACrefDOI}
\PrintBackRefs{\CurrentBib}

\bibitem [\protect \citeauthoryear {%
Zhang%
\ \protect \BOthers {.}}{%
Zhang%
\ \protect \BOthers {.}}{%
{\protect \APACyear {2022}}%
}]{%
zhang_forecasting_2022}
\APACinsertmetastar {%
zhang_forecasting_2022}%
\begin{APACrefauthors}%
Zhang, B.%
, Dreksler, N.%
, Anderljung, M.%
, Kahn, L.%
, Giattino, C.%
, Dafoe, A.%
\BCBL {}\ \BBA {} Horowitz, M\BPBI C.%
\end{APACrefauthors}%
\unskip\
\newblock
\APACrefYearMonthDay{2022}{}{}.
\newblock
{\BBOQ}\APACrefatitle {Forecasting {AI} Progress: Evidence from a Survey of
  Machine Learning Researchers} {Forecasting {AI} progress: Evidence from a
  survey of machine learning researchers}.{\BBCQ}
\newblock
\APACjournalVolNumPages{arXiv preprint arXiv:2206.04132}{}{}{}.
\PrintBackRefs{\CurrentBib}

\bibitem [\protect \citeauthoryear {%
Zwetsloot%
\ \BBA {} Dafoe%
}{%
Zwetsloot%
\ \BBA {} Dafoe%
}{%
{\protect \APACyear {2019}}%
}]{%
zwetsloot_thinking_2019}
\APACinsertmetastar {%
zwetsloot_thinking_2019}%
\begin{APACrefauthors}%
Zwetsloot, R.%
\BCBT {}\ \BBA {} Dafoe, A.%
\end{APACrefauthors}%
\unskip\
\newblock
\APACrefYearMonthDay{2019}{}{}.
\newblock
\APACrefbtitle {Thinking About Risks From {AI}: Accidents, Misuse and
  Structure.} {Thinking about risks from {AI}: Accidents, misuse and
  structure.}
\newblock
\APAChowpublished {Lawfare}.
\newblock
\begin{APACrefURL} \url{https://perma.cc/H3CQ-SEQ9} \end{APACrefURL}
\PrintBackRefs{\CurrentBib}

\end{thebibliography}

\newpage
\section*{Appendix: Other popular risk assessment techniques}\label{Appendix}

\begin{table}[ht!]
  \centering
  \small
  \begin{tabularx}{\textwidth}{>{\raggedright}p{0.2\linewidth}>{\raggedright}p{0.4\linewidth}>{\raggedright\arraybackslash}p{0.3\linewidth}}
    \toprule
        \textbf{Technique} & \textbf{Explanation} & \textbf{Reason(s) for excluding} \\
    \midrule
        Brainstorming & Brainstorming is a group exercise to generate and explore ideas. It works best when facilitated to ensure stimulation, avoid common fallacies (digressions, group-think, some participants not speaking up), and make sure that ideas are captured. Brainstorming can be used on its own or within other techniques. & Included in many other techniques \\
        Interviews & In structured interviews, interviewees are asked a set of prepared questions. In unstructured interviews, there is freedom to explore issues which arise. Semi-structured interviews are a mix of both pure forms. Interviews can be used on their own or within other techniques. & Similar to surveys, but often more time-consuming \\
        Surveys & Surveys are questionnaires that provide a way to get responses from a large number of people. Surveys can be used on their own or within other techniques. Note that surveys and interviews are overlapping and not necessarily distinct. & Included in many other techniques which gather expert opinions \\
        Nominal group technique & The Nominal group technique is a procedure to collect, explore, and decide on ideas. Participants answer questions independently and then each present their ideas. Next, participants discuss and aggregate these ideas, and finally vote on the ideas anonymously. & Somewhat similar to Delphi technique, but not fully anonymous and most useful for making decisions about risks (which already is part of risk treatment) \\
        Cindynic approach & The Cindynic approach involves semi-structured interviews that are conducted with different stakeholders. Their answers are put together in a matrix. This helps to identify dissonances, inconsistencies, ambiguities, omissions, and ignorances between stakeholders. & Somewhat similar to Delphi technique, but much less established and does not comprise exchange of reasoning between stakeholders \\
    \bottomrule
  \end{tabularx}
  \vspace{0.5em} \caption{Techniques for eliciting views from experts and stakeholders} \label{table5}
\end{table}
\newpage

\begin{table}[ht!]
  \centering
  \small
  \begin{tabularx}{\textwidth}{>{\raggedright}p{0.2\linewidth}>{\raggedright}p{0.4\linewidth}>{\raggedright\arraybackslash}p{0.3\linewidth}}
    \toprule
        \textbf{Technique} & \textbf{Explanation} & \textbf{Reason(s) for excluding} \\
    \midrule
        Hazard identification (HAZID)	 & HAZID is a high-level technique. It involves conducting a workshop to identify risks, for example, through brainstorming (see above). & Needs to be concretized through other techniques (e.g. brainstorming) \\
        Preliminary hazard analysis (PHA) & PHA involves a preliminary collection of hazards through analyzing historical data, gathering expert forecasts, and brainstorming. & Relatively superficial; does not provide a very specific or structured approach \\
        Failure modes and effects analysis (FMEA) and failure modes, effects and criticality analysis (FMECA) & A system or process is divided into elements, for each of which failure modes, their causes and effects (FMEA), and optionally their criticality (FMECA) are identified through brainstorming. If quantified, a risk priority number can be calculated. All information is captured in a worksheet. & Traditional hardware reliability technique; may be too simplistic for catastrophic risks from AI which involve complex interactions between various events (techniques like causal mapping or STPA may be more adequate); overlaps with fishbone method and bow tie analysis, which additionally provide visualizations \\
        Hazard and operability (HAZOP) & Within a facilitated workshop, a system or process is divided into elements, for each of which possible deviations from design intent, their causes and effects are identified. The facilitator prompts the participants with a set of guide words like “no”, “less”, etc. All information is captured in a worksheet. & Traditional hardware reliability technique; may be too simplistic for catastrophic risks from AI which involve complex interactions between various events (techniques like causal mapping or STPA may be more adequate); overlaps with fishbone method and bow tie analysis, which provide more freedom (no predetermined set of guide words) and visualizations \\
        What-if analysis & Within a facilitated workshop, a system or process is examined for potential risks and risk sources. The facilitator prompts the participants with “what if?” questions. & Implicitly performed in fishbone method, which provides more structure (through categories of causes) as well as a visualization \\
        Structured what-if technique (SWIFT) & Within a facilitated workshop, a system or process is examined for potential risks. The facilitator prompts the participants with combinations of “what if?” and a set of guide words on timing, amount, etc. SWIFT is relatively little time consuming. It can be used on its own or within other techniques. & Similar to fishbone method, which provides more freedom (no predetermined set of guide words) as well as a visualization \\
        Business impact analysis (BIA) & BIA is a method to identify critical business processes and functions. It aims to enable appropriate planning for disruptive events. BIA is undertaken using surveys, interviews, workshops or a combination of all of these. & Limited to financial risks (does not include risks to the organization’s environment for their own sake) \\
        Process mapping & Process mapping involves drawing a graphical representation of the steps involved in industrial or other processes and identifying the associated risks. & Relatively superficial; does not provide a very specific approach for identifying risks once the process has been mapped \\
  \end{tabularx}
\end{table}
\newpage

\begin{table}[ht!]
  \centering
  \small
  \begin{tabularx}{\textwidth}{>{\raggedright}p{0.2\linewidth}>{\raggedright}p{0.4\linewidth}>{\raggedright\arraybackslash}p{0.3\linewidth}}
        Job hazard analysis (JHA) / job safety analysis (JSA) & JHA/JSA are conducted before conducting a specific task. The task is broken down into its smallest sub-tasks, for each of which potential risks are brainstormed. & Best for routine safety-critical tasks or individual but not very complex safety-critical tasks (such as evals – the technique is usually not used for these kinds of tasks because it would be extremely comprehensive and might not add much as safety is already paramount when designing them) \\
        Change analysis & Change analysis aims to identify risks by comparing a known system with an unknown system. It can be done when considering or observing changes. & Implicit within risk assessment (generally, risks should be assessed again whenever changes are considered or observed) \\
        Gap analysis & Similar to change analysis, gap analysis compares two states of a system. However, it starts from an ideal system and aims to identify shortcomings in the current system. & Not very helpful if it is unclear what the ideal system would look like \\
        Root cause analysis (RCA) & RCA is a high-level technique. It involves investigating a risk for its ultimate sources, especially after an incident has occurred. This is done, for example, through what-if analysis (see above). & Needs to be concretized through other techniques (e.g. what-if analysis) \\
    \bottomrule
  \end{tabularx}
  \vspace{0.5em} \caption{Techniques for identifying risks} \label{table6}
\end{table}
\newpage

\begin{table}[ht!]
  \centering
  \small
  \begin{tabularx}{\textwidth}{>{\raggedright}p{0.2\linewidth}>{\raggedright}p{0.4\linewidth}>{\raggedright\arraybackslash}p{0.3\linewidth}}
    \toprule
        \textbf{Technique} & \textbf{Explanation} & \textbf{Reason(s) for excluding} \\
    \midrule
        Bayesian networks & Bayesian networks are graphical representations of events and include their probability based on other events and the probability of those. There is readily available software to build Bayesian networks. & Similar to cross-impact analysis (cross-impact matrices and interdependency maps), which takes into account more complex interactions between events \\
        Influence diagrams & Bayesian networks are called influence diagrams when they include actions and uncertainties. There is readily available software to build influence diagrams. & Similar to causal mapping and parts of cross-impact analysis (cross-impact matrices and interdependency maps), which take into account more complex interactions between events \\
        Event tree analysis (ETA) & ETA involves drawing a graphical representation of the sequence of consequences resulting from a given initiating event (in binary terms of their failure / success). It is developed using forward reasoning to identify these consequences. ETA can be quantified to provide the probabilities of the identified possible outcomes. & Traditional hardware reliability technique; may be too simplistic for catastrophic risks from AI which involve complex interactions between non-binary events (techniques like causal mapping or STPA may be more adequate); overlaps with several other techniques (event trees can help to develop scenarios and constitute the right side of bow tie diagrams); see also Section \ref{3} \\
        Fault tree analysis (FTA) & FTA involves drawing a graphical representation of the sequence of causes leading to a given undesired event, making use of Boolean logic (such as AND / OR). It is developed using backwards reasoning to identify these causes. FTA can be quantified to provide the probabilities of the identified possible failures. & Traditional hardware reliability technique; may be too simplistic for catastrophic risks from AI which involve complex interactions between non-binary events (techniques like causal mapping or STPA may be more adequate); overlaps with several other techniques (fault trees are implicit in fishbone diagrams and constitute the left side of bow tie diagrams); see also Section \ref{3} \\
        Cause-consequence analysis (CCA) & CCA is a combination of ETA and FTA. & Same reasoning as for ETA and FTA \\
        Master logic diagram (MLD) & MLD is similar to FTA (see above), but not limited to binary failure/success events. It can also be understood to be a summary FTA. & Same reasoning as for FTA (see above) \\
        Human reliability analysis (HRA) & HRA is a family of techniques that aim to establish the likelihood of human error within a system. It involves identifying the steps and substeps of an activity, estimating the probability of human error and identifying the factors influencing this probability. & Best for routine safety-critical tasks, for which data on human reliability exists \\
        Markov analysis & Markov analysis can be made of any system that has discrete, independent states. It uses the probabilities of the transitions between these states to estimate the long-run probability of the system being in a specified state, the expected time before its first failure and the expected time before its return to a specified state. & May be too simplistic for catastrophic risks from AI which involve complex interactions between various events (techniques like causal mapping or STPA may be more adequate) \\
  \end{tabularx}
\end{table}
\newpage

\begin{table}[ht!]
  \centering
  \small
  \begin{tabularx}{\textwidth}{>{\raggedright}p{0.2\linewidth}>{\raggedright}p{0.4\linewidth}>{\raggedright\arraybackslash}p{0.3\linewidth}}
        Monte Carlo simulations & Monte Carlo simulations are a stochastic method that use random sample values to provide a probability distribution of in principle deterministic outcomes. They can be used on their own or within other techniques. Monte Carlo simulations tend to de-emphasize high consequence/low probability risks. & Neglects low-probability risks; included in cross-impact analysis \\
        Value at risk (VaR) & VaR is a measure that indicates the amount of possible loss of financial assets under normal market conditions over a specific time period. It can be developed using Monte Carlo simulations, historical simulations, analytical methods or a combination of all of these. VaR calculations for the tails are often unstable. & Limited to financial risks (does not include risks to the organization’s environment for their own sake); neglects low-probability risks \\
        Conditional value at risk (CVaR) or expected shortfall (ES) & In response to the issues of VaR calculations for the tails, CVaR/ES is the VaR that focuses on the outcomes generating the greatest loss. & Limited to financial risks (does not include risks to the organization’s environment for their own sake) \\
    \bottomrule
  \end{tabularx}
  \vspace{0.5em} \caption{Techniques for analyzing causes, consequences, and likelihood of risks} \label{table7}
\end{table}
\newpage

\begin{table}[ht!]
  \centering
  \small
  \begin{tabularx}{\textwidth}{>{\raggedright}p{0.2\linewidth}>{\raggedright}p{0.4\linewidth}>{\raggedright\arraybackslash}p{0.3\linewidth}}
    \toprule
        \textbf{Technique} & \textbf{Explanation} & \textbf{Reason(s) for excluding} \\
    \midrule
        Hazard analysis and critical control points (HACCP) & HACCP is a structure for identifying sources of risks and putting controls in place at all relevant parts of a process to protect against them. After identifying hazards, influencing factors, and possible preventive measures, the points in the process where monitoring and control are possible are determined. Then, a system of critical limits and corrective actions is established. & May be too simplistic for catastrophic risks from AI which involve complex interactions between various events (techniques like causal mapping or STPA may be more adequate); already constitutes transition into risk treatment \\
        Layers of protection analysis (LOPA) & LOPA is a simplified version of ETA (see above), namely a single cause-consequence pair. It involves analyzing the reduction in risk that is achieved by controls by identifying independent protection layers and estimating their individual and the overall probability of failure. & Same reasoning as for ETA (see above) \\
    \bottomrule
  \end{tabularx}
  \vspace{0.5em} \caption{Techniques for analyzing controls} \label{table8}
\end{table}
\newpage

\begin{table}[ht!]
  \centering
  \small
  \begin{tabularx}{\textwidth}{>{\raggedright}p{0.2\linewidth}>{\raggedright}p{0.4\linewidth}>{\raggedright\arraybackslash}p{0.3\linewidth}}
    \toprule
        \textbf{Technique} & \textbf{Explanation} & \textbf{Reason(s) for excluding} \\
    \midrule
        Risk indices & Risk indices are indices that allow comparisons of different risks. Factors which are believed to influence the magnitude of risk are identified, scored and combined. In the simplest formulations, factors that increase the level of risk are multiplied together and divided by those that decrease the level of risk. & Similar to risk matrices, which are more common \\
        As low as reasonably practicable (ALARP) and so far as is reasonably practicable (SFAIRP) & ALARP/SFAIRP involve establishing criteria and categories accounting for both the acceptability of risks and the practicability to reduce them. These help to decide whether it is reasonably practicable to further reduce risk. & Similar to risk matrices, which are more common \\
        Frequency-number (F-N) diagrams & F-N diagrams are a special case of a risk matrix, namely one that focuses on fatalities. They are a graphical representation with x-axis as the cumulative number of fatalities and the y-axis as their frequency. & Similar to risk matrices, which are more comprehensive \\
        S-curves & S-curves are a graphical display of the severity of consequences of a risk as a probability distribution function (PDF) or cumulative distribution function (CDF). They allow representing the significance of a risk where there is a distribution of consequences. & Neglects low-probability risks \\
        Reliability centered maintenance (RCM) & RCM is a version of FMECA (see above) that focuses on situations where potential failures can be eliminated or reduced in frequency and/or consequence through maintenance of equipment. It enables decisions to be made based on the significance of risk. RCM spans all three risk assessment steps. & Only for hardware reliability \\
    \bottomrule
  \end{tabularx}
  \vspace{0.5em} \caption{Techniques for evaluating risks} \label{table9}
\end{table}

\end{document}